\documentclass[pre,superscriptadress,showpacs,twocolumn]{revtex4}
\usepackage{natbib}
\usepackage{epsfig,amsmath,amsfonts}
\usepackage{graphicx}

\begin{document}

%
\newcommand{\fig}[2]{\epsfxsize=#1\epsfbox{#2}}
%
%
%
\pagestyle{myheadings}
%
%
%
%
%


%

\newcommand{\ArcTanh}
{\operatorname{ArcTanh}}
\newcommand{\ArcTan}
{\operatorname{ArcTan}}
\newcommand{\ArcCoth}
{\operatorname{ArcCoth}}
\newcommand{\Erf}
{\operatorname{Erf}}
\newcommand{\Erfi}
{\operatorname{Erfi}}

\title{Exact multilocal renormalization on the effective action : application
to the random sine Gordon model statics and non-equilibrium dynamics}

\author{Gregory Schehr and Pierre Le Doussal}
\address{CNRS-Laboratoire de Physique Th\'eorique de l'Ecole Normale Sup\'erieure, 24 rue Lhomond, F-75231
Paris}


\begin{abstract}
We extend the exact multilocal renormalization group (RG) method 
to study the flow of the effective action functional. This important
physical quantity satisfies an exact RG equation which is then
expanded in multilocal components. Integrating the nonlocal
parts yields a closed exact RG equation for the local part, to a given order
in the local part. The method is illustrated on the $O(N)$ model by
straightforwardly recovering the $\eta$ exponent and scaling functions.
Then it is applied to study the glass phase of the Cardy-Ostlund, random phase sine
Gordon model near the glass transition temperature.
The static correlations and equilibrium dynamical exponent $z$ are recovered
and several new results are obtained. The equilibrium
two-point scaling functions are obtained. The nonequilibrium,
finite momentum, two-time $t,t'$ response and correlations are computed. 
They are shown to exhibit scaling forms, characterized by
novel exponents $\lambda_R \neq \lambda_C$, as well
as universal scaling functions that we compute. The
fluctuation dissipation ratio is found to be non trivial
and of the form $X(q^z (t-t'), t/t')$. Analogies and
differences with pure critical models are discussed.

\end{abstract}
\maketitle

\section{Introduction}

Recently a method was devised, the exact multilocal renormalization group
(EMRG) \cite{chauvepld}, to obtain perturbative renormalization
group equations from first principles, in a controlled way to any order, and
for arbitrary smooth cutoff function. It starts, as numerous previous exact RG
studies
\cite{wegner_houghton,morijmp,revue,GRN,NCS1,morrisderiv,hughes_liu,
hubbard,shukla,rudnick,tissier},
from
the exact Polchinski-Wilson renormalization group equation
\cite{polchinski,wilson_kogut} for the action functional 
${\cal S}(\phi)$. The next step however consists in splitting
it onto local and higher multilocal components \cite{scheidl}, and integrating
exactly all multilocal components in terms of the local part. This yields
an exact and very general RG flow equation for the local part of the action,
i.e. a function, expressed in an expansion in powers of the local part.

The aim of this paper is first to develop a similar method using instead the
effective action functional $\Gamma(\phi)$. This is needed because
$\Gamma(\phi)$ is a very important physical object, both as the generating
function of proper vertices, and related to the probability distribution of
an arbitrary macroscopic mode $\phi_q$ \cite{zinn-justin}.
A multilocal expansion is also performed and yields a RG equation again
in terms of the local part. The major advantage compared to the previous method
\cite{chauvepld}
is that one actually follows directly physical observables
and that correlations are immediately obtained
(while in the previous method one had to use a second formula
to compute correlations from the flowing action). The price to pay is
a slightly more involved RG equation, but this inconvenience arises only
at higher orders. As a simple check the $\eta$ exponent of the
$O(N)$ model will be recovered to lowest order.

A motivation to develop such EMRG methods comes from disordered
models. The physics of these being more complex than standard field
theories for pure systems, it is useful to be able to control the RG
procedure. This is
crucial for instance in the functional RG (FRG) which describes pinned
elastic manifolds  \cite{fisher_functional_rg,balents_fisher,giamarchi_vortex_long},
relevant for e.g. superconductors and density waves
\cite{blatter-vortex-review,nattermann_scheidl_revue,giam_vortex_revue,
giam_vortex_revue2,gruner-revue-cdw}, and the EMRG has been applied to study
that problem \cite{chauvepld,scheidl}.
 Here, and this is the second aim of this paper, we
will study another instance of a glass phase, arising in the random
phase sine Gordon model excluding vortices, as discovered by Cardy
and Ostlund \cite{cardy-desordre-rg}.
This model has been studied extensively, in its statics
\cite{villain-cosine-realrg,toner-log-2,giamarchi_vortex_long,hwa_fisher,
ludwig_co,rsb,num} 
and its dynamics 
\cite{goldschmidt-dynamics-co,tsaishapir},
as one of
the simplest but non trivial example of a topologically ordered glass, a
continuation to two dimension \cite{carpentier,nattermann_scheidl_revue}
of the fixed point describing the Bragg glass phase in
three dimension \cite{giamarchi_vortex_long}.
We first show that the present method allows to
recover very simply and in a controlled way previous results for
correlation functions in the statics and in the equilibrium dynamics. 
Next we obtain new results, such as the full scaling functions for both
equilibrium and non-equilibrium dynamics. We obtain 
the corresponding exponents $\lambda$ and $\theta$. 
We also obtain the full and non trivial behavior of the fluctuation
dissipation ratio in the glass phase. 

The outline of the paper is as follows. First in Section II we derive the
EMRG method for the effective action, and give the explicit general
lowest order RG equations. In Section III we apply these RG equations
to the pure $O(N)$ model, as a test of the method. In Section IV we consider the
Cardy Ostlund model statics. In Section V we study the CO model equilibrium
dynamics. Section VI is devoted to the non-equilibrium
dynamics of the CO model. All calculational
details are contained in the appendices.

\section{Method}
\label{method}

\subsection{Exact RG method}
\label{exactrgpro}

We want to study interacting bosonic degrees of freedom described by a set
of fields denoted $\phi \equiv \phi^{i}_{x}$ where $x$ is the
position in space and $i$ a general label denoting any
quantity which will not undergo the coarse-graining (e.g. fields
indices, spin, replica indices, additional coordinate). The problem is
defined by an action functional:
\begin{eqnarray}
{\cal S}(\phi) = \frac{1}{2} \phi : G^{-1} : \phi + {\cal V}(\phi)
\label{startaction}
\end{eqnarray}
and by the functional integral (i.e. the partition function) $Z=\int ~
D\phi ~ e^{- {\cal S}(\phi)}$.
The action consists of a quadratic part ($G_{ij}^{xy} =G_{ji}^{yx}$ is
a symmetric invertible matrix) and ${\cal  V} (\phi)$ the interaction, a
functional of $\phi$. The notation $:$ denotes full contractions over
$x,i$ (i.e. $\phi :G^{-1}:\phi=\sum_{ij}
\int_{xy}  \phi_{x}^{i} (G^{-1})_{ij}^{xy} \phi_{y}^{j} $). We will denote
$\int_x \equiv \int d^d x$ where $d$ is the space dimension,
and $\int_q \equiv \int \frac{d^d q}{(2 \pi)^d}$
for integration in Fourier. Our aim is to compute
the effective action $\Gamma(\phi)$, i.e. the
generating function of proper vertices, since once it is known, all correlation functions
are known being simply obtained as sums of all tree diagrams drawn using $\Gamma$.
For all observables to be well defined one usually requires
both an ultraviolet UV cutoff (e.g. $\Lambda_0$ in momentum space)
and an infrared IR cutoff (noted here $\Lambda_l=e^{-l} \Lambda_0$).
For example, in a single scalar theory one chooses $G \equiv G_l$
with:
\begin{eqnarray}
&& G_l^q =q^{-2} c(\frac{q^2}{2 \Lambda_l^2}, \frac{q^2}{2 \Lambda_0^2})
\label{gencutoff}
\end{eqnarray}
in Fourier. Here $c(z,s)$ is a cutoff function which decreases to
zero as $z \to 0$ or $s \to \infty$ and for convenience, see below,
we choose $c(z,z)=0$.
To study finite momentum observables in a massless theory one is also interested
in the zero IR cutoff limit, $\Lambda_l=0$ with
$G \equiv G_{l=\infty}$ denoting $c(z)=c(\infty,z)$.

In this paper we will use that $\Gamma(\phi)$ satisfies the
following exact RG functional equation
when the quadratic part $G$ is varied (for a fixed ${\cal V}(\phi)$):
\begin{equation}
 \partial \Gamma(\phi) = \frac{1}{2} Tr
\partial G^{-1} : [ \frac{\delta^2 \Gamma(\phi)}{\delta \phi \delta \phi } ]^{-1}
+ \frac{1}{2} \phi : \partial G^{-1} : \phi
\label{exactrgeff}
\end{equation}
Derivations and more details are given in Appendix A. This can be used to
express how the effective action
$\Gamma(\phi) \equiv \Gamma_l(\phi)$ of the model (\ref{startaction})
with $G \equiv G_l$ depends on the IR cutoff $\Lambda_l$.
Indeed the following property holds:
\begin{eqnarray}
\Gamma_l(\phi) = - \frac{1}{2} Tr \ln G_l + \frac{1}{2} \phi : G_l^{-1} : \phi + {\cal U}_l(\phi)
\label{gammaexact}
\end{eqnarray}
with ${\cal U}_l(\phi) \equiv {\cal U}_{G_l}(\phi)$ satisfies the
exact flow equation:
\begin{equation}
 \partial_l {\cal U}_l(\phi) =  \frac{1}{2}Tr \partial_l G_l : ( G_l^{-1} - G_l^{-1}  (1+G_l :
\frac{\delta^2 {\cal U}_l}{\delta \phi\delta \phi})^{-1}) \label{ERGG}
\end{equation}
with the initial condition ${\cal U}_{l=0}(\phi) = {\cal V}(\phi)$, simply reflecting that
the effective action equals the action when all fluctuations are
suppressed (at $l=0$
where the running propagator satisfies $G_{l=0} = 0^+$ from the
property $c(z,z)=0$).
The above equation
(\ref{exactrgeff}) simply expresses how $\Gamma(\phi)$ in (\ref{gammaexact}) depends on
the final value $G \equiv G_{l}$. The zero IR cutoff limit
$\Lambda_l=0$ can then studied by integrating the
above equation up to $l=\infty$.

For actual calculations, simpler and useful choices read, in momentum space:
\begin{eqnarray}
G^q_{l} = \frac{1}{q^2} (c(q^2/2 \Lambda_0^2) - c(q^2/2 \Lambda_l^2)) \label{choixpropag1}
\end{eqnarray}
where the cutoff function $c(x)$ satisfies $c(0)=1$ and $c(\infty)=0$. With the
choice $c(x)=1/(1+ 2 x)$ one finds the massive, Pauli-Villars like, propagator:
\begin{eqnarray}
G^q_{l} = \frac{1}{q^2 + m^2}  - \frac{1}{q^2 + M^2} \label{Pauli}
\end{eqnarray}
with $m=\Lambda_l^2$, $M=\Lambda_0^2$, where the
IR cutoff mass $m \equiv m_l$ is lowered from $m=\infty$ ($l=0$) to $m=0$ ($l=\infty$)
(and $\partial_l \to - m \partial_m$). Whenever one needs a stronger
UV cutoff and one may use:
\begin{eqnarray}
&& G^q_m = \frac{1}{q^2 + m^2} c(\frac{q^2}{2 \Lambda_0^2}) \label{choixpropag2}
\end{eqnarray}
a different choice.

The full exact RG equation (\ref{ERGG}) can also be expanded in series
of ${\cal U}_l$ as:
\begin{eqnarray}
&&\partial_l {\cal U}_l(\phi) = \frac{1}{2}Tr \partial_l G_l : \frac{\delta^2
{\cal U}_l(\phi)}{\delta \phi \delta \phi} \nonumber \\
&& - \frac{1}{2} Tr \partial_l G_l
: \frac{\delta^2 {\cal U}_l(\phi)}{\delta \phi \delta \phi} : G_l : \frac{\delta^2
{\cal U}_l(\phi)}{\delta \phi \delta \phi} +
O({\cal U}_l^3) \label{perturbERG}
\end{eqnarray}
which admits the graphical representation given in
Fig.1.

To summarize, the philosophy of the method is, in a sense
the exact opposite of the Wilson one, since it amounts to start from the
action with no fluctuations $\Lambda_l=\Lambda_0$, and then add modes and
their fluctuations until one reaches the desired theory $\Lambda_l \ll \Lambda_0$.
In that limit one expects that the effective action
reaches a fixed point form, given by the asymptotic solution of
(\ref{ERGG}) at large $l$.

\begin{figure}[h]
\centering
\centerline{\includegraphics[width=7cm]{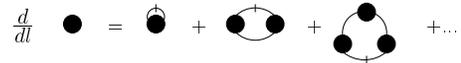}}
\label{graphicerg}
\caption{
Representation of the exact RG equation (\ref{exactrgeff}), (\ref{perturbERG}). The dot is the vertex ${\cal U}_l$, the
solid line a propagator $G_l$ and the crossed solid line the on shell propagator
$\partial G_l$. The sum is over all one loop graphs with a factor $(-1)^{p-1}/2$ for each
$p$ vertex graph represented.}
\end{figure}

\begin{figure}[h]
\centering
\centerline{\includegraphics[width=8cm]{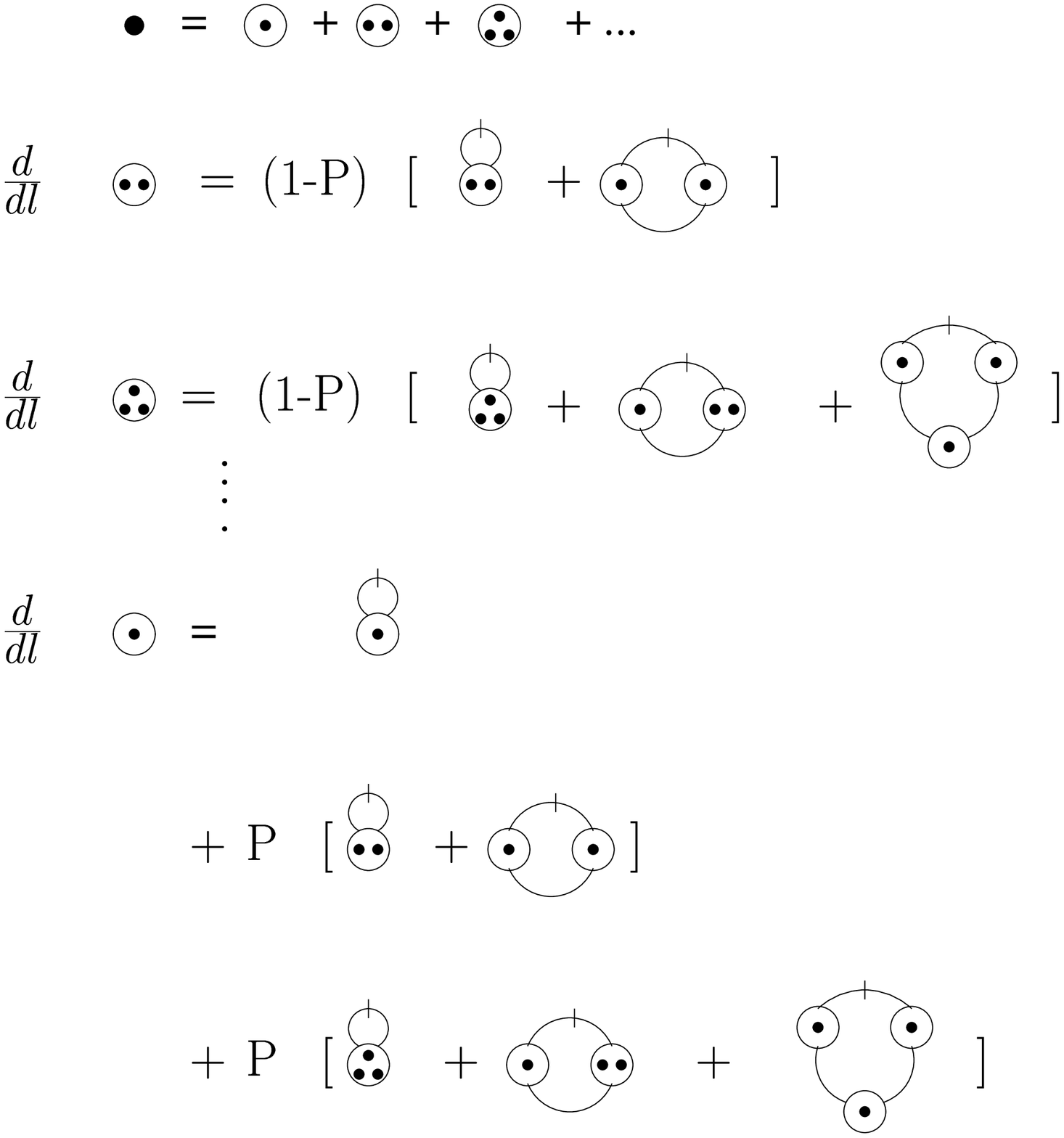}}
\caption{
\label{graphicexpansion}
Schematic representation of the splitting of the functional ${\cal U}$ vertex
into local, bilocal, trilocal etc.. parts respectively (top line). Representation
of the exact RG equation for the bilocal, trilocal, etc.. as well as local vertex
(last several lines). Note that by definition the ERG equation for the bilocal
part contains only exactly two feeding terms, trilocal three etc...  The
solid lines represent a propagator $G_l$ and the crossed soli lines the on shell propagator
$\partial G_l$. Combinatorial factors are not represented.
$P$ here denotes the projection operator on the local part (denoted $\overline{P}_1$ and
$P_1$) in the text).}
\end{figure}

\subsection{Multilocal expansion}
\label{multi}

To handle the formidably complicated functional equation (\ref{ERGG})
we follow the method introduced in \cite{chauvepld} and expand
the interaction functional ${\cal U}_l$ in local, bilocal, trilocal
etc.. components as:
\begin{eqnarray}
&& {\cal U}_l(\phi) = \int_x U_l(\phi_x) + \int_{x y}
V_l(\phi_x,\phi_y,x-y) \nonumber \\
&&  +
\int_{x y z} W_l(\phi_x,\phi_y,\phi_z,x,y,z) + ..
\label{expmult}
\end{eqnarray}
The local part depends only on the function $U_l(\phi)$,
uniquely defined from the projection operator $\bar{P_1}$. This operator is
fully defined in \cite{chauvepld} (see also Appendix \ref{app:expansion}).
We recall here only its action on a bilocal operator $F(\phi_x,\phi_y,x-y)$,
namely $(\bar{P_1}F)(\phi) = \int_y F(\phi,\phi,y)$. It can be used to
split an action depending only on two points into:
\begin{eqnarray}
&& \int_{xy} F(\phi_x,\phi_y,x-y) = \int_x (\bar{P}_1F)(\phi_x)
\nonumber \\
&& +
\int_{xy} ((1-P_1)F)(\phi_x,\phi_y,x-y)
\end{eqnarray}
where, by definition, $(P_1F)(\phi,\psi,z) = \delta(z) \int_y F(\phi,\psi,y)$, in such
a way that the second part is properly bi-local
(i.e. $(\bar{P}_1(1-P_1)F)(\phi) = 0$). A similar construction holds for
higher multilocal operators .

The idea is then to project the functional equation (\ref{ERGG}) so that the
bilocal, trilocal etc.. can be expressed exactly in terms of the local part $U_l$
only. One notices that there is a simplest way to do it so that the
bilocal part is $V \sim O(U^2)$, trilocal $W \sim O(U^3)$ etc..
This determines one possible splitting of the higher multilocal components
(e.g. bilocal vs trilocal) as is represented in the Fig. \ref{graphicexpansion},
and further explained in \cite{chauvepld}. This expansion is clearly suited to the
situations where the flowing functional ${\cal U}_l$ becomes "small" and dominated by its
local part (e.g. in the context of a dimensional
expansion), but it has a more general validity, since in all cases
it is an exact expansion in
series of the local part of the full effective action functional.

We now pursue the analysis exactly to order $O(U_l^2)$, sufficient to a number of one loop
applications. Details are given in Appendix \ref{app:expansion}.
The bilocal part is exactly given by:
\begin{equation}
V_l(\phi_1,\phi_2,x) = \frac{1}{2} (F_l(\phi_1,\phi_2,x)
- \delta(x) \int_y F_l(\phi_1,\phi_2,y)) \label{resultbiloc}
\end{equation}
with:
\begin{eqnarray}
&& F_l(\phi_1,\phi_2,x)  =
-\int_0^l dl' (\partial^1.\partial G^x_{l'}.\partial^2)(\partial^1.G^x_{l'}.\partial^2) \nonumber \\
&& e^{- \frac{1}{2} \partial^1.G^{x=0}_{l' l}.\partial^1
-\frac{1}{2}\partial^2.G^{x=0}_{l' l}.\partial^2 - \partial^1. G^x_{l' l}.\partial^2}
U_{l'}(\phi_1) U_{l'}(\phi_2)  \nonumber \\
&& \label{resultbiloc2}
\end{eqnarray}
to all orders (by definition), and the resulting exact RG equation for the
local part of the effective action (i.e. the exact $\beta$-function up
to $O(U_l^3)$) 
is:
\begin{widetext}
\begin{eqnarray}
&& \partial_l U_l(\phi) =   \frac{1}{2} \partial G^{x=0}_{l,ij} \partial_i
\partial_j U_l(\phi)  -  \frac{1}{2} \int_x \partial G^x_{l,ij} \partial_j
\partial_k U_l(\phi) (G^x_{l})_{km} \partial_m \partial_i
U_l(\phi)\label{RGorder2}  \\
&& - \frac{1}{2} \int_x \partial^1.(\partial G_l^x -
\partial G_l^0).\partial^2 \int_0^l dl' (\partial^1.\partial G^x_{l'}.\partial^2)
(\partial^1.G^x_{l'}.\partial^2) e^{-\frac{1}{2} \partial^1.G^{x=0}_{l' l}.\partial^1
-\frac{1}{2}\partial^2.G^{x=0}_{l' l}.\partial^2 - \partial^1. G^x_{l' l}.\partial^2}
U_{l'}(\phi_1) U_{l'}(\phi_2) |_{\phi_1=\phi_2=\phi} \nonumber
\end{eqnarray}
\end{widetext}
We use the following notations: $\partial_i \equiv \partial_{\phi^i}$,
$\partial^1_i$ (resp. $\partial^2_i$) denotes derivation with respect to the first argument
(resp. second argument) of a function of two vectors $\phi_1$, $\phi_2$,
$\partial^1 \cdot G^{xy} \cdot \partial^2 \equiv \sum_{ij}
G^{xy}_{ij} \partial^1_i \partial^2_j$ etc.. Also one notes
in real space $G^{xy} \equiv G^{x-y}$ and
$G^x_{l'l} = G_{l'}^x - G_{l}^x = - \int_{l'}^{l} \partial G_{l''}^x dl''$.
Note that the first line contains two one loop diagrams (tadpole and
bubble) with one ''on shell'' propagators, and
the second line represents a sum over diagrams with at least two loops.

\section{Application to the O(N) model.}

We first illustrate the method on the O(N) model defined by
(\ref{startaction}) with
\begin{eqnarray}
&& {\cal V}(\phi) = \frac{g_2}{2} \int_x \phi_x^2 + \frac{g_4}{4!}
\int_x (\phi_x^2)^2 \label{start_action_on}
\end{eqnarray}
$\phi_x$ being a N-component vector, $\phi_x^2 = \sum_i (\phi_x^i)^2$.
The propagator is diagonal, and using an infrared cutoff
$\Lambda_l$, it reads $G\equiv G_l$ with:
\begin{eqnarray}
&& G^q_{l,ij} = \delta_{ij} G_l^q
\end{eqnarray}
with $G_l^q$ as in (\ref{gencutoff}).
We study this model near the dimension 4, in $d=4-\epsilon$,
and compute the effective action to order $O(\epsilon^2)$.
For some explicit calculations, we will further use the
form (\ref{choixpropag1}) with the following convenient
parametrization and notation for the cut-off function $c(z)$:
\begin{eqnarray}
c(z) = \int_0^{+ \infty} da \hat{c}(a) e^{-az} \equiv  \int_a e^{-ax} \label{decomp_cutoff}
\end{eqnarray}
The condition $c(0)=1$ imposes $\int_a = 1$.

\subsection{Derivation of the $\beta$-functions and fixed points.}

The local part of the running effective action admits the polynomial
expansion:
\begin{eqnarray}
&& U_l(\phi) = g_{0,l} + \frac{g_{2,l}}{2!} \phi^2 +
\frac{g_{4,l}}{4!} (\phi^2)^2 +
\frac{g_{6,l}}{6!} (\phi^2)^3 + .. \nonumber \\
&&  \label{polyexp}
\end{eqnarray}
From power counting, it is more convenient to introduce
the dimensionless couplings $\tilde g_{2 n,l}$ defined
from:
\begin{eqnarray}
&& g_{2,l} = \Lambda_l^{2} \tilde g_{2,l} \nonumber \\
&& g_{4,l} = \Lambda_l^{\epsilon} \tilde g_{4,l}
 \label{local_on}
\end{eqnarray}
and more generally $\tilde g_{2 n,l}=g_{2 n,l} \Lambda_l^{(d-2) n - d}$
which flows to some fixed point values $\tilde g_{2 n}^*$, as
discussed below. Since $\tilde g_{6}^* \sim O(\epsilon^3)$ and
$\tilde g_{2 n}^* \sim O(\epsilon^{n})$ for $n \geq 3$ (see Appendix
\ref{app:on} for the RG equation of $\tilde g_{6,l}$ and the
free energy $\tilde g_{0,l}$), we drop
from now on these higher monomials and study only the coupled RG equation for
$\tilde g_4$ and $\tilde g_2$ easily obtained by inserting
(\ref{polyexp}) into
(\ref{RGorder2}) as detailed in Appendix \ref{app:on}:
\begin{eqnarray}
&&\partial_l {\tilde{g}_{4, l}} =  \epsilon {\tilde{g}_{4,l}}
- \frac{N+8}{3} \tilde I_l^{(1)} {{\tilde{g}^2_{4,l}}} =
-\beta[\tilde{g}_{4,l}]
\label{RG_Eq_On_g4}    \\
&&\partial_l {\tilde{g}_{2l}} = 2{\tilde{g}_{2l}} + \frac{N+2}{6}
 \tilde I_l^{(0)}
 \tilde{g}_{4,l}
 - \frac{N +2}{3}  \tilde I_l^{(1)} {\tilde{g}_{2,l}}{\tilde{g}_{4,l}}   \nonumber \\
&& - \frac{N+2}{3} \int_0^l dl' \tilde I_{l ,l'}^{(2)} {{\tilde{g}^2_{4,l'}}}
\label{RG_Eq_On_g2}
\end{eqnarray}
with the integrals:
\begin{eqnarray}
\tilde I_l^{(0)} = \Lambda_l^{-2+\epsilon} \int_q \partial_l G_l^q  \quad , \quad
\tilde I_l^{(1)} = \Lambda_l^{\epsilon} \int_q \partial_l G_l^q G_l^q   \label{integrals}
\end{eqnarray}
$\tilde I_{l , l'}^{(2)}$ is given in (\ref{I2_ll'}) where we show that the
coefficient of the term
proportional to $\tilde{g}_{4,l'}^2$ in (\ref{RG_Eq_On_g2}) is well
defined in the limit $l \to \infty$. One finds that $\tilde I_l^{(0)} =
\tilde I^{(0)}$ is $l$-independent and that $\lim_{l\to \infty} \tilde I_l^{(1)}
= I^{(1)}$ is universal (independent of $c(s)$) in dimension $d=4$:
\begin{eqnarray}
&&\tilde I^{(0)} = -\frac{1}{4\pi^2} \int_0^{\infty} ds s c'(s) + O(\epsilon)
\label{integrals4} \\
&&\tilde I^{(1)} = S_d \int_{s>0} (2 s)^{-\epsilon/2} c'(s) (c(s)-1)
= \frac{1}{16 \pi^2} + O(\epsilon) \nonumber
\end{eqnarray}
where $S_d$ is the unit sphere area divided by $(2 \pi)^d$
and we recall $c'(s) < 0$. Finally
eq. (\ref{RG_Eq_On_g4},\ref{RG_Eq_On_g2}) together with
(\ref{integrals}) lead to the
fixed point values
\begin{eqnarray}
&&{\tilde{g}_4}^* = \frac{48 \pi^2}{N+8} \epsilon + O(\epsilon^2)
\label{g4_fp} \\
&&{\tilde{g}_2}^* = -\frac{N+2}{12}\tilde I^{(0)}{\tilde{g}_4}^* +
O(\epsilon^2)
\end{eqnarray}
This fixed point describes the standard $O(N)$ critical
system exactly at the critical temperature $T=T_c$. The initial
conditions which end up for $l=\infty$ exactly at the fixed point
describe the critical manifold.

Besides we obtain the correction of the critical exponent
characterizing the divergence of the magnetic susceptibility near the
critical temperature from the positive eigenvalue $\lambda_l$
(corresponding to the instable direction)
\begin{eqnarray}
&&\partial_l ({\tilde{g}_{2l}} - {\tilde{g}_2}^*) = \lambda_l
({\tilde{g}_{2l}} - {\tilde{g}_2}^*) \\
&& \lambda_l = 2(1 - \frac{N+2}{2(N+8)}\epsilon)
\end{eqnarray}
which gives correctly \cite{blz} the exponent $\gamma$ to order $\epsilon$
\begin{eqnarray}
\gamma = \frac{2}{\lambda^*} = 1 + \frac{N+2}{2(N+8)}\epsilon + O(\epsilon^2)
\end{eqnarray}

\subsection{Computation of the $2$ and $4$ points proper vertices.}

We now compute the effective action on the critical manifold,
up to order $O(\epsilon)$ for the local part, and
$O(\epsilon^2)$ for the bi-local part (i.e. the $q$
dependent part), in the limit of large $l$.
Eq. (\ref{resultbiloc}) allows to construct the bilocal term in the effective
action by inserting (\ref{polyexp}) in (\ref{resultbiloc2}). As we
restrict our analysis to order $\epsilon^2$, we do not consider
monomials higher than $(\phi^2)^2$ in (\ref{polyexp}), and therefore
we expand the exponential in (\ref{resultbiloc2}) to order one. Using the
combinatorics already explained for the local part in Appendix
\ref{app:on}, one gets
\begin{eqnarray}
&&V_l(\phi_1,\phi_2,q) = \frac{1}{2}\int_x
(e^{iqx}-1)F_l(\phi_1,\phi_2,x) \label{bilocal_On} \\
&&F_l(\phi_1,\phi_2,x) = \frac{N+2}{3} \phi_1 \cdot \phi_2 \int_0^l dl'
\partial_{l'} G^x_{l'} G^x_{l'} G^x_{l'l}{{\tilde{g}^2_{4l'}}}
\Lambda_{l'}^{2 \epsilon} \nonumber \\
&& - ( \frac{N+4}{(3!)^2}\phi_1^2 \phi_2^2 +
\frac{4}{(3!)^2} (\phi_1 \cdot \phi_2)^2) \int_0^l dl' \partial_{l'}
G^x_{l'}  G^x_{l'}{{\tilde{g}^2_{4l'}}} \Lambda_{l'}^{2 \epsilon}
\nonumber
\end{eqnarray}
where we have not written terms of the form $f(\phi_i,x)$ (i.e. which depend
only on one field argument) as they
cancel out from the effective action. To this order in
$\epsilon$ ($O(\epsilon^2)$) there are no other contributions.
The explicit expressions
of $G_l^x$ and $\partial_l G_l^x$ using (\ref{decomp_cutoff})
are given in Appendix \ref{app:on} (\ref{propag_x}).
This bi-local term (\ref{bilocal_On}) allows
to treat the renormalization of
the wave function and compute the exponent $\eta$ to order
$\epsilon^2$. A natural way to obtain it, within this method, is to
compute directly the 1 particle irreducible (1PI) two-points function and then take
the limit $l \to
\infty$ (directly at $T_c$). Its local part comes from the
quadratic contribution of (\ref{polyexp}) and the bi-local part is the
sum of $G_l^{-1}(q)$ (\ref{gammaexact}) and the quadratic contribution of
(\ref{bilocal_On})
\begin{eqnarray}
&&\Gamma^{(2)}_{l,ij}(q) = \frac{\delta^2 \Gamma_l}{\delta \phi^i_q
\delta \phi^j_{-q}}\bigg{|}_{\phi=0}  = \delta_{ij}\Gamma^{(2)}_{l}(q)
\label{gamma2_on_gene} \\
&&\Gamma^{(2)}_{l}(q) =G_l^{-1}(q) + \Lambda_l^2 \tilde{g}_{2,l} -
\frac{N+2}{18}
\tilde{g}_{4,l}^2 \int_x (e^{iqx} -1)(G^x_l)^3 \nonumber
\end{eqnarray}
In the appendix \ref{app:on}, we show that it has the form, up to
terms of order $(\Lambda_l/\Lambda_0)^{2}$
\begin{eqnarray}
&&\Gamma^{(2)}_{l}(q) = G_l^{-1}(q) + \Lambda_l^2 \tilde{g}_{2,l}  -
q^2 \eta[\tilde{g}_{4,l}] (\ln{\frac{\Lambda_l}{\Lambda_0}} +
\chi^{(2)}(\frac{q}{\Lambda_l})) \nonumber \\
&&\eta[\tilde{g}_{4,l}] = \frac{N+2}{18 (4 \pi)^4}  \label{gamma2_on}
\tilde{g}_{4,l}^2
\end{eqnarray}
with the following asymptotic behaviors
\begin{eqnarray}
&& \chi^{(2)}(k) \sim a k^2 \quad k \ll 1 \nonumber \\
&& \chi^{(2)}(k) \sim \ln{k} \quad k \gg 1
\label{fonc_gamma2}
\end{eqnarray}
with $a$ some non universal (i.e. dependent of the cutoff
function (\ref{decomp_cutoff})) coefficient. The two-point scaling
function $\chi^{(2)}(k)$ which is computed here (see Appendix \ref{app:on})
for an arbitrary infrared cutoff function $c(x)$, is
up to an additive constant,
independent of the UV cutoff \footnote{note that in the particular choice used here the
same function $c(x)$ appears both as IR and UV cutoff}. For the particular
choice (\ref{Pauli}) one recovers the result of
\cite{aharony_fisher}.

The large argument
behavior of $\chi^{(2)}(k)$ allows to take the limit $l \to \infty$,
using the fixed point value $\tilde{g}_4^*$ (\ref{g4_fp}), we have
(for $q \ll \Lambda_0$):
\begin{eqnarray}
\lim_{l \to \infty} \Gamma^{(2)}_{l}(q) = (q^2 - q^2
\frac{N+2}{2(N+8)^2}\epsilon^2\ln{\frac{q}{\Lambda_0}}) \label{gamma2_fp}
\end{eqnarray}
which coincides with the expansion of $\lim_{l \to \infty}
\Gamma^{(2)}_{l}(q)
\sim q^2(q/\Lambda_0)^{-\eta}$ to order $\epsilon^2$  with
the universal value of the $\eta$ exponent to this order
\begin{eqnarray}
\eta = \eta[\tilde{g}_4^*] = \frac{N+2}{2(N+8)^2}\epsilon^2 \label{exp_eta}
\end{eqnarray}
in agreement with standard results \cite{blz}.

Let us focus on the construction of the quartic term in
$\Gamma_l(\phi)$, obtained from the quartic contribution of (\ref{polyexp})
and (\ref{bilocal_On}). After combinatorial manipulations, we obtain
\begin{eqnarray}
&&\Gamma_l^{\text{quart}} = \frac{{\tilde{g_4}}_{l}\Lambda_l^{\epsilon}}{4!}
\int_{q_i}'
(\phi_{q_1} \cdot \phi_{q_2})( \phi_{q_3} \cdot \phi_{q_4}) \nonumber \\
&&- {\tilde{g_4}}_{l}^2 \frac{1}{(3!)^2}
\int_{q_i}'\big((\frac{N+4}{4})(\phi_{q_1} \cdot \phi_{q_2})( \phi_{q_3} \cdot
\phi_{q_4})  \nonumber \\
&&+ (\phi_{q_1} \cdot \phi_{q_3})( \phi_{q_2} \cdot
\phi_{q_4}) \big)
\chi^{(4)}_l(q_3+q_4) \label{gamma4_on}
\end{eqnarray}
with $\chi^{(4)}_l(q)$ defined by
\begin{eqnarray}
\chi^{(4)}_l(q) = \int_x (e^{iqx}-1)(G^x_l)^2 + O(\Lambda_0^{-2})
\end{eqnarray}
and where we used the notation $\int_{q_i}' \equiv
\int_{q_1,q_2,q_3,q_4} (2 \pi)^d \delta^{(d)}(q_1+q_2+q_3+q_4)$.
The local term, i.e. the first line in (\ref{gamma4_on}),
contains a contribution of order $\epsilon^2$ which is
divergent in the limit $l \to \infty$. Indeed, expanding it
to second order gives
$\tilde{g}_{4,l}\Lambda_l^{\epsilon} = \tilde{g}_{4,l}(1 +
\epsilon \ln{\Lambda_l}) + O(\epsilon^3)$ and at first sight this
term would lead to a divergent contribution in the limit $l \to
\infty$. However, the analysis of $\chi^{(4)}_l(q)
=\chi^{(4)}(q/\Lambda_l)$ shows the following asymptotic behaviors
\begin{eqnarray}
&&\chi^{(4)}(k) \sim b k^2 \quad k \ll 1 \label{gamma4_petit}\\
&&\chi^{(4)}(k) \sim -\frac{1}{16 \pi^2} \ln(k^2) \quad k \gg 1
\label{gamma4_grand}
\end{eqnarray}
with $b$ a non universal constant.
When considering the large $l$ limit of the effective action, we are
interested in the large argument behavior of $\chi^{(4)}(k)$
(\ref{gamma4_grand}). Using the fixed point value $\tilde{g}_{4}^*$
(\ref{g4_fp}), one gets that this
cancels exactly the divergence when $l \to \infty$ due to the local
term. Thus we obtain:
\begin{eqnarray}
&&\lim_{l \to \infty} \Gamma_l^{\text{quart}}
= \frac{2 \pi^2 \epsilon}{(N+8)}
[ \Lambda_0^{\epsilon}
\int_{q_i}'
(\phi_{q_1} \cdot \phi_{q_2})( \phi_{q_3} \cdot \phi_{q_4}) \nonumber \\
&& + \frac{4 \epsilon}{(N+8)}
\int_{q_i}'\big((\frac{N+4}{4})(\phi_{q_1} \cdot \phi_{q_2})( \phi_{q_3} \cdot
\phi_{q_4})  \nonumber \\
&&+ (\phi_{q_1} \cdot \phi_{q_3})( \phi_{q_2} \cdot
\phi_{q_4}) \big) \ln(\frac{|q_3+q_4|}{\Lambda_0}) ] \label{gamma4_on2}
\end{eqnarray}
which is independent of $\Lambda_0$ to order $O(\epsilon^2)$.
Note that in the large $N$ limit one recovers correctly the ''screened'' \cite{bray}
four point renormalized vertex $\sim \epsilon q^\epsilon$
(where $q$ is the transfer momentum).

The result of this analysis is that we have constructed the
large scale theory by obtaining directly a fixed point for the
effective action,  keeping the UV
cut-off $\Lambda_0$ finite, which is the relevant object for
statistical physics, and for an arbitrary cutoff function.

\subsection{Relation with field-theoretical methods}

It is interesting to make the connection with standard
field-theoretical methods for critical phenomena. There one is usually interested
in the limit $\Lambda_0 \to \infty$. Note that in this
limit (\ref{gamma2_fp}) diverges. It is however possible
to define a ``renormalized'' effective action $\Gamma_R(\phi_R)$
which is well defined in that limit.

One can first check directly on (\ref{gamma2_fp}) the standard
Callan-Symanzik (CS) ''bare'' RG equation \cite{zinn-justin} for the physical correlation function
of the massless theory at the fixed point
\begin{eqnarray}
(\Lambda_0 \frac{\partial}{\partial \Lambda_0} - \eta)
(\lim_{l \to \infty}\Gamma^{(2)}_{l})(q) = 0 + O(\epsilon^3)
\end{eqnarray}
One can also connect to the CS equation for the renormalized
theory. One defines:
\begin{eqnarray}
&& \Gamma_{R}(\phi) = \Gamma_l(\sqrt{Z} \phi)
\end{eqnarray}
where $Z\equiv Z(\frac{\Lambda_l}{\Lambda_0},\tilde{g}_{4,l})$ is the
so called ''wave-function renormalization'' factor such that
\begin{eqnarray}
&& \Gamma_{R}^{(2)}(q) = m_R^2 + q^2 + O(q^4)
\end{eqnarray}
Using (\ref{gamma2_on_gene}) and noting that
$G_l^{-1}(q) = - \frac{2 \Lambda_l^2}{c'(0)} + A q^2 + O((\frac{\Lambda_l}{\Lambda_0})^2)$,
with $A=\frac{c''(0)}{2 c'(0)^2}$ one finds the renormalized
mass $m_R^2 = \frac{1}{A} \Lambda_l^2 (\frac{2}{|c'(0)|} + \tilde g_{2,l})$
and $Z=\frac{1}{A} (1 + \eta(\tilde g_{4,l}) \ln (\frac{\Lambda_l}{\Lambda_0}))$.
One can see that up to higher order terms,
$\Lambda_l$ plays the role of the renormalized mass.
From (\ref{gamma2_on}) one finds, to order $(\epsilon
\tilde{g}_{4,l}^2,\tilde{g}_{4,l}^3)$
\begin{eqnarray}
m_R \partial_{m_R}|_{\Lambda_0}
\ln Z(\frac{\Lambda_l}{\Lambda_0},\tilde{g}_{4,l}) = - \partial_l \ln Z
= \eta[\tilde{g}_{4,l}]
\end{eqnarray}
these derivatives being taken at fixed $\tilde{g}_{4,l}$. This is the
standard definition for the $\eta(g)$ function. One can
go further, define a renormalized coupling $g_R$
, e.g. through $\Gamma_{R}^{(4)}(q=0) = m_R^\epsilon g_R$,
with $g_R = \tilde g_{4,l}$ up to higher order terms,
and derive the CS equations for the renormalized
vertices. Here, we just mention one such equation \cite{zinn-justin}
for the ``renormalized'' two-point vertex function
in the critical regime
$\Lambda_l \ll \Lambda_0$ but {\it finite}
\begin{eqnarray}
(\partial_l + \eta[\tilde{g}_{4,l}] ) \Gamma_l^{R(2)}
(q) \simeq 0 \quad q/\Lambda_l \gg 1
\end{eqnarray}
obtained using
the large $k$ behavior of $\chi^{(2)}(k)$ (\ref{fonc_gamma2}).
We get again the universal value of the
$\eta$ exponent form $\eta = \eta[\tilde{g}_4^*]$ (\ref{exp_eta}).

The connection between the EMRG method and the
standard field theoretical methods in the {\it massless}
scheme (i.e. imposing $\Gamma^{(2)}_R(q=0)=0$)
is more subtle here (since one should use $l = \infty$ strictly).

\section{Cardy Ostlund Model: Statics}

In this section, we show how this EMRG method can be used to study
perturbatively the Cardy Ostlund model \cite{cardy-desordre-rg} near its glass transition.

\subsection{Model, choice of propagator}

This model is a random phase Sine-Gordon model which can represent
an XY model in a random  magnetic field where the vortices are
excluded by hand. As mentioned in the introduction, the statics of
this model has been extensively studied using various methods
\cite{villain-cosine-realrg,toner-log-2,giamarchi_vortex_long,hwa_fisher,ludwig_co,rsb,num}.
The system at equilibrium is described by the
partition function $Z = \int D\phi e^{-H^{CO}[\phi]/T}$, $T$ being the
temperature with the hamiltonian
\begin{equation}
H^{CO}[\phi] = \frac{1}{2}\int d^2 x (\nabla \phi_x)^2 - \int d^2x
(h^1_x \cos{\phi_x} + h^2_x \sin{\phi_x})  \label{HamCO}
\end{equation}
whith $\phi_x \in ]-\infty, + \infty[$ as there are no vortices,
where ${\bf h}_x = (h^1_x,h^2_x)$ is a $2d$ random Gaussian vector
of zero average with
fluctuations decorrelated from site to site:
\begin{eqnarray}
&&\langle h^i_x h^j_{x'}  \rangle = 2 g_0  \Lambda_0^2 \delta_{ij} \delta^{(2)}(x-x')
\end{eqnarray}
The quenched average over this random variable is performed by the
means of replicas,
which is used here as a simple trick to restore
translational invariance and to organize perturbation theory. After
averaging over the disorder, one obtains
\begin{eqnarray}
&& \overline{\ln {Z}} = \lim_{n \to 0} \frac{\overline{Z^n} -1}{n}, \qquad
\overline{Z^n} = \int D\phi^a e^{-\frac{H^{\text{rep}}[\phi^a]}{T}}
\nonumber \\
&&\frac{H^{\text{rep}}[\phi^a]}{T} = \frac{1}{2 T} \sum_{a,b} \int d^2x \nabla
\phi^a_x \nabla \phi^b_x \delta_{a,b}  \nonumber \\
&&- \frac{ g_0 \Lambda_0^2}{T^2} \sum_{ab} \int_x \cos{(\phi^{a}_x -
\phi^{b}_x)} \label{originXYRF}
\end{eqnarray}
where $a,b = 1,..,n$ are replica indices.
We use the same propagator as for the O(N) model, the Gaussian part of
(\ref{originXYRF}) being diagonal in replicas, one has
\begin{eqnarray}
{G^q_{l}}_{ab} = \delta_{ab}\frac{T}{q^2}
(c(q^2/2 \Lambda_0^2) - c(q^2/2 \Lambda_l^2))
\end{eqnarray}
with the same decomposition of the cutoff function $c(x)$
(\ref{decomp_cutoff}). Notice that the hamiltonian $H^{\text{rep}}$ possesses the
statistical
tilt symmetry (STS) \cite{brezin_sts}: the last term in
(\ref{originXYRF}) is invariant
under the change of variable
$\phi^a_x \to \phi^a_x + u_x$ which protects the diagonal (in replica space)
quadratic term in the effective action to all orders in perturbation theory
\cite{hwa_fisher,carpentier}.

\subsection{$\beta$-functions and fixed point}

For this model, the Fourier representation in the fields
(\ref{def_Fourier}) is more
natural. Although only one harmonic is present in the starting
hamiltonian ($\ref{originXYRF}$),
higher harmonics are generated by perturbation theory and
we write the local interacting part of the effective action (\ref{expmult}) as
\begin{eqnarray}
U_{l}(\phi) = - \Lambda_l^2 \sum_{K \neq 0} \frac{g^K_l}{T^2}
e^{i K . \phi}
\label{COint}
\end{eqnarray}
where $K=(K_1,...,K_n)$, $\phi = (\phi^1,...,\phi^n)$ are
$n$-components vectors and one defines $K.K' = \sum_a K_a K'_a$.
The sum is over all $K$ such that $K_a$ are
integers not all zero with $\sum_{a}
K_a=0$. $U_l(\phi)$ is real,
imposing $g^K_l = g^{-K}_{l}$, and the symmetry under replica indices
permutation, which is assumed here, imposes $g^{K}_l =
g^{\sigma(K)}_l$, $\sigma(K)$ being
any vector obtained from $K$ by a permutation of the $K_a$.
By inserting (\ref{COint}) in (\ref{RGorder2})
(see also (\ref{RGorder2_Fourier}) in Appendix \ref{app:expansion})  one obtains the
RG equation
for the local part to second order in $g_K$:
\begin{eqnarray}
&&\partial_l g^K_l = (2 - \frac{TK^2}{4 \pi})g^K_l +
\frac{\tilde{J}^{(1)}_l}{2 T^2} \sum_{P,Q,P+Q=K} g^P_l g^Q_l (P.Q)^2
\nonumber \\
&&-\frac{1}{2 T^2} \sum_{P,Q,P+Q=K} (P.Q)^3 \int_0^l dl'
\tilde{J}^{2}_{l,l'} g^P_{l'} g^Q_{l'} \label{beta_co_gene}
\end{eqnarray}
with the integrals
\begin{eqnarray}
&&\tilde{J}^{1}_l = \Lambda_l^2\int_x \partial G_l^x
G_{l}^x   \\
&&\tilde{J}^{2}_{l,l'}=\Lambda_l^{-2}\int_x
(\partial G_l^x - \partial G_l^{x=0}) \partial G_{l'}^x
G_{l'}^x  \Lambda_{l'}^4 \nonumber \\
&& \times e^{(\frac{P^2}{2}+\frac{Q^2}{2})G_{l'l}^{x=0}+P.Q G_{l'l}^x}
\end{eqnarray}
The glass transition temperature $T_c$ below which the charges of minimal modulus such that
$K_{1,-1} = (0,..,1,..,-1,..,0)$, $K_{1,-1}^2 = 2$ become relevant
is
\begin{eqnarray}
T_c = \frac{8 \pi}{K_{1,-1}^2} = 4\pi \label{Tc}
\end{eqnarray}
and a small parameter $\tau = (T_c - T)/T_c > 0$ can be defined, which allows to
construct perturbatively the effective
action of this model (\ref{originXYRF}) in its glass phase. Indeed
just below $T_c$ the
higher harmonics are
irrelevant (the eigenvalues $(2 - \frac{TK^2}{4 \pi})$ are negative
and of order one). Such irrelevant higher harmonics include for instance
$3$ replicas term \footnote{The higher replica operators used here are defined
in terms of excluded replica sums. They thus form a different basis than the one used in,
e.g. FRG studies of zero temperature fixed point in higher dimension \cite{fisher_functional_rg,balents_fisher,giamarchi_vortex_long}. These
are defined in terms of unrestricted sums directly related to
the cumulant of the disorder, which is not the case here}
$g_l^{1,-2,1}\sum_{a \neq b\neq c}e^{i(\phi^a_x -
2\phi^b_x + \phi^c_x)}$, corresponding to $K_{1,-2,1}^2 = 4$.
We denote $g_l =
g^{1,-1}_l$ the coupling constant associated to
$K_{1,-1}$, and obtain its RG flow from (\ref{beta_co_gene}) by taking
into account the $2(n-2)$
possible fusions such that $P + Q = K_{1,-1}$, $P,Q$ being themselves obtained
by a permutation of the components of $K_{1,-1}$ ($g^P_l = g^Q_l =
g_l $)
with $P.Q = -1$ \cite{cardy-desordre-rg}.  After some transformations detailed in the Appendix
\ref{app:CO} one obtains
\begin{eqnarray}
&&\partial_l g_l = (2 - \frac{T}{2 \pi}) g_l - {\cal B}_l g_l^2 \\
&& {\cal B}_l =
2 \partial \gamma_0(0)\int_{\tilde{x}} \gamma_l(\tilde{x}) \nonumber \\
&& + \frac{2}{T_c}
\int_{\tilde{x}} (\partial \gamma_0(\tilde{x}) - \partial \gamma_0(0))
(e^{T_c \gamma_l(\tilde{x})} -1) + O(\tau)
 \label{RGlocco}
\end{eqnarray}
where we used the dimensionless variable $\tilde{x} = x \Lambda_l$
and defined:
\begin{eqnarray}
&& \partial G_{l'}^x = T \partial \gamma_{\mu=l-l'}(\tilde{x}) \nonumber \\
&& G_{l' l}^x = - T \gamma_{\mu=l-l'}(\tilde{x}) \label{rescal_propagCO}
\end{eqnarray}
where the two functions $\partial \gamma_{\mu}(x)$ and $\gamma_{\mu}(x)$
are given in (\ref{propag_app}).

As shown in the appendix, we can transform the
integral over $\tilde{x}$ in (Eq. \ref{RGlocco}) and express its cutoff
dependence in a simple way. One finds ${\cal B}_\infty =
\frac{4 \pi}{T_c^2} e^{-(\gamma_E - \int_a \ln{2a} )}$
yielding for $T<T_c$, the stable fixed point
of the RG flow is given by
\begin{eqnarray}
g^* = 8 \pi e^{(\gamma_E - \int_a \ln{2a} )} ~~ \tau + O(\tau^2)
\label{co_fp}
\end{eqnarray}
with $\tau = (T_c - T)/T_c$ and $\gamma_E = 0.577216$ the Euler constant.

\subsection{Bilocal term and 2-point correlation function.}

Eq. (\ref{resultbiloc_Fourier}) allows to construct the bilocal term
in the effective action to lowest order (i.e. $O(\tau^2)$)
using a Fourier representation (\ref{def_Fourier}):
\begin{eqnarray}
&& V_l(\phi, \psi, x) = \sum_{K,P} \hat V^{KPx}_{l} e^{i K \cdot \phi + i P \cdot \psi}
\end{eqnarray}
Just below $T_c$, only the charges of minimal modulus $K_{1,-1}^2 =
2$ are relevant, therefore to this order
the sums in (\ref{resultbiloc_Fourier}) are restricted to
such harmonics.
By inserting (\ref{COint}) into (\ref{resultbiloc_Fourier}) one has
\begin{eqnarray}
&&\hat V^{KPq}_{l} = \frac{1}{2}\int_x (e^{iqx} -1) \hat F_{l}^{KPx} \nonumber\\
&&\hat F_l^{KPx} = - \frac{(K.P)^2}{T^4} \int_0^l dl' \partial
G_{l'}^x G_{l'}^x \nonumber \\
&&e^{\frac{K^2+P^2}{2}G_{l'l}^{x=0}} e^{K.P G^x_{l'l}} \Lambda_{l'}^4
g^K_{l'} g^P_{l'} \label{Fl_co_gen}
\end{eqnarray}
where $K,P$ are of the form $K_{1,-1}$, and thus $g^K_l = g^P_l = g_l$.
Performing the integral over $l'$ as explained in
Appendix~\ref{app:CO} we have: 
\begin{equation} \label{Fl_co}
\hat F_l^{KP\tilde{x}} = - \frac{\Lambda_l^4}{T^2} g_l^2
\left(\frac{1}{T^2} (e^{-T_c
K.P \gamma_l(\tilde{x})} -1) + K.P \frac{\gamma_l(\tilde{x})}{T}
\right)
\end{equation}
with $\tilde{x} = \Lambda_l x$. For the charges $K,P$ we are
considering here, there are a priori $5$ different cases of $K.P =
-2,-1,0,1,2$ to consider. However we see immediately on the previous
expression (\ref{Fl_co}) that the charges such that $K.P = 0$ do not
contribute to the bilocal part of the effective action
 (they correspond to four replicas terms
$g_l^2 \sum_{a \neq b\neq c \neq
d} e^{i(\phi_a - \phi_b) + i (\psi_c - \psi_d)}$).
We show in Appendix~\ref{app:CO} that
$\hat V_{l}^{K,P,q}$ takes the form, up to terms of order
$(\Lambda_l/\Lambda_0)^2$
\begin{eqnarray}
&&\hat V_l^{K,P,q} = - A_l q^2 ( \delta_{K,-P}\ln{\frac{\Lambda_l}{\Lambda_0}}
+ \chi^{K.P}(\frac{q}{\Lambda_l})) \label{Vl_CO} \\
&&A_l = \frac{\pi g_l^2}{4 T_c^4} e^{-2 \gamma_E + 2 \int_a \ln{2a}}
 \label{amplitude_CO}
\end{eqnarray}
where $\chi^{K.P}(k)$ behaves aymptotically at small argument as
\begin{eqnarray}\label{gamma_co_small}
\chi^{K.P}(k) \sim
\begin{cases}
a_{K.P}  \quad &K.P \neq -2
\\
a_{-2} k^2 \quad &K.P=-2
\end{cases}
\quad k \ll 1
\end{eqnarray}
and at large argument (relevant for the limit $l \to \infty$) as
\begin{eqnarray}\label{gamma_co_large}
\chi^{K.P}(k) \sim
\begin{cases}
b_{K.P}\frac{1}{k^2} \quad &K.P = 1,2
\\
b_{-1} \frac{\ln{k}}{k^2} \quad &K.P = -1
\\
\ln{k}   \quad &K.P = -2
\end{cases}
\quad k \gg 1
\end{eqnarray}
The large argument behavior of $\chi^{K.P}(x)$ allows to take the
limit $\Lambda_l \to 0$ of (\ref{Vl_CO}) as the logarithmic divergence
(which only exists for $K=-P$) is cancelled. We notice also that only
such terms whith $K=-P$ survive in this limit : in particular,
three replicas term as $g_l^2 \sum_{a\neq b\neq c} e^{i(\phi_a
- \phi_b) + i (\psi_b - \psi_c)}$  do not exist in the effective action
to order $\tau^2$ at the fixed point for $\Lambda_l=0$. Besides, by inserting the fixed
point value $g^*$ (\ref{co_fp}) in $A_l$ (\ref{amplitude_CO}), we
see that the cut-off dependence (encoded in the factor $e^{\int_a
\ln{2a}}$) disappears in $\lim_{l \to \infty} A_l$ leading to
\begin{eqnarray}
\lim_{l \to \infty} \hat V^{KPq}_l = - \delta_{K,-P} \frac{\tau^2}{16 \pi} q^2
\ln{\frac{q}{\Lambda_0}} \label{Vl_fp_CO}
\end{eqnarray}
Eq. (\ref{gammaexact}) together with (\ref{Vl_fp_CO})
allow to construct the bilocal term as
\begin{eqnarray}
&&\lim_{l\to\infty} \Gamma_{l}^{biloc}(\phi) = \frac{1}{2 T} \sum_a \int_q \frac{q^2}{
c(\frac{q^2}{2\Lambda_0^2})} \phi^a_q \phi^a_{-q} \\
&& + \sum_{a,b} \int_{x,x'} \int_q
\lim_{l\to \infty} \hat V_{l}^{K,-Kq} e^{i q(x-x')} e^{i(\phi^a_x -
\phi^b_x)} e^{-i(\phi^a_{x'} - \phi^b_{x'})} \nonumber
\end{eqnarray}
from which we extract the two-point 1 PI function $\lim_{l \to
\infty} \Gamma_{l,ab}^{(2)}(q)$ :
\begin{eqnarray}
&&\Gamma_{l,ab}^{(2)}(q) = \frac{\delta^2 \Gamma_{l}(\phi)}{\delta \phi^a_q
\delta \phi^b_{-q}}\Big{|}_{\phi=0} \\
&&\lim_{l \to \infty} \Gamma_{l,ab}^{(2)}(q) = \frac{q^2}{T
c(q^2/2\Lambda_0^2)}
 \delta_{ab} +
\frac{\tau^2}{4 \pi} q^2
\ln{\frac{q}{\Lambda_0}}
\end{eqnarray}
from which we extract the correlation function at the fixed point (up
to terms of order $\Lambda_0^{-2}$)
\begin{eqnarray}
&&\overline{\langle(\phi_x - \phi_0)^2\rangle} =\lim_{l \to \infty} 2
\int_q (1 - e^{i
q x}) [\Gamma^{(2)}_{l}]^{-1}_{aa}\\
&& = 2T_c \int_q \frac{1-e^{iqx}}{q^2} c(\frac{q^2}{2\Lambda_0^2})(1 - \tau
-\tau^2 \ln{\left(\frac{q}{\Lambda_0}\right)}
c(\frac{q^2}{2\Lambda_0^2})) \nonumber \\
&&\sim 2 \tau^2 \ln^2{(|x| \Lambda_0)} + 4 (1 - \tau + O(\tau^2))
\ln{(|x| \Lambda_0)})
\end{eqnarray}
which shows that the amplitude of these anomalous fluctuations in
$\ln^2{(|x| \Lambda_0)}$ is universal (\cite{carpentier}).

We finally mention that, due to the STS, the connected correlation function
$\overline{\langle(\phi_x - \phi(0))^2\rangle} -
\overline{\langle(\phi_x - \phi(0))
\rangle\langle(\phi_x - \phi(0)) \rangle}$ is the same as in the pure system.

\section{Cardy Ostlund: equilibrium dynamics}

We now turn to dynamics, which, within the EMRG framework can be conveniently studied
by introducing an infrared cutoff on
space only, keeping the full time dependence. 

\subsection{Model and propagator.}

Within this EMRG framework we want to study the dynamics of the
model (\ref{HamCO})
\cite{goldschmidt-dynamics-co,tsaishapir},
described by a Langevin type equation :
\begin{eqnarray}
\eta \frac{\partial}{\partial t} u_{xt} = - \frac{\delta H^{\text{CO}}}{\delta
u_{xt}} + \zeta(x,t) \label{Eq_Langevin}
\end{eqnarray}
where $\langle \zeta(x,t) \rangle = 0$ and
$\langle \zeta(x,t) \zeta(x',t') \rangle = 2\eta T \delta(x-x')
\delta(t-t') $ is the thermal noise and $\eta$ the friction
coefficient. A convenient way to study the dynamics is to use the
Martin-Siggia-Rose \cite{msr} generating functional, on which perturbation
theory can be done. Moreover, using the Ito presription, it can be
readily averaged over
the disorder. The disorder averaged generating functional reads
\begin{eqnarray} \label{dynstartaction}
&&Z[j,\hat{j}] = \int {\cal D} u {\cal D} i\hat{u} e^{-S[u,i\hat{u}] +
j:u + \hat{j} : i\hat{u}} \\
&&S[u,i\hat{u}] = S_0[u,i\hat{u}] + S_{\text{int}}[u,i\hat{u}]
\nonumber \\
&&S_0[u,i\hat{u}] = \int_{qt} i\hat{u}_{-qt} (\eta \partial_t + cq^2)
u_{qt} - \eta T \int_{xt}i\hat{u}_{xt} i\hat{u}_{xt} \nonumber \\
&&S_{\text{int}}[u,i\hat{u}] = -  g_0 \Lambda_0^2 \int_{xtt'} i\hat{u}_{xt}
i\hat{u}_{xt'}
\cos{(u_{xt}-u_{xt'})} \nonumber
\end{eqnarray}
where $\int_t = \int_{t_i}^\infty dt$, where in this section the initial
time $t_i$ is sent to $t_i = - \infty$ before taking $\Lambda_l/\Lambda_0$
large, in order to describe equilibrium dynamics.

In our formulation (\ref{startaction}), the field $\phi$ is now a two
components vector
\begin{eqnarray} \label{vector_dyn}
\phi_{xt} = \left (
\begin{array}{c}
u_{xt}  \\ i \hat{u}_{xt}
\end{array}
\right )
\end{eqnarray}
and from $S_0$ in (\ref{dynstartaction}), we compute
the inverse bare propagator $G_l^{-1}$
\begin{eqnarray}\label{propag_dyn}
&&G_l^{-1}(q) =
\left(
\begin{array}{c c}
0 & \delta(t-t')(- \eta \partial_t + cq^2) \\
\delta(t-t') (\eta \partial_t + cq^2) & -2 \eta T \delta(t-t')
\end{array}
\right) \nonumber \\
&&\times \frac{1}{c(q^2/2\Lambda_0^2) - c(q^2/2\Lambda_l^2)   }
\end{eqnarray}
By inverting this matrix we obtain the bare response and
correlation functions
\begin{eqnarray} \label{barefunc}
&&{C_l}^q_{tt'} = {C_l}^q_{t't}= \overline{\langle u_{qt} u_{-qt'}
\rangle} \\
&& = \frac{T}{q^2}e^{-q^2 |t-t'|} (c(q^2/2\Lambda_0^2)
-  c(q^2/2\Lambda_l^2)) \nonumber \\
&& {R_l}^q_{tt'} = \overline{\frac{\delta \langle u_{qt}
\rangle}{\delta h_{-qt'}}} =
\overline{\langle u_{qt} i\hat{u}_{-qt'} \rangle} \nonumber \\
&& = \theta(t-t') e^{-q^2(t-t')}(c(q^2/2\Lambda_0^2)
-  c(q^2/2\Lambda_l^2))
\end{eqnarray}
where we have set the bare $\eta = 1$.
As we consider here the equilibrium dynamics of the system, the time
translation invariance (TTI)
and the fluctuation dissipation theorem (FDT) hold. These properties hold to
all orders
in perturbation theory and, as we will see has strong consequences on the
structure of the effective action $\Gamma_l(u,i\hat{u})$.
This means for the dressed (i.e. exact) response and correlation
functions:
\begin{eqnarray}
&&{{\cal C}_l}^q_{tt'} = {{\cal C}_l}^q_{t-t'} \\
&& {{\cal R}_l}^q_{tt'} = {{\cal R}_l}^q_{t-t'} \\
&& {{\cal R}_l}^q_{t-t'} = -\theta(t-t') \frac{1}{T} \partial_t {{\cal C}_l}^{q}_{t-t'}
\end{eqnarray}

\subsection{Response function and dynamical exponent.}

We will study the dynamics near the transition temperature $T_c$ (\ref{Tc}),
below which the lowest harmonic of the disordered pontential becomes
relevant. Near $T_c$, we showed
previously that the higher harmonics, although generated by
perturbation theory,
are irrelevant.

As we considered here static disorder, the average over the disorder
generates an effective interaction $S_{\text{int}}[u,i\hat u]$ in
(\ref{dynstartaction}) which is non
local in time, so we
expect the
friction coefficient to be renormalized by the disorder.
We therefore
construct the effective action to order one in $\tau = (T-T_c)/T_c$,
and extract
the dynamical exponent $z$ from the response function.
In the starting dynamical action (\ref{dynstartaction}) the
interacting part
is purely local in space, so to order one the interacting part of the
associated effective action $\Gamma_l(u,i\hat u)$ will remain so. We therefore
search a perturbative solution of the equation for $\Gamma_l(u,i\hat u)$ of
the form (\ref{expmult}):
\begin{eqnarray} \label{local_dyn}
&& {\cal U}_l (u,i\hat u) = \int_{x}  U_l (u_x,i\hat u_x) \\
&& = \int_{xt} i\hat{u}_{xt}{F_l}_t(u_x) -
\frac{1}{2}\int_{xtt'} i\hat{u}_{xt} {\Delta_l}_{tt'}(u_x)
i\hat{u}_{xt'}
\end{eqnarray}
where ${F_l}_t(u_x)$ and ${\Delta_l}_{tt'}(u_x)$ are functionals only with respect to the
time dependence, i.e. functions of the ''vector''
$u_x\equiv\{u_{xt}\}$ at a given point $x$ in space. In addition these
will acquire an {\it explicit} time dependence, indicated by their $t$ and $t'$ indices.
One has the initial conditions
\begin{eqnarray} \label{CI_flow}
&& {\Delta_{l=0}}_{tt'}(u) = 2 g_0 \Lambda_0^2 \cos{(u_{t} -
u_{t'})} \nonumber \\
&& {F_{l=0}}_t(u) = 0
\end{eqnarray}
The ${F_l}_t(u)$ term is indeed generated by perturbation
theory and is related - in the case
of equilibrium
dynamics - to ${\Delta_l}_{tt'}(u)$ by a generalized FDT relation,
namely a Ward Identity,
which can be written to lowest order:
\begin{eqnarray}
\frac{\delta {F_l}_t(u)}{\delta u_{xt'}} = - \frac{1}{T} \partial_{t'}
{\Delta_l}_{tt'}(u) \quad t > t'
\label{genFDT}
\end{eqnarray}
where $\partial_{t'}$ acts only on the explicit time dependence
(i.e. not on $u_{t'}$).
Notice finally that terms containing higher powers of the field $i\hat{u}$, ie
 $(i\hat{u})^{p+2}$ are of order $\tau^{p+1}$. They correspond to
higher cumulant of the disorder (i.e. higher number of replica
terms in the statics).
The exact RG equation to order one (\ref{RGorder2}) then reads
 (see Appendix \ref{app_dyncoeq})
\begin{eqnarray}
&& \partial_l {\Delta_l}_{tt'}(u) = \int_{t_1t'_1} k^{(1)}_{lt_1t'_1}
{\Delta_l}_{tt'}(u) \label{ERGE_dyn_order_one} \\
&&\partial_l {F_l}_t(u) = \int_{t_1t'_1} k^{(1)}_{lt_1t'_1} {F_l}_t(u)
- \int_{t_1>t'_1}k^{(2)}_{lt_1t'_1} {\Delta_l}_{tt'_1}(u) \nonumber
\end{eqnarray}
with
\begin{eqnarray}
&&k^{(1)}_{lt_1t'_1} = \frac{1}{2} \frac{\delta}{\delta u_{xt_1}}
\partial_l {C_l}^{x=0}_{t_1t'_1} \frac{\delta}{\delta u_{xt'_1}}
\nonumber \\
&&k^{(2)}_{lt_1t'_1} = \frac{\delta}{\delta u_{xt_1}}
\partial_l {R_l}^{x=0}_{t_1t'_1}
\end{eqnarray}
The solution of this coupled set of equations
(\ref{ERGE_dyn_order_one}) together with (\ref{CI_flow}) is given  by
\begin{eqnarray}\label{loc_co_dyn_eq}
&& {\Delta_l}_{tt'}(u) = 2 \Lambda_l^2 g_l e^{{C_l}_{t-t'}^{x=0}}
\cos(u_{xt} - u_{xt'})  \\
&& \frac{\delta{F_{lt}}(u)}{\delta u_{xt'}} = - 2 \Lambda_l^2 g_l
e^{{C_l}_{t-t'}^{x=0}}
{R_{l}}_{t-t'}^{x=0} \cos(u_{xt} - u_{xt'}) \quad t>t' \nonumber
\end{eqnarray}
where we can check explicitly the previously mentioned generalized FDT relation (\ref{genFDT}).
Finally, as we consider here static disorder, the flow of $g_l$ is given by the
previous study, the fixed point value $g^*$ being given by (\ref{co_fp}).

From $\Gamma_l[u,i\hat{u}]$, we obtain the response function in the
following way
\begin{eqnarray}\label{Exprreponse}
{\cal R}^q_{l t-t'} = \langle u_{qt} i\hat u_{-qt'} \rangle =
\left( \frac{\delta^2 \Gamma_{l}}{\delta i\hat u_{qt_1} \delta
u_{-qt'_1}}\Big|_{u=i\hat u = 0}  \right)^{-1}
_{q,t,t'}
\end{eqnarray}
We define
\begin{eqnarray} \label{def_sigma_d}
&&D_{ltt'} = \Delta_{ltt'}(u=0) \nonumber \\
&&\Sigma_{ltt'} = \frac{\delta
{F_l}_t(u)}{\delta u_{xt'}}\Big|_{u=0}
\end{eqnarray}
notice that in the case of equilibrium dynamics $D_{ltt'} =
D_{lt-t'}$ and $\Sigma_{ltt'} =\Sigma_{lt-t'}$.
One gets
\begin{equation}
\frac{\delta^2 \Gamma_l}{\delta i\hat{u}_{qt} \delta
u_{-qt'}}\Big|_{u = i\hat u =0} = \delta(t-t')(q^2 + \partial_t) +
\Sigma_{lt-t'} \label{def_sigma}
\end{equation}
When considering equilibrium dynamics, the use of Fourier transform
allows to compute
this matrix element (\ref{Exprreponse}) in a simple way
\begin{eqnarray}
{\cal R}^q_{l\omega} = \frac{1}{q^2 - i\omega + \Sigma_{l\omega}}
\end{eqnarray}
where
${\cal R}^q_{lt} = \int \frac{d\omega}{2\pi} e^{- i \omega t} {\cal
R}^q_{l\omega}$
and $\Sigma_{l\omega}$ is the Fourier transform,  of
$\Sigma_{lt-t'}$.
In Appendix \ref{app_dyncoeq} we show that it has the following form
(up to terms of
order $\Lambda_l^2/\Lambda_0^2$)
\begin{eqnarray}\label{expr_sigma}
&&\Sigma_{l\omega} = i\omega B_l
( \ln{\frac{\Lambda_l^2}{\Lambda_0^2}} +
\chi^{\text{(dyn)}}(\frac{\omega}{\Lambda_l^2})   ) \\
&&B_l = \frac{g_l e^{\int_a \ln{2a}}}{2 T_c}
\end{eqnarray}
whith the following asymptotic behaviors
\begin{eqnarray}
&& \chi^{\text{(dyn)}}(\nu) \sim a_{\text{dyn}} \ln{\nu} \quad \nu \ll 1
\label{chi_dyn_small} \\
&& \chi^{\text{(dyn)}}(\nu) \sim \ln{\nu} \quad \nu \gg 1
\label{chi_dyn_large}
\end{eqnarray}
where $a_{\text{dyn}}$ is a non universal constant. The large argument
behavior of $\chi^{\text{(dyn)}}$ (\ref{chi_dyn_large}) allows to take
the large $l$ limit in (\ref{expr_sigma}) as the logarithmic
divergence is cancelled, which gives
\begin{eqnarray} \label{reponse_fp}
&&\lim_{l \to \infty} {\cal R}^q_{l\omega} = \frac{1}{q^2 - i\omega + i\omega
B^* \ln{\frac{\omega}{\Lambda_0^2}} } \\
&& B^* = \lim_{l \to \infty} B_l = e^{\gamma_E} \tau
\end{eqnarray}
where we have used (\ref{co_fp}) to compute $B^*$ which is universal~:
the cut-off dependence encoded in $e^{2\int_a \ln{2a}}$ has disappeared.
On the other hand we expect that the scaling function in Fourier should
read:
\begin{eqnarray} \label{reponse_fourier}
&&\lim_{l \to \infty} {\cal R}^q_{l\omega} =
\frac{1}{ q^2 - i \omega (\frac{\omega}{\Lambda_0^2})^{2/z - 1} }
\end{eqnarray}
from scaling. If the initial model possess STS then
the coefficient of $q^2$ is fixed to unity. The $q$-independence
of the self-energy is expected to hold only to the
order in $\tau$ that we are working at, and it should be corrected
by higher loops. Expansion of the denominator of (\ref{reponse_fp}) coincides
with the expansion to order $\tau$ of the denominator of
(\ref{reponse_fourier}) and yields the universal
value of the dynamical exponent $z$
\begin{eqnarray}\label{expz}
z - 2 = 2 B^* = 2 e^{\gamma_E} \tau + O(\tau^2)
\end{eqnarray}
in agreement with previous studies.

It is interesting in view of later applications
to non-equilibrium dynamics, and a useful check,
to compute this response function in the time domain.
Indeed, writing simply the identity
$\Gamma_l^{(2)}{\Gamma_l^{(2)}}^{-1} = \mathbb{I} $, where $\Gamma_l^{(2)}$ is
the matrix of the second functional derivatives of the effective action
with respect ot the fields $u_{xt}$  and $i\hat u_{xt}$, we obtain a
system of closed equations for the exact response and correlation
functions ${{\cal R}_l}^{xx'}_{tt'}$ and ${{\cal C}_l }^{xx'}_{tt'}$
to  order one (more generally, ${F}_{lt}(u)$ and ${\Delta_l}_{tt'}$ can
be bilocal in space)
\begin{widetext}
\begin{eqnarray}
&&\partial_{t} {{\cal R}_l}_{tt'}^{xx'} - \nabla^2 {{\cal R}_l}_{tt'}^{xx'} +
\int_{t_i}^t d {t_1} \Sigma_{ltt_1}
{{\cal R}_l}_{t_1t'}^{xx'} =
\delta(t-t')\delta(x-x')  \label{eqR} \\
&&\partial_t {{\cal C}_l}_{tt'}^{xx'} - \nabla^2 {{\cal
C}_l}_{tt'}^{xx'} +
\int_{t_i}^t dt_1 \Sigma_{ltt_1}
{{\cal C}_l}_{t_1 t'}^{xx'}  = 2 \eta T {{\cal R}_l}_{t't}^{xx'} +
\int_{t_i}^{t'}d t_1 D_{ltt_1} {\cal R}_{t't_1}^{xx'}
 \label{eqC}
\end{eqnarray}
\end{widetext}
We remind that we have chosen the Ito presription, which
fixes the following initial condition for the response function
\begin{eqnarray}
&& \lim_{\epsilon \to 0} {{\cal R}_l}_{t,t-\epsilon} = 1 \nonumber \\
&& {{\cal R}_l}_{t,t} = 0 \label{Ito}
\end{eqnarray}
Before using these equations to study non-equilibrium dynamics, we
show how the equation for the response (Eq. \ref{eqR}) function allows
to recover the dynamical exponent $z$. Using (\ref{def_sigma_d})
together with (\ref{loc_co_dyn_eq}) and TTI (which holds for
equilibrium dynamics),
the equation for the response function reads
\begin{eqnarray}
&&(\partial_t + q^2) {{\cal R}_l}_{t-t'}^{q} = \label{eqResp} \\
&&2 g_l
\Lambda_l^2\int_{-\infty}^{t} dt_1 R_{lt-t_1}^{x=0} e^{C^{x=0}_{lt-t_1}}
  ({{\cal R}_l}_{t_1 - t'}^{q} - {{\cal R}_l}_{t-t'}^{q}) \nonumber
\end{eqnarray}
The limit $l \to \infty$ is taken
as explained in the Appendix \ref{app_dyncoeq} (\ref{sigma_t_fp}), and
a way to solve this
equation is simply to say that in the rhs, we may replace ${{\cal
R}_l}_{t_1t'}^{q}$, by its bare value, which is simply $\theta(t_1 -
t') e^{-q^2(t_1 - t')}$  as this term is already of order $\tau$.

One expects that the response function
can be written as:
\begin{eqnarray}
{\cal R}_{t-t'}^{q} = \lim_{l\to \infty} {\cal R}_{l,t-t'}^{q} =
\tilde{q}^{z-2} F_R^{\text{eq}}(\tilde q^{z} (\tilde t- \tilde t')) \label{scalingf}
\end{eqnarray}
where $\tilde q = q/\Lambda_0$, $\tilde t = t \Lambda_0^2$, $\tilde t' = t' \Lambda_0^2$,
with $F_R^{\text{eq}}$ a universal scaling function (up to an
overall non universal scale) such that
$F_R^{\text{eq}}(v) \sim v^{(2-z)/z}$ for $v \to 0$. As a function it
admits an expansion in powers of $\tau$, obtained as:
\begin{eqnarray}\label{perturbscaleqexpl}
&& F_R^{\text{eq}}(v) = F_R^0(v) + \tau F_R^{1\text{eq}}(v) + O(\tau^2) \\
&& F_R^{0}(v) = e^{- v} \nonumber \\
&& F_R^{1\text{eq}}(v) = e^{\gamma_E} ((v-1) Ei(v) e^{- v} + e^{- v} - 1)
\nonumber \\
&&  v = \tilde q^{z} (\tilde t - \tilde t')
\nonumber
\end{eqnarray}
as shown in the appendix \ref{app_dyncoeq}. This is
established by identifying the direct expansion
of (\ref{scalingf}) in terms of the argument 
$v' = \tilde q^{2} (\tilde t - \tilde t')$:
\begin{eqnarray}\label{perturbscaleq}
&& {\cal R}_{\tilde{t}}^{\tilde{q}} = F_R^0(v') +
(z-2) \ln \tilde q ( F_R^0(v')
+ v'  F_R^{0'}(v') ) \\
&& + \tau
F_R^{1\text{eq}}(v') + O(\tau^2)  \nonumber
\end{eqnarray}
with the result of solving (\ref{eqResp}). Note
that the term proportional to $\ln \tilde q$ has
precisely the expected $v$ dependence, a check
of the calculation. Since there is an overall
non-universal scale $\tilde{q} \to \lambda \tilde{q}$,
$F_R^{1\text{eq}}(v)$ is defined up to a
change in the constant $\rho$
defined in the Appendix \ref{app_dyncoeq} (\ref{rho2}).

One can check explicitly that the scaling function
in the time domain obtained by this second method coincides
with the inverse Fourier transform of (\ref{reponse_fourier}) to the
lowest order in $\tau$. The asymptotic behavior of the scaling function
in the time domain is:
\begin{eqnarray}
&&  F_R^{1\text{eq}}(v) \approx e^{\gamma_E} \ln( 1/(e^{\gamma_E} v))  \quad ,
\quad v \to 0 \label{FReq_asymp_small} \\
&&  F_R^{1\text{eq}}(v) \approx e^{\gamma_E} v^{-2} \quad , \quad v \to
\infty \label{FReq_asymp_large}
\end{eqnarray}
the slow time decay $1/t^{1 + \frac{2}{z}}$, for $z>2$, arises from
the disorder. Notice that a similar power law tail for large
$\tilde{q}^z\tilde{t}$ has already been obtained for the diluted Ising
model \cite{kissner_random_mass}.

Using the FDT we also obtain the equilibrium correlation function
in the scaling regime as:
\begin{eqnarray}
&& {{\cal C}}_{t t'}^{q} = T \tilde{q}^{-2} F_C^{\text{eq}}(\tilde{q}^z (\tilde t - \tilde t')) \\
&& F_C^{\text{eq}}(v) = \int_v^{+\infty} dw F_R^{\text{eq}}(w)
\end{eqnarray}

We conclude this section on equilibrium dynamics by noticing a
few interesting properties. The first one is an exact consequence of
the scaling form (\ref{scalingf}) combined with the
STS. Indeed, the STS imposes:
\begin{eqnarray}\label{STS_resp}
\lim_{t \to \infty} \int_{\tilde{t_i}}^{\tilde{t}} dt'{\cal
R}^{\tilde{q}}_{\tilde{t}\tilde{t'}} =
\frac{1}{\tilde{q}^2}
\end{eqnarray}
Using the scaling property we showed previously, this symmetry
(\ref{STS_resp}) implies
\begin{eqnarray}
\int_0^{\infty} dt {\tilde{q}}^{z-2} F_R^{\text{eq}}(\tilde{q}^z t) =
\frac{1}{\tilde{q}^2} \Rightarrow \int_0^{\infty} du F_R^{\text{eq}}(u) = 1
\end{eqnarray}
from which it follows that
\begin{eqnarray}\label{auto_func_STS}
&&{\cal R}^{x=0}_{tt'} = \int_q {\tilde{q}}^{z-2}
F_R^{\text{eq}}(\tilde{q}^z
(\tilde{t}-\tilde{t'}))  = \frac{1}{2\pi z (t- t')}
\nonumber \\
&&{\cal C}^{x=0}_{\tilde{t}\tilde{t'}} = \frac{T}{2 \pi z}
\ln{(t - t')}
\end{eqnarray}
where we have used FDT in the last line. Note that 
the unrescaled time $t$ appears in these formulae. 
Although the 
scaling form (\ref{scalingf}) is only valid for small $\tilde{q}$
we believe that the behaviors (\ref{auto_func_STS})
may actually be the exact leading ones in the
large $t-t'$ limit, their coefficients being fixed
(non-perturbatively) by the STS. This would
be interesting to check numerically.

The second property is a comparison with the so-called Porod's
law \cite{bray_porods}.
If the form (\ref{reponse_fourier}) were to hold to all orders,
the scaling functions would decay at large
arguments as $F^{\text{eq}}_R(v) \sim 1/v^{1 + \frac{2}{z}}$ and
$F^{\text{eq}}_C(v) \sim 1/v^{\frac{2}{z}}$. That yields
\begin{eqnarray} \label{auto_func_STS} 
&& {\cal C}^{q}_{\tilde{t} \tilde{t'}} \sim
\frac1{(t - t')^{2/z} q^4}
\end{eqnarray}
as in the Porod's law with $d=2$ and $n=2$ \cite{bray_porods}.
Here this property holds to the order of our calculation
$O(\tau)$.

\section{Non-equilibrium dynamics of the CO model}

Applying standard scaling arguments, we expect
${\cal R}^q_{tt'}$ and ${\cal C}^q_{tt'}$
to be functions of the scaling variables $\tilde{q}^z \tilde{t}$ and
$\tilde{q}^z \tilde{t'}$ where $\tilde{q} = q/\Lambda_0$ and
$\tilde{t} = \Lambda_0^2 t$ and
$z$ is the dynamical exponent. As is
the case for pure systems at a critical point one
can write from RG arguments \cite{janssen_noneq_rg}
with little restriction:
\begin{eqnarray}
&&{\cal R}^{\tilde{q}}_{\tilde{t}\tilde{t'}} =
\tilde{q}^{-2+z+\eta}\left( \frac{\tilde{t}}{\tilde{t'}} \right)^{\theta}
F_R(\tilde{q}^z(\tilde{t}-\tilde{t'}),\tilde{t}/\tilde{t'})
\label{janssenscalingresp} \\
&&{\cal C}^{\tilde{q}}_{\tilde{t}\tilde{t'}} = T
\tilde{q}^{-2+\eta}\left( \frac{\tilde{t}}{\tilde{t'}} \right)^{\theta}
F_C(\tilde{q}^z(\tilde{t}-\tilde{t'}),\tilde{t}/\tilde{t'})
\label{janssenscalingcorr}
\end{eqnarray}
where the exponent $\theta$ is defined by imposing the following
behavior of the response scaling function
$F_{R}(v,u)$ when $u~\to~\infty$~:
\begin{eqnarray}\label{asymptexpr}
&&F_{R}(v,u) = F_{R,\infty}(v) + O(u^{-1})
\end{eqnarray}
This has been checked for pure systems
\cite{janssen_noneq_rg,
godreche_spherique_fdr,henkel_noneq_conforme,
calabrese_on_twoloops}
and, partially for one
case of a disordered system (only for the response function in
\cite{kissner_random_mass} and for the Fourier mode $q=0$ for both
functions in \cite{calabrese_fdr_randommass}). It
was found in all the pure cases that
one also has
\begin{eqnarray}
F_C(v,u) = \frac{F_{C,\infty}(v)}{u} + O(u^{-2}) \label{Fc}
\end{eqnarray}
These forms (\ref{asymptexpr},\ref{Fc}) yield a non trivial Fluctuation Dissipation
Ratio (FDR) characterizing the violation of the Fluctuation
Dissipation Theorem (FDT) \cite{cugliandolo_fdr,cugliandolo_leshouches}.
It has been 
computed exactly for the spherical model in $d>2$
\cite{godreche_spherique_fdr}, using dynamical RG methods for the
pure $O(N)$ model at criticality up to two loops in an
$\epsilon = 4-d$ expansion
\cite{calabrese_on_twoloops}, and up to one loop
for the critical diluted Ising model in a $\sqrt{\epsilon}$ expansion
\cite{calabrese_fdr_randommass}.

Another standard definition for the autocorrelation exponent
$\lambda_C$ \cite{fisher_huse_noneq,bray_lambda,huse_lambda} and for
the autoresponse exponent $\lambda_R$ \cite{picone_autoresponse} is:
\begin{eqnarray}
&& {\cal C}^{\tilde{q}}_{\tilde{t}\tilde{t'}} =
\tilde t^{\frac{d- \lambda_C}{z}} \phi_C(\tilde{q}^z \tilde{t})
\label{asymptexprla_C} \\
&& {\cal R}^{\tilde{q}}_{\tilde{t}\tilde{t'}} =
\tilde t^{\frac{d- \lambda_R}{z}} \phi_R(\tilde{q}^z \tilde{t})
\label{asymptexprla_R}
\end{eqnarray}
in the limit $\tilde{t} \to \infty$, $\tilde{q} \to 0$ with
$\tilde{t'}$ fixed and $\tilde{q}^z \tilde{t}$ fixed, with
$\phi_{R,C}(0)=Cst$. Assuming the
behaviors (\ref{asymptexpr},\ref{Fc}) one finds the connection:
\begin{eqnarray}\label{connection_lambda}
&& (d - \lambda_C)z^{-1} = \theta - 1 + (2 - \eta)z^{-1}  \\
&& \lambda_R = \lambda_C \nonumber \\
&& \phi_C(v) = T v^{\frac{\eta -2}{z}} F_{C,\infty}(v) (t')^{1-
\theta} \nonumber \\
&& \phi_R(v) =  v^{\frac{\eta -2 + z}{z}} F_{R,\infty}(v) (t')^{
\theta} \nonumber
\end{eqnarray}
which seems to hold for pure models, together with the inequality $d/2
\leq \lambda_C = \lambda_R$ \cite{fisher_huse_noneq,yeung_bound_lambda}.

For the nonequilibrium dynamics of the CO model, we obtain similar
scalings  (\ref{janssenscalingresp}, \ref{janssenscalingcorr}),
($\eta = 0$ in this case because of STS) but with a {\it different
asymptotic behavior} at large $u$ of the
scaling function $F_{C}(v,u)$. As we will see, this has strong
consequences on the FDR.
Note that although ${\cal C}^{\tilde{q}}_{\tilde{t}\tilde{t'}}$
is the full correlation function, to this order in
the $\tau$ expansion it coincides with the connected one
(which is the correct one to consider e.g. to obey FDT in the
equilibrium regime), the difference between the two being
of order $g^2 = O(\tau^2)$.

\subsection{General framework}

We want to study the dynamics of the system described by
(\ref{Eq_Langevin}) which, at the initial time $t_i = 0$, is in a non equilibrium configuration
$u_{xt_i} = u_x^0$, whose statistical weight is given by
$e^{-H_0[u^0]}$ (where $H_0[u_0] \neq H_{CO}[u_0]$). The
general framework to incorporate this feature in the MSR formalism has
been developped in \cite{janssen_noneq_rg}, and it amounts to describe
the system in terms
of the generating functional $S[u,i\hat{u}] \to S[u,i\hat{u}] +
H_0[u_x^0]$. If the system is prepared in a high temperature state,
with short range correlations $\langle u^0_{x} u^0_{x'}\rangle =
m_0^{-2} \delta^d(x-x')$, the corresponding $H_0[u^0]$ is given by
\begin{eqnarray}
H_0[u^0] = \frac{m_0^2}{2} \int_x (u_x^0)^2
\end{eqnarray}
any addition of anharmonic terms in $H_0[u^0]$ is irrelevant as long
$m_0^2 \neq 0$. Moreover by power counting one has that $m_0^{-2}$ is
irrelevant \cite{janssen_noneq_rg}, so that to study the leading
scaling behavior it is
sufficient to assume $m_0^{-2} = 0$, i.e. $u_x^0=0$. The effect of this nonequilibrium
initial condition is then completely encoded in the lower bound $t_i =
0$ on the time integrals in the MSR functional (\ref{dynstartaction}).
The running bare response and correlation
functions are given by \cite{janssen_noneq_rg}:
\begin{eqnarray}\label{Dirichlet}
&&{R_l}^{q}_{tt'} = \theta(t-t')
e^{-q^2(t-t')}\left(c(\frac{q^2}{2\Lambda_0^2})
-  c(\frac{q^2}{2\Lambda_l^2})\right) \\
&&{C_l}^q_{tt'} =
\frac{T}{q^2}\left(e^{-q^2 |t-t'|} - e^{-q^2(t+t')}
\right)\left(c(\frac{q^2}{2\Lambda_0^2})
-  c(\frac{q^2}{2\Lambda_l^2})\right) \nonumber
\end{eqnarray}

\subsection{Nonequilibrium response function}

In order to compute the response function, we solve
perturbatively the equation for ${\cal R}^q_{tt'}$ (\ref{eqR})
using the trick explained above, i.e. replacing the exact ${\cal
R}^q_{tt'}$ in the rhs of (\ref{eqR}) by its bare value. Doing
this, we obtain a perturbative expansion of the exponents $z$ (already
obtained previously (\ref{expz})), $\theta$ and of the scaling function
$F_R(v,u)$
in the same spirit as (\ref{perturbscaleqexpl}). Indeed, as shown in
Appendix \ref{app_dyncononeq}, one has the scaling (\ref{janssenscalingresp}),
in terms of the scaling variables $v =
\tilde{q}^z (\tilde{t}-\tilde{t'})$ and  $u = \tilde{t}/\tilde{t'}$ with
\begin{eqnarray}\label{expr_FR_gen}
&&F_R(v,u) = F_R^{0}(v) + \tau F^1_R(v,u) \\
&&F^1_R(v,u) = F^{1\text{eq}}_R(v) + F^{1\text{noneq}}_R(v,u) \nonumber \\
&&\theta = e^{\gamma_E} \tau + O(\tau^2) \nonumber 
\end{eqnarray}
which is established by comparison with the direct perturbative expansion of
(\ref{janssenscalingresp}) in powers of $\tau$:
\begin{eqnarray}
&&{\cal R}^{\tilde{q}}_{\tilde{t}\tilde{t'}} = F^{0}_R(v') + (z-2)
\ln{\tilde{q}} (F^{0}_R(v') + v'
F^{0'}_R(v')) \nonumber \\
&&+ \theta \ln{u}F^{0}_R(v') + \tau F_{R}^{1}(v',u)
+ O(\tau^2)  \label{pertresp} 
\end{eqnarray}
with $v'=\tilde{q}^2 (\tilde{t}-\tilde{t'})$ and
$F_{R}^{1\text{eq}}(v)$ is given 
by (\ref{perturbscaleqexpl}) and $F_{R}^{1\text{noneq}}(v,u)$
given in (\ref{exprF1R})
has a complicated expression left in the Appendix
(\ref{exprF1RR}). However its asymptotic behaviors, which
we now focus on have remarkably simple forms. First,
in order to compare with the prediction for pure critical systems
 one is interested in the limit of large $u$,
keeping $v$ fixed. This defines
$F_{R,\infty}(v)$ (\ref{asymptexpr}) which, we find to be:
\begin{eqnarray}\label{FRlarge_u}
&&F_{R,\infty}(v) = e^{-v} + e^{\gamma_E}\tau \Big \{ -
\sqrt{\pi v}\Erf{\sqrt{v}}  \nonumber \\
&&- e^{-v}\Big( (1-v)\ln{(4ve^{\gamma_E})}
 \nonumber \\
&&- 2
v(v-\frac{1}{2})_2F_2(\{1,1\},\{\frac{3}{2},2\},v) \Big) \Big \} + O(\tau^2)
\end{eqnarray}
where $\Erf(z)$ is the error function and
$_2F_2(\{1,1\},\{\frac{3}{2},2\},z)$ is a generalized
hypergeometric series \cite{gradstein,wolfram,bateman}.
This shows that the response function has a scaling behavior as
predicted for pure systems at a critical point
(\ref{asymptexpr}). The small $v$ behavior of $F_{R,\infty}(v)
\sim 1 - e^{\gamma_E}\tau \ln{v}$ shows that $\phi_R(v)$
(\ref{connection_lambda}) has a
good limit when $v \to 0$, $\phi_R(0) = Cst$, and
this gives the autocorrelation exponent
$\lambda_R$ (\ref{asymptexprla_R}):
\begin{eqnarray}
\lambda_R = 2 + O(\tau^2)
\end{eqnarray}
It is also interesting to analyze the asymptotic behavior in the
limit of large $v$
(and in particular $v \gg u$ ), keeping $u$ fixed. This limit is
relevant e.g. to study the behavior at fixed $q$, large
$t,t'$ with $u=\tilde{t}/\tilde{t'}$ fixed. It is
obtained from (\ref{exprF1R}) as explained in the Appendix
\ref{app_dyncononeq}. The behavior of the response function in this
limit is then given by
\begin{eqnarray}\label{FRlarge_v}
&&\lim_{v \to \infty, u \text{fixed}} F_R(v,u)
 \sim e^{-v} + \frac{\tau}{v^2}
P_R(u) + O(\tau v^{-3},\tau^2) \nonumber \\
&&P_R(u) = e^{\gamma_E} \frac{u+1}{2 \sqrt{u}}
\end{eqnarray}
Notice that in the limit $u \to 1$ we recover the result of the
equilibrium dynamics (\ref{FReq_asymp_large}). This is more
general as one can check from (\ref{exprF1RR}) 
that $F^{\text{noneq}}_R(v,u) = O((u-1)^2)$ as 
$u \to 1$. Finally, one must keep in mind that the
limits $v \to \infty$ and $u\to \infty$ do not commute,
indeed one expects that a scaling function of $v/u \sim q^z t'$
interpolates between these limits, left for future investigation.

Another interesting behavior is the limit of vanishing
momentum $\tilde{q} = 0 $, the so called diffusion mode.
Although well defined, this limit is a bit
peculiar due to the prefactor $\tilde{q}^{z-2}$ in the scaling function
(\ref{janssenscalingresp}). However, the function $F_R(v,u)$ behaves
when $v \to 0$ in such a way to cancel this divergence as in
(\ref{FReq_asymp_small}) and
leads to well a defined response function ${\cal
R}^{\tilde{q}=0}_{\tilde{t}\tilde{t'}}$ which has the scaling form
\begin{eqnarray}
&&{\cal R}^{\tilde{q}=0}_{\tilde{t}\tilde{t'}} =
\frac{1}{(\tilde{t}-\tilde{t'})^{(z-2)/z}}
\left( \frac{\tilde{t}}{\tilde{t'}} \right)^{\theta} F_R^{\text{diff}}\left(
\frac{\tilde{t}}{\tilde{t'}}\right) \\
&&F_R^{\text{diff}}(u) = F_R^{\text{diff} 0}(u) + \tau
F_R^{\text{diff} 1}(u) \nonumber \\
&&F_R^{\text{diff} 0}(u) = 1 \nonumber \\
&&F_R^{\text{diff} 1}(u) = 2 e^{\gamma_E} \ln
\left(\frac{1+ \sqrt{u}}{2 \sqrt{u}}  \right) \nonumber
\end{eqnarray}
which is identified with the perturbative expansion of ${\cal
R}^{\tilde{q}=0}_{\tilde{t}\tilde{t'}}$ straighforwardly obtained from
the general expression (\ref{exprF1RR}).

\subsection{Nonequilibrium correlation function.}

To compute the correlation function, instead of solving the equation
for ${\cal C}^q_{ltt'}$ (\ref{eqC}), we obtain it using the following
formal solution for $\tilde{t} > \tilde{t'}$
\begin{eqnarray}\label{start_correl}
&&{\cal C}^{\tilde{q}}_{\tilde{t}\tilde{t'}} = \lim_{l \to \infty} {\cal
C}^{\tilde{q}}_{l\tilde{t}\tilde{t'}} \\
&& = 2T\int_0^{\tilde{t'}} dt_1 {{\cal R}}^{\tilde{q}}_{\tilde{t}t_1}{{\cal
R}}^{\tilde{q}}_{\tilde{t'}t_1}
+ \int_0^{\tilde{t}} dt_1 \int_0^{\tilde{t'}} dt_2{{\cal
R}}^{\tilde{q}}_{\tilde{t}t_1} D_{t_1t_2}
{{\cal R}}^{\tilde{q}}_{\tilde{t'} t_2} \nonumber
\end{eqnarray}
where $D_{t_1t_2} = \lim_{l \to \infty} D_{lt_1t_2}$ is defined in
(\ref{def_sigma_d}) and explicitly given in
(\ref{d_noneq}),
that we expand perturbatively using the expression we obtained for
${\cal R}^{\tilde{q}}_{\tilde{t}\tilde{t'}}$ (\ref{pertresp}).
In the Appendix, we show that
${\cal C}^{\tilde{q}}_{\tilde{t}\tilde{t'}}$ has the following scaling form
(\ref{janssenscalingcorr}) with
\begin{eqnarray}
&&F_C(v,u) = F_C^0(v,u) + \tau F_C^1(v,u) \\
&&F_C^0(v,u) = e^{-v} - e^{-v \frac{1+u}{u-1}} \nonumber
\end{eqnarray}
and $F_C^1(v,u)$ given in Appendix. Again, this is established by
identifying the direct perturbative exansion of (\ref{janssenscalingcorr}):
\begin{eqnarray}\label{pert_corr}
&&{\cal C}^{\tilde{q}}_{\tilde{t}\tilde{t'}} =
\frac{T}{\tilde{q}^2}\Big(F_C^0(v',u) + (z-2)\ln{(\tilde{q})}
v'\frac{\partial F_C^0(v',u)}{\partial v'}   \nonumber \\
&&+ \theta \ln{u}F_C^0(v',u) +
\tau F_C^1(v',u)\Big)
\end{eqnarray}
with $v'=\tilde q^2(\tilde t - \tilde t')$, 
which is similar to the scaling form expected for pure systems at a
critical point (\ref{janssenscalingcorr}). However, the large $u$
behavior is {\it different}, indeed
one has in the large $u$ limit, keeping $v$ fixed
\begin{eqnarray}\label{FC_large_u}
&&\lim_{u \to \infty} F_{C}(v,u) \sim \frac{2e^{-v}v}{u} + \tau
\frac{F^1_{C,\infty}(v)}{\sqrt{u}}
  + O(u^{-2},\tau u^{-1},\tau^2) \nonumber \\
&&F^1_{C,\infty}(v) = e^{\gamma_E} e^{-v} \sqrt{\pi v} \Erfi{\sqrt{v}}
\end{eqnarray}
which decays more slowly than the predicted scaling for pure system at
a critical point (\ref{Fc}). Besides, using (\ref{erfi_asymp_small})
$F^1_{C,\infty}(v) \sim
v + O(v^2)$, $\phi_C(0) = Cst$ (\ref{connection_lambda}), this defines
the autocorrelation exponent $\lambda_C$:
\begin{eqnarray}
\lambda_C = d - \frac{z}{2} + O(\tau^2) = 1 - e^{\gamma_E} \tau + O(\tau^2)
\end{eqnarray}
where we have used the explicit expressions of the exponents $z$
(\ref{expz}),
$\theta$ (\ref{expr_FR_gen}) and the relation $(d - \lambda_c)z^{-1} =
\theta - 1/2 + (2 - \eta)z^{-1}$ arising from $1/\sqrt{u}$ decay of
$F^1_{C}(v,u)$ (\ref{FC_large_u}). Note first that $\lambda_C$ is
different from its
trivial value $\lambda_C \neq d$. Besides we note that $\lambda_C \neq
\lambda_R$ and finally that it violates the bound $\lambda_C < d/2$
predicted for pure systems. 

For the correlation function it is also instructive to look at the
asymptotic behavior $v \gg 1$, $u$ fixed. As detailed in the Appendix
\ref{app_dyncononeq} one has
\begin{eqnarray}\label{FC_large_v}
&&\lim_{v\to \infty}F_C(v,u) \sim
F_C^0(v,u) + \frac{\tau}{v} P_C(u) +
O(\tau v^{-2},\tau^2 ) \nonumber \\
&&P_C(u) = P_R(u)
\end{eqnarray}

Finally we also study the correlation function in the limit of
vanishing momentum $\tilde{q} = 0$. As mentionned previously for the
response function this limit is a bit peculiar due to the $q^{-2}$
prefactor in (\ref{janssenscalingcorr}). The small $v$ behavior of
$F_C(v,u)$ leads to the scaling form (up to a non universal scale)
\begin{eqnarray}\label{Cq=0}
&&{\cal C}^{\tilde{q}=0}_{\tilde{t}\tilde{t'}} = 2\tilde{t'} T
\frac{1}{(\tilde{t}-\tilde{t'})^{(z-2)/z}}
\left( \frac{\tilde{t}}{\tilde{t'}} \right)^\theta
F_C^{\text{diff}}(u)  \\
&&F_C^{\text{diff}}(u) = F_C^{\text{diff}0}(u) + \tau F_C^{\text{diff}1}(u)
\nonumber \\
&&F_C^{\text{diff}0}(u) = 1 \nonumber \\
&&F_C^{\text{diff}1}(u) = \frac{1}{2} e^{\gamma_E} \Big( 4 \sqrt{u} + (u+1)
\ln{(u-1)} \nonumber \\
&&- 2(u-1)\ln{(1+\sqrt{u})} - 2 \ln{u} + 6 - 8\ln{4} \Big) \nonumber
\end{eqnarray}
with the asymptotic behaviors
\begin{eqnarray}
&&F_C^{\text{diff}}(u) \sim 1 + \tau e^{\gamma_E} \ln{(u-1)} \quad u
\to 1^+  \\
&&F_C^{\text{diff}}(u) \sim 1 + \tau e^{\gamma_E} \sqrt{u} \quad u
\gg 1 \nonumber
\end{eqnarray}
These behaviors are such that the singularity as $\tilde t - \tilde t' \to 0$
cancels and one finds that the diffusion of the zero mode become anomalous
at large time:
\begin{eqnarray}\label{Cqtt=0}
&& 
{\cal C}^{\tilde{q}=0}_{\tilde{t} \tilde{t'}} \sim
A \tilde{t'}^{2/z}  \\ 
&& A = 2 T_c + O(\tau) \nonumber
\end{eqnarray}
this formula being valid for $\tilde{t} - \tilde{t'} \ll \tilde{t'}$,
the random walk result being recovered when $z=2$. 

\subsection{Fluctuation Dissipation Ratio.}

We now give the results for 
The Fluctuation Dissipation Ratio (FDR) $X^q_{tt'}$ defined by
\cite{cugliandolo_fdr}
\begin{eqnarray}\label{defFDR}
\frac{T}{X^q_{tt'}} = \frac{\partial_{t'} {\cal C}^q_{tt'}}{{\cal
R}^q_{tt'}}
\end{eqnarray}
Starting from the scaling laws that we established above, we can compute the FDR $
X^q_{tt'} \equiv X^{\tilde{q}}_{\tilde{t}\tilde{t'}} $ as a function
of the scaling
variables $\tilde{q}^z(\tilde{t}-\tilde{t'})$ and
$\tilde{t}/\tilde{t'}$. As we saw previously, both the
exponent $z$ and the scaling function associated with the FDR will
have an expansion in powers of $\tau$, i.e.
\begin{eqnarray}
&&\frac{T}{X^{\tilde{q}}_{\tilde{t}\tilde{t'}}} =
F_X(\tilde{q}^z(\tilde{t}-\tilde{t'}),\frac{\tilde{t}}{\tilde{t'}}) \\
&&F_X(v,u) = F_X^0(v,u) + \tau F_X^1(v,u) + O(\tau^2) \nonumber
\end{eqnarray}
the expansion of $z$ to order $\tau$ being given by
(\ref{expz}). $F_X^0(v,u)$ corresponds to the Gaussian model and
from the perturbative expansions that we obtained for ${\cal
R}^{\tilde{q}}_{\tilde{t}\tilde{t'}}$
(\ref{pertresp}) and ${\cal C}^{\tilde{q}}_{\tilde{t}\tilde{t'}}$
(\ref{pert_corr}) one can
identify (perturbatively) this
scaling form, with
\begin{eqnarray}
&&F_X^0(v,u) = 1 + e^{-2 \frac{v}{u-1}} \\
&&F_X^1(v,u) = - \frac{\theta}{\tau} \frac{u-1}{v} (1 - e^{-2
\frac{v}{u-1}}) - e^v F_R^1(1+e^{-2 \frac{v}{u-1}}) \nonumber \\
&&- e^v\left(\frac{\partial F_C^1(v,u)}{\partial v} + \frac{u(u-1)}{v}
\frac{\partial F_C^1(v,u)}{\partial u} ) \right) \nonumber
\end{eqnarray}
Inserting the formulae for $F_R^1$ and $F_C^1$ obtained in 
the Appendix yields the general result for $F_X$ as a non trivial function
of the two variables $u,v$.
Here we only give the behaviors of this scaling function in the
different asymptotic limits studied previously. First, we note that
this formula gives back the FDT result $F_X=1$ for $u=1$.

Second, focusing on
the limit $u \gg 1$, keeping $v$ fixed one has
\begin{eqnarray}\label{FDR_large_u}
&&F_X(v,u) = 1 + e^{-2\frac{v}{u-1}} + \frac{\sqrt{\pi}}{2} e^{\gamma_E} \tau
\sqrt{\frac{u}{v}}\Erfi{\sqrt{v}} \nonumber \\
&& + O(\tau u^0,\tau^2)
\end{eqnarray}
Thus in this regime $X$ decreases below its FDT value $X_{FDT}=1$.
Looking at this result one is tempted to conclude that
$X^q_{t t'}$ vanishes as $t/t' \to \infty$ when $q^z (t-t')$ is kept fixed.
In particular for $q=0$ (see below the direct calculation in this 
case) one finds the analogous quantity $X^{q=0}_\infty$ computed in
\cite{calabrese_on_twoloops,calabrese_fdr_randommass} 
for several models. However one must keep in mind that
(\ref{FDR_large_u}) is perturbative in $\tau$ and the
divergence of the coefficient of $\tau$ could also be a sign of
a non analyticity in $\tau$ of the $u=\infty$ result. 
Elucidation of this point is left for future study.

In the other limit that we studied previously, corresponding to $v \gg 1$,
keeping $u$
fixed, we obtain straightforwardly the following behavior
\begin{eqnarray}\label{FDR_large_v}
&&F_X(v,u) = 1 + e^{-2\frac{v}{u-1}} - \frac{e^{\gamma_E} \tau e^v}{2 v^2}
\frac{(u-1)^2}{2  \sqrt{u}} \nonumber \\
&&+ O(\tau e^v v^{-3},\tau^2)
\end{eqnarray}
This limit is relevant to study fixed $q$. It shows that there
is still aging behavior in a given non zero mode, and
appears to contradict some claims \cite{calabrese_on_twoloops} that only the zero mode (diffusion)
exhibits interesting aging behavior. Note also that
in this regime one has $X > X_{FDT}$, a feature found in
other disordered models \cite{rfim}. 

Finally in the limit of vanishing momentum $\tilde{q} = 0 $, the FDR
is a function of the scaling variable $\tilde{t}/\tilde{t'}$
whose perturbative expansion is given by
\begin{eqnarray}\label{FDR_q=0}
&&\frac{T}{X^{\tilde{q}=0}_{\tilde{t}\tilde{t'}}} =
F_X^{\text{diff}}\left(\frac{\tilde{t}}{\tilde{t'}}\right) \\
&&F_X^{\text{diff}}(u) = F_X^{\text{diff}0}(u) + \tau
F_X^{\text{diff}1}(u) + O(\tau^2) \\
&&F_X^{\text{diff}0}(u) = 2 \nonumber \\
&&F_X^{\text{diff}1}(u) = 2 F_C^{\text{diff}1}(u) - 2u \frac{ d
F_C^{\text{diff}1}(u)}{du} - 2 F_R^{\text{diff}1}(u) \nonumber \\
&&+ \frac{2(z-2)}{z \tau(u-1)}
- 2 \frac{\theta}{\tau} \nonumber
\end{eqnarray}
Using the results of previous Sections we find:
\begin{eqnarray}\label{FDR_q=0}
&&\frac{T}{X^{\tilde{q}=0}_{\tilde{t}\tilde{t'}}} = 2 + \tau e^{\gamma_E} (
\sqrt{u} + \ln(\frac{\sqrt{u} -1}{\sqrt{u} +1}) + \sigma )
\end{eqnarray}
where $\sigma$ is a numerical constant. This constant
depends on additive constants to respectively 
$F_R^1$ and $F_C^1$, each of them being nonuniversal
as discussed above (see Appendix). However, a
distinct possibility is that $F_X^{\text{diff}}(u)$
is universal (i.e. that the  non universal
parts cancel). Checking this can be done 
with the present method, and is left for future study.
The value obtained here, $\sigma=5 - 12 \ln 2$, may only be
indicative since we did not keep track of all additive
constants. 
In particular, in the scaling regime $\tilde{t} \gg \tilde{t'} \gg 1$,
one obtains
\begin{eqnarray}\label{FDR_q=0_large_u}
\frac{T}{X^{\tilde{q}=0}_{\tilde{t}\tilde{t'}}} \sim 2 + \tau
e^{\gamma_E} \sqrt{u} + O (\tau u^0,\tau^2)
\end{eqnarray}
Notice that taking the limit $v \to 0$ (using
(\ref{erfi_asymp_small})) on the asymptotic expression
(\ref{FDR_large_u}) where we have taken the limit $u \gg 1$ {\it before} $v$
small one recovers the same result (\ref{FDR_q=0_large_u}).

One way to understand the result (\ref{FDR_q=0}), i.e. the
divergence of $X^{\tilde{q}=0}_{tt'}$ when $t' \to t$ is to note that
the same divergence occurs for a simple diffusion process
with the same close times asymptotic behaviors:
\begin{eqnarray} 
&& {\cal C}^{\tilde{q}=0}_{t t'} \sim {t'}^{2/z} \\
&& {\cal R}^{\tilde{q}=0}_{t t'} \sim (t-t')^{(2-z)/z}
\end{eqnarray}
which yields straigforwardly $X^{\tilde{q}=0}_{t t'} \sim A (u-1)^{(2-z)/z}$
as $u \to 1$. Note however that to obtain the correct
{\it amplitude} $A$ one needs to take into account further
corrections to ${\cal C}^{\tilde{q}=0}_{t t'}$, specifically
we note that one can rewrite (\ref{Cq=0}) as:
\begin{eqnarray} 
&& {\cal C}^{\tilde{q}=0}_{t t'} = {t'}^{2/z} {\cal A}(u)
\end{eqnarray}
and that the detailed asymptotics of ${\cal A}(u)$ near
$u=1$ determines the amplitude of the divergence.

\section{conclusion}

In this paper we have developped a novel EMRG method to perform 
first principle perturbative calculations based on exact RG.
Contrarily to previous works it is based on a multilocal
expansion of the effective action functional. It allows
to perform conveniently calculations with an arbitrary cutoff
function in a fully controlled way 
and check explicitly the universality of the observables.

We have tested the method on the standard $O(N)$ model. We have shown
that the exponent $\eta$ to order $O(\epsilon^2)$ 
can be simply recovered within the
exact RG multilocal expansion. This is interesting
since previous approaches relied on approximations such as polynomial and derivative
expansions, which are not needed here.
We have also obtained several two-point scaling functions
and explicitly checked universality. Finally,
we explained how the method compares with more standard
field theoretical approaches. In a sense the present
method directly yields the renormalized theory.

We have applied the EMRG method to study the glass phase
of the two dimensional random Sine-Gordon model (Cardy-Ostlund)
near the glass transition temperature.
We have first recovered known results for the statics
and for the equilibrium dynamical exponent $z$ which we showed 
to be universal. The method
of derivation however is quite different from previous ones, since
it yields directly the self energy $\Sigma_l(\omega)$ 
as a scaling function of $\omega/\Lambda_l$ where
$\Lambda_l$ is the infrared cutoff. We have 
given for the first time the scaling functions
associated to finite momentum equilibrium
response and correlation. 

Next we studied the out of equilibrium dynamics
of the Cardy-Ostlund model. We obtained the two time
response and correlations at finite momentum. These were found to take a scaling form
and we computed analytically the corresponding scaling functions
which depend on two arguments $v=\tilde{q}^z (\tilde{t}-\tilde{t'})$ and
$u=\tilde{t}/\tilde{t'}$. We showed that they exhibit aging behavior 
characterized by a non trivial fluctuation dissipation
ratio $X$, itself a universal function of $u,v$
that we obtained. We also obtained the off equilibrium exponents
$\theta$ and $\lambda$. Interestingly we found that, at variance
with pure systems, one must introduce two
distinct exponents $\lambda_R$ and $\lambda_C$
for response and correlation respectively. 
Our study raises the question of whether this could be a more general
property of glassy dynamics in disordered systems.

Our method is promising for further RG studies of disordered systems,
as it allows to attack the problem with few assumptions. 
Other situations where it can
be applied are elastic manifolds in random media, where
it can be used to put the so-called Functional RG on
a more solid basis \cite{pld,pldlb}. Concerning the
results of the present paper, a numerical simulation
of the Cardy Ostlund glass phase can be performed
\cite{gsnumsim} and should provide an interesting
test of the predictions of our RG calculation.
In particular, some points require further examination,
e.g. the asymptotic value $X_\infty$ of the FDR. 
This would be interesting especially in the light of
the present activity on FDR in mean field models,
and interpretations in terms of effective temperatures.
Indeed developing real space, RG type methods 
beyond mean field remains a challenge in the
theory of glasses.

\newpage

\appendix

\section{exact RG equation for the effective action}
\label{app:erg}

Here we present a simple derivation of the exact RG equation satisfied by the
effective action, denoted here $\Gamma_G(\phi)$ (and $\Gamma(\phi)$ in the text), for the theory
of action given in (\ref{startaction}), when the propagator
$G$ is varied, for a fixed interacting functional ${\cal V}(\phi)$.
One first introduces the generating functional:
\begin{eqnarray}
Z_G(j) = \int D \phi ~ e^{ - \frac{1}{2} \phi : G^{-1} : \phi - {\cal V}(\phi) + j : \phi}
\end{eqnarray}
i.e. the partition function in presence of a set of sources denoted
$j\equiv j^i_x$. For any variation $\partial G$ of $G$, its variation $\partial Z_G(j)$
satisfies:
\begin{eqnarray}
&& \partial Z_G(j) = - \frac{1}{2} Tr \partial G^{-1}  \int D \phi \phi \phi
e^{ - \frac{1}{2} \phi : G^{-1} : \phi + {\cal V}(\phi) + j : \phi } \nonumber \\
&&
= - \frac{1}{2} Tr \partial G^{-1} \frac{\delta^2 Z_G(j) }{\delta j \delta j }
\end{eqnarray}
where $\partial G^{-1} = - G^{-1} \partial G G^{-1}$ and $Tr$ denotes a trace
over all spatial coordinates and indices $x,i$.
Next, one introduces the generating functional $W_G(j) = \ln  Z_G(j)$ of connected correlations,
which varies as:
\begin{equation}
\partial W_G(j) = - \frac{1}{2} Tr \partial G^{-1} ( \frac{\delta^2 W_G(j) }{\delta j \delta j }  +
\frac{\delta W_G(j) }{\delta j  } \frac{\delta W_G(j) }{\delta j  } )
\label{eqw}
\end{equation}
an exact RG equation for this quantity.
From there it is simple to obtain the RG equation obeyed by its
Legendre transform
$\Gamma_G(\phi) = {\text{min}}_j ( \phi : j - W_G(j)) $. We will
assume that no problem arise from the
convexity condition and that
$\Gamma_G(\phi)$ can be obtained using only the saddle point conditions:
\begin{eqnarray}
&& \frac{\delta W_G }{\delta j  } (j_G(\phi)) = \phi \\
&& \frac{\delta \Gamma_G(\phi) }{\delta \phi }  = j_G(\phi)
\label{saddles}
\end{eqnarray}
For the variation of $\Gamma_G(\phi) = \phi : j_G(\phi) - W_G(j_G(\phi))$,
this yields:
\begin{eqnarray}
&& \partial \Gamma_G(\phi) = - \partial W_G(j_G(\phi)) \\
&& = \frac{1}{2} Tr
\partial G^{-1} ( \frac{\delta^2 W_G}{\delta j \delta j }(j_G(\phi))  +
\frac{\delta W_G}{\delta j }(j_G(\phi)) \frac{\delta W_G }{\delta j  }(j_G(\phi)) )
\nonumber
\end{eqnarray}
since the term proportional to $\partial_G j_G$ cancels as usual from the
saddle point conditions (\ref{saddles}). Using (\ref{saddles})
once more, as well as the standard relation
$\frac{\delta W_G }{\delta j  \delta j} (j_G(\phi)) =
[\frac{\delta \Gamma_G }{\delta \phi  \delta \phi}]^{-1}$,
gives the equation (\ref{ERGG}) of the text.

Writing then
\begin{eqnarray}
\Gamma_G(\phi) = \frac{1}{2} \phi : G^{-1} : \phi + {\cal U}_G(\phi)
- \frac{1}{2} Tr \ln G
\end{eqnarray}
this is equivalent to the equation for ${\cal U}_G$:
\begin{eqnarray}
&& \partial {\cal U}_G(\phi) =  \frac{1}{2} Tr \partial G : ( G^{-1} - G^{-1}  (1+G :
\frac{\delta^2 {\cal U}_G}{\delta \phi\delta \phi})^{-1}) \label{ERGG2}
\end{eqnarray}
or its equivalent form given in the text.

Now that we have an exact equation for $\Gamma_G(\phi)$, we can relate the
effective action in theories with the same ${\cal V}(\phi)$ but different $G$.
All we need to fully determine the effective action is an initial condition.
It is provided by the action itself. Indeed one has the following perturbative
loop expansion:
\begin{eqnarray}
\Gamma(\phi) = - \frac{1}{2} Tr \ln G + {\cal S}(\phi) + \sum_{k \geq 1} \Gamma^k(\phi)
\end{eqnarray}
where $\Gamma^k(\phi)$ is the sum of all $k$ loop 1PI graphs using
${\cal V}(\phi)$ as interaction and $G$ as propagator.
Thus if the initial condition for the propagator $G_{l=0}$
is such that all $\Gamma^k(\phi)$ graphs vanish when computed with $G_{l=0}$, then
one can choose the initial condition as ${\cal U}_{l=0}(\phi) = {\cal V}(\phi)$.
This is the case for the choice (\ref{choixpropag1},\ref{choixpropag2})
made in the text (similarly the initial condition for
$W_G(j)$ in (\ref{eqw}) is the Legendre transform of
the initial action ${\cal S}(\phi)$).

Finally let us note that the RG equation can also be written as:
\begin{eqnarray}
&& \frac{d {\cal U}_G(\phi) }{d G}  = \frac{\partial }{\partial G}
\frac{1}{2} Tr \ln
( 1 + G : \frac{\delta^2 {\cal U}_G (\phi)}{\delta \phi \delta \phi} ) \\
&& = \frac{1}{2}
\frac{\delta^2 {\cal U}_G (\phi)}{\delta \phi \delta \phi} :
( 1 + G : \frac{\delta^2 {\cal U}_G (\phi)}{\delta \phi \delta \phi} )^{-1}
\end{eqnarray}
where the derivative $\frac{\partial }{\partial G}$ in the r.h.s. of the first equation
is restricted to the {\it explicit} $G$ dependence (i.e. not the one implicit in
${\cal U}_G(\phi)$).

\section{multilocal expansion to $O(U^2)$}
\label{app:expansion}

To $O(U^2)$ one needs only $U$ and $V \sim O(U^2)$ in the expansion
(\ref{expmult}) of $\tilde{U}$. The functional derivative reads:
\begin{eqnarray}
&& \frac{\delta \tilde{U} }{\delta \phi_x^i \delta \phi_y^j} =
\delta_{xy} [ \partial_i \partial_j U( \phi_x ) +
\int_z ( \partial^1_i \partial^1_j V( \phi_x , \phi_z , x-z) \nonumber \\
&& +
\partial^2_i \partial^2_j V( \phi_z , \phi_x , z-x) ] +
2 \partial^1_i \partial^2_j V( \phi_x , \phi_y , x-y) +  O(W) \nonumber \\
&&
\end{eqnarray}
using parity $V(\phi,\psi,-x)=V(\phi,\psi,x)$. Inserting in
(\ref{perturbERG}) and keeping only terms up to order $O(U^2)$ one finds the
resulting RG equation:
\begin{widetext}

\begin{eqnarray}
&&\partial_l U_l(\phi) = \frac{1}{2} \partial G^{x=0}_{ij} \partial_i
\partial_j U_l(\phi) + \int_x \partial G^x_{ij} \partial^1_i
\partial^2_j V_l(\phi,\phi,x) - \frac{1}{2} \int_x \partial G^x_{ij} \partial_j \partial_k
U_l(\phi) (G^x_l)_{km} \partial_m \partial_i U_l(\phi) \\
&& \partial_l V_l (\phi,\psi,x) = - \frac{1}{2} \partial G^x_{ij}
\partial_j \partial_k U_l(\phi) (G^x_l)_{km} \partial_m \partial_i
U_l(\psi) \nonumber \\
&& + \frac{1}{2} \partial G^{x=0}_{ij} (\partial^1_i
\partial^1_j + \partial^2_i \partial^2_j)V_l(\phi,\psi,x)
+ \partial_i^1 \partial_j^2 (\partial G^x_{ij} V_l(\phi,\psi,x) \nonumber \\
&& -
\delta(x)\int_y\partial G^y_{ij} V_l(\phi,\psi,y))
+ \frac{1}{2} \delta(x) \int_y \partial G^y_{ij} \partial_j \partial_k U_l(\phi)
(G^y_l)_{km} \partial_m \partial_i U_l(\psi) \label{RG_eq_bilocal}
\end{eqnarray}
\end{widetext}

where the local projection $\overline{P}_1$ operator has been applied to
obtain the first equation, and the operator $1-P_1$ to obtain the
second. This is illustrated in Fig. 2 (dropping all terms of order
$O(U^3)$ and higher). Note that $\int_x V(\phi,\psi,x)=0$.
The differential equation for the bilocal part $V_l$ is linear,
a general property which allows
to solve all higher multilocal components (here $V_l$) as a function
of the local part $U_l$ only.
The equation for $V_l$ can be integrated in the forms
(\ref{resultbiloc}), (\ref{resultbiloc2}) 
given in the text.
The method is similar to \cite{chauvepld} to which we refer for
further details.
Inserting this solution in the equation for $U_l$ one obtains (\ref{RGorder2}) in the text.
We have assumed that no bilocal term exists in the original action. Near the
fixed point form at large $l$ these assumptions are not strictly necessary, a statement which
can be checked using the present method.

It can be useful, in particular for the Cardy Ostlund model that we study
in the text, to introduce a Fourier
representation in the fields:
\begin{eqnarray}
&&U_l^K = \int d\phi e^{-iK.\phi} U_l(\phi) \nonumber \\
&&V_l^{KPx} = \int d\phi d\psi e^{-iK.\phi-iP.\psi} V_l(\phi,\psi,x) \label{def_Fourier}
\end{eqnarray}
Using this representation, we obtain the RG equations (\ref{resultbiloc}-\ref{RGorder2})
in Fourier space
\begin{widetext}
\begin{eqnarray}
&&V_l^{KPx} = \frac{1}{2}(F_l^{KPx} - \delta(x) \int_y F_l^{KPy})
\label{resultbiloc_Fourier} \\
&&F_l^{KPx} = - \int_0^l dl' (K.\partial G^x_{l'}.P)(K.G^x_{l'}.P)
e^{\frac{1}{2}K.G^{x=0}_{l'l}.K +\frac{1}{2}P.G^{x=0}_{l'l}.P
+ K.G^x_{l'l}.P } U^K_{l'} U^P_{l'} \nonumber \\
&&\partial_l U_l^K =  \frac{-1}{2}K. \partial G^{x=0}.K U_l^K
 - \frac{1}{2} \int_{P,Q,P+Q=K}\int_x (P.\partial G^x.Q) (P.G^x_l.Q)  U_l^P U_l^Q  \nonumber \\
&& + \frac{1}{2} \int_{P,Q,P+Q=K} \int_x P.(\partial G_l^x -
\partial G_l^0).Q \int_0^l dl' (P.\partial G^x_{l'}.Q)(P.G^x_{l'}.Q)
 e^{\frac{1}{2}P.G^{x=0}_{l'l}.P +\frac{1}{2}Q.G^{x=0}_{l'l}.Q
+ P.G^x_{l'l}.Q }U^P_{l'}U^Q_{l'} \label{RGorder2_Fourier}
\end{eqnarray}
\end{widetext}
where $\int_{P,Q,P+Q=K} \equiv \int \frac{d^N P d^N Q}{(2 \pi)^N} \delta(K-P-Q)$
where $N$ is the number of components of $\phi$. In the text we have used
$\hat V_l^{KPx}$ to distinguish the Fourier series coefficients from the Fourier
transform.

\section{Detailed calculations for the $O(N)$ model}

\label{app:on}

\subsection{$\beta$-function}

Let us insert (\ref{polyexp}) into the ERG equation (\ref{RGorder2}), keeping
only $g_0$, $g_2$ and $g_4$ for now, and first focus on
the first line in (\ref{RGorder2}) which reads:
\begin{eqnarray}
&& \partial_l [ g_{0,l} + \frac{\tilde g_{2, l}}{2!} \Lambda_l^{2} \phi^2
+ \frac{\tilde g_{4, l}}{4!} \Lambda_l^{\epsilon} (\phi^2)^2 ] =  \\
&&\frac{1}{2} \int_q \partial G^q_l \partial_i\partial_i U_l(\phi)
- \frac{1}{2} \int_q \partial G^q_l G^q_l
(\partial_i\partial_j U_l(\phi))^2 \nonumber
\end{eqnarray}
with implicit sums on repeated indices. Using that:
\begin{eqnarray}
&&\partial_i \partial_j U_l(\phi)
= g_{2,l} \delta_{ij} + \frac{g_{4,l}}{3!}
(\delta_{ij} \phi^2 + 2 \phi^i  \phi^j )
\end{eqnarray}
which yields:
\begin{eqnarray}
&&\partial_i \partial_i U_l
= N g_{2,l} + \frac{N+2}{3!} g_{4,l} \phi^2 \nonumber \\
&& \partial^1_i \partial^1_j U_l(\phi_1) \partial^2_i \partial^2_j U_l(\phi_2)
= N g_{2,l}^2 \label{double_somme} \\
&&+ \frac{N+2}{3!} g_{2,l} g_{4 ,l} (\phi_1^2
+ \phi_2^2   )  \nonumber \\
&& + \frac{1}{(3!)^2} g_{4 ,l}^2 ((N+4)\phi_1^2\phi_2^2  + 4
(\phi_1 \cdot \phi_2)^2) \nonumber
\end{eqnarray}
Setting $\phi_1=\phi_2$ in (\ref{double_somme}) and
identifying the coefficients of $\phi^2$ and $(\phi^2)^2$,
one then easily obtains all terms in
(\ref{RG_Eq_On_g2},\ref{RG_Eq_On_g4}) apart from
the last one, with the scaled integrals defined in (\ref{integrals}).

Inserting now (\ref{polyexp}) into the second line of the
ERG equation (\ref{RGorder2}) one obtains only a correction
to $g_2$ from the term with the lowest number of derivatives (six).
Noting that:
\begin{eqnarray}
\partial_i \partial_j \partial_k U_l(\phi)
= \frac{g_{4,l}}{3} (\delta_{ij} \phi^k + \delta_{ik} \phi^j + \delta_{jk} \phi^i )
\end{eqnarray}
one finds:
\begin{eqnarray}
(\sum_i \partial_i^1 \partial_i^2)^3 U_l(\phi_1) U_l(\phi_2)
= \frac{g_{4,l}^2}{3} (N+2) \phi_1 \cdot \phi_2 \label{triple_somme}
\end{eqnarray}
yielding the last term in (\ref{RG_Eq_On_g2})
\begin{eqnarray}
- \frac{N+2}{3} \int_0^l dl' \tilde I_{l ,l'}^{(2)}
  {{\tilde{g}^2_{4,l'}}} \label{last_term}
\end{eqnarray}
with
\begin{eqnarray}
\tilde I_{l , l'}^{(2)} = \Lambda_l^{-2} \int_x (\partial_l G_l^{x}
- \partial_l G_l^{x=0}) \partial_{l'} G_{l'}^{x} G_{l'}^{x}
\Lambda_{l'}^{2 \epsilon}  \label{I2_ll'}
\end{eqnarray}
This term does not modify the fixed
point value $\tilde{g}_2^*$ to order $\epsilon$, provided it remains
finite in the limit
$l \to \infty$.
A way to compute it is to make an integration by part to treat
the integral over $l'$:
\begin{eqnarray}
&&\int_0^l dl' \partial_{l'} G_{l'}^{x} G_{l'}^{x}
\tilde{g}_{4,l'}^2 \Lambda_{l'}^{2 \epsilon} \label{IPP} \\
&&= \frac{1}{2} (G^x_l)^2 \tilde{g}_{4,l}^2 \Lambda_{l}^{2 \epsilon} -
\int_0^l dl' (G^x_{l'})^2 \tilde{g}_{4,l'} \Lambda_{l'}^{\epsilon}
\partial_{l'}(\tilde{g}_{4,l'} \Lambda_{l'}^{\epsilon}) \nonumber \\
&& = \frac{1}{2} (G^x_l)^2 \tilde{g}_{4,l}^2 + O(\epsilon
\tilde{g}_{4,l}^2,\tilde{g}_{4,l}^3) \nonumber
\end{eqnarray}
as from Eq. (\ref{RG_Eq_On_g4})
$\partial_{l'}(\tilde{g}_{4,l'} \Lambda_{l'}^{\epsilon})$ is of order
$\tilde{g}_{4,l'}^2$ and where we have used $G^x_{l=0} = 0^+$. The
terms we dropped are of order $\epsilon^3$ in the
limit $l \to \infty$.
Finally, in the large $l$ limit we are left with
\begin{equation}
\int_0^l dl' \tilde I_{l , l'}^{(2)} \tilde{g}_{4,l'}^2 =
\frac{\tilde{g}_{4,l}^2}{2} \Lambda_l^{-2}
\int_x (\partial_l G_l^{x} - \partial_l G_l^{x=0}) (G^x_l)^2 +
O(\epsilon^3)
\end{equation}
which is already of order $\epsilon^2$, so that the integral over
$x$ can be performed in $d=4$ exactly.
Using the decomposition of the cutoff (\ref{decomp_cutoff}), we
compute the following integrals exactly in $d=4$
\begin{eqnarray}
&& G^x_l = \frac{1}{4 \pi^2} \int_a \frac{1}{x^2}(e^{-x^2
\Lambda_l^2/2a} - e^{-x^2 \Lambda_0^2/2a}) \nonumber \\
&& \Lambda_l^{-2} \partial_l G_l^x = \frac{1}{4 \pi^2} \int_a
\frac{1}{a} e^{-x^2 \Lambda_l^2/2a} \label{propag_x}
\end{eqnarray}
Eq. (\ref{last_term}) can finally be written as an integral over the
rescaled variable $\tilde{x} = \Lambda_l x$
\begin{eqnarray}
&&\frac{N+2}{3}\int_0^l dl' \tilde I_{l , l'}^{(2)} \tilde{g}_{4,l'}^2
\propto \tilde{g}_{4,l}^2 \int_{\tilde{x}} \int_a
(e^{-\tilde{x}^2/2a} -1) \frac{1}{a \tilde{x}^4} \nonumber \\
&&\left(\int_a e^{-(\Lambda_0/\Lambda_l)^2\tilde{x}^2/(2a) } -
e^{-\tilde{x}^2/2a}  \right)^2
\end{eqnarray}
In the limit $l \to \infty$, this integral is well defined. Indeed,
there is no UV divergence (due to the term $(e^{-\tilde{x}^2/2a} -1)$ which
behaves as $\tilde{x}^2$) nor IR divergence due to the term
$e^{-\tilde{x}^2/a}$.

By the same calculation
one obtains the flow of the free energy:
\begin{eqnarray}
&& \partial_l g_{0,l} =
\frac{N}{2} \Lambda_l^d (\tilde I^{(0)}_l \tilde g_{2,l}
- \tilde I^{(1)}_l \tilde g_{2,l}^2 )\nonumber  \\
&&
+ \frac{N+2}{3} \Lambda_l^d  \int_0^l \tilde I^{(3)}_{l,l'} g_{4,l'}^2
\end{eqnarray}
with:
\begin{eqnarray}
&&\tilde I_{l , l'}^{(3)} = \Lambda_l^{-d} \int_x (\partial_l G_l^{x} - \partial_l G_l^{x=0})
\partial_{l'} G_{l'}^{x} G_{l'}^{x} G_{l' l}^{x}  \Lambda_{l'}^{2
\epsilon} \nonumber
\end{eqnarray}

We finally obtain the flow for $\tilde{g}_{6,l}$ in (\ref{polyexp})
with the same kind of manipulations, and using furthermore
\begin{eqnarray}
&&\partial_i \partial_j ((\phi^2)^3) = 6 \delta_{ij} (\phi^2)^2 + 24
\phi^i \phi^j \phi^2  \\
&&\partial_i \partial_j \partial_k ((\phi^2)^3) = 24 \phi^2
(\delta_{ij}\phi^k +\delta_{ik}\phi^j +  \delta_{jk}\phi^i) + 48
\phi^i \phi^j \phi^k \nonumber
\end{eqnarray}
one gets
\begin{eqnarray}
&&\partial_l \tilde{g}_{6,l} = (2\epsilon - 2)\tilde{g}_{6,l} -
(N+14)
\tilde{I}^{(1)}_l \tilde{g}_{4,l}\tilde{g}_{6,l} \nonumber \\
&& - \frac{8}{5}(3N+16)
\int_0^l dl' \tilde{I}^{(2)}_{l,l'} \tilde{g}_{6,l'}^2 +
O(\tilde{g}_{4,l}^3)
\end{eqnarray}
which shows that $\tilde{g}_{6}^* \sim \epsilon^3$. Similarly there
is a term proportional to $\tilde I^{(0)} \tilde{g}_{6,l}$ in
the flow equation of $\tilde{g}_{4,l}$ which affect the
fixed point value $\tilde{g}_{4}^*$ only to next order in $\epsilon$.

\subsection{Computation of the exponent $\eta$.}

The quadratic term in (\ref{bilocal_On}) is obtained by inserting
(\ref{polyexp}) in (\ref{resultbiloc2}) and expanding the exponential
in (\ref{resultbiloc2}) to order one. One gets, using
(\ref{triple_somme})
\begin{eqnarray}
&&F_l(\phi_1,\phi_2,x) = \int_0^l dl' \partial G_{l'}^x G_{l'}^x
G_{l'l}^x  (\sum_i \partial_i^1 \partial_i^2)^3 U_l(\phi_1)
U_l(\phi_2) \nonumber \\
&&= \frac{N+2}{3} \phi_1 \cdot \phi_2  \int_0^l dl'
\partial_{l'} G^x_{l'} G^x_{l'} G^x_{l'l}{{\tilde{g}^2_{4l'}}}
\Lambda_{l'}^{2 \epsilon} \nonumber \\
&& \label{bilocal_On_quad}
\end{eqnarray}
which is the second line of (\ref{bilocal_On}). We have dropped terms
of the form $f(\phi_i,x)$ such as $\tilde{g}_{2,l}\tilde{g}_{4,l}
\phi_1^2$, $\tilde{g}_{2,l}\tilde{g}_{4,l}
\phi_2^2$ (resulting from the expansion of the exponential in
(\ref{resultbiloc2}) to order 0), or $\tilde{g}_{4,l}^2 \phi_1^2$,
$\tilde{g}_{4,l}^2 \phi_2^2$ (resulting from the expansion of the
exponential in
(\ref{resultbiloc2}) to order 1 but acting respectively with
$\partial^2.G_{l'l}^{x=0}\partial^2$ or $\partial^1.G_{l'l}^{x=0}\partial^1$)
because they do not give any contribution to the effective action.
Indeed, the
contribution of such terms to the interaction functional
${\cal U}_l(\phi)$ (\ref{expmult}), will be
\begin{eqnarray}
&&V_l(\phi_x,\phi_y,x-y) = f(\phi_x,x-y) - \delta(x-y) \int_z
f(\phi_x,z)\nonumber \\
&&{\cal U}_l(\phi) \sim \int_{x,y} (f(\phi_x,x-y) - \delta(x-y)\int_z
f(\phi_x,z) ) \nonumber \\
&&= \int_{x,y} f(\phi_x,x-y) - \int_{x,z}f(\phi_x,z) = 0 \nonumber
\end{eqnarray}
where we assumed parity $f(\phi_i,x) = f(\phi_i,-x)$ (which is the
case here) and translational invariance. To treat the integral
over $l'$ in (\ref{bilocal_On_quad}), we use an integration by part as
in (\ref{IPP}), one gets
\begin{eqnarray}
&&\frac{N+2}{3} \phi_1 \cdot \phi_2  \int_0^l dl'
\partial_{l'} G^x_{l'} G^x_{l'} G^x_{l'l}{{\tilde{g}^2_{4l'}}}
\Lambda_{l'}^{2 \epsilon} \nonumber \\
&&= - \frac{N+2}{18} \tilde{g}_{4,l}^2 \phi_1 \cdot \phi_2
(G^x_l)^3 + O(\epsilon \tilde{g}_{4,l}^2,\tilde{g}_{4,l}^3) \nonumber
\end{eqnarray}
which leads to (\ref{gamma2_on_gene}).
Using (\ref{propag_x}), the last term in (\ref{gamma2_on_gene}) reads
(forgetting for the
discussion the numerical prefactor)
\begin{eqnarray}
&&H(q,\Lambda_0,\Lambda_l) = \int_x (e^{iqx} -1)(G^x_l)^3  \\
&& = \frac{1}{(4 \pi^2)^3}\int_x (e^{iqx} - 1)\frac{1}{x^6}
 \left(\int_a e^{-x^2 \Lambda_l^2/2a} - e^{-x^2 \Lambda_0^2/2a} \right)^3 \nonumber
\end{eqnarray}
where the integral over $x$ is evaluated in $d=4$ (as this term is
already of order
$\tilde{g}_{4,l}^2$). For any $\Lambda_0$, $\Lambda_l$, this integral
is well defined
but in the limit $\Lambda_0 \to \infty$, the integrand is not any more
regularized
at small $x$ and there is a logarithmic divergence. We are interested
in the limit $q, \Lambda_l \ll \Lambda_0$. A simple way to
isolate this
divergence is to rewrite it as
\begin{eqnarray}
&&H(q,\Lambda_0,\Lambda_l) = \\
&&\frac{1}{(4\pi^2)^3}\Big\{-\frac{1}{2} \int_x (qx)^2\frac{1}{x^6}
\left(\int_a e^{-x^2 \Lambda_l^2/2a} - e^{-x^2 \Lambda_0^2/2a}
\right)^3 \nonumber \\
&& + \int_x (e^{iqx} - 1 + \frac{1}{2}(qx)^2)\frac{1}{x^6}
\left(\int_a e^{-x^2 \Lambda_l^2/2a} - e^{-x^2 \Lambda_0^2/2a}
\right)^3\Big \}
 \nonumber
\end{eqnarray}
The limit $\Lambda_0 \to \infty$ can be taken safely in the second
term, the UV divergence
coming only from the first one which can be written
\begin{eqnarray}
&&-\frac{1}{2} \int_x (qx)^2 \frac{1}{x^6}
\left(\int_a e^{-x^2 \Lambda_l^2/2a} - e^{-x^2 \Lambda_0^2/2a} \right)^3
= h\left(\frac{\Lambda_0^2}{\Lambda_l^2}\right) \nonumber \\
&&h(\lambda) = -\frac{\tilde{S}_4}{8}q^2 \int_0^{\infty} \frac{dx}{x}
(\int_a e^{-x^2/2a } - e^{-\lambda x^2/2a})^3 \nonumber
\end{eqnarray}
where in the second line we performed the change of variable $x \to
\Lambda_l x$
and denoted $\tilde{S}_4 = 2\pi^2$ the unit sphere area in
dimension $d=4$.
Interestingly, we have (using the variable $u = \lambda x^2$), up to terms
of order $\lambda^{-2}$:
\begin{eqnarray}
&&h'(\lambda) =  -\frac{3\tilde{S}_4 q^2}{16 \lambda} \int_0^{\infty} du
\int\frac{1}{2a} e^{-u/2a}
\left(\int_a 1 - e^{-u/2a}\right)^2  \nonumber \\
&&=- \frac{\tilde{S}_4 q^2}{16\lambda} \left[\left(\int_a 1 -
e^{-u/2a}\right)^3\right]^{u=\infty}_{u=0} + O(\lambda^{-2})
= -\frac{\tilde{S}_4 q^2}{16 \lambda} \nonumber
\end{eqnarray}
where we have used $c(0) = \int_a = 1$, which leads to $h(\lambda) \sim
-\frac{\pi^2 q^2}{8}
\ln{\lambda} + O(\lambda^{-1})$.
Finally one obtains
\begin{eqnarray}
&&H(q,\Lambda_0,\Lambda_l) =  \frac{q^2}{(4\pi)^4}(
\ln{\frac{\Lambda_l}{\Lambda_0}} + \chi^{(2)}(q/\Lambda_l)) +
O\left(\frac{\Lambda_l^2}{\Lambda_0^2}\right) \nonumber \\
&& \chi^{(2)}(\tilde{q}) = \frac{4}{\pi^2\tilde{q}^2} \int_x
(e^{i\tilde{q}x} - 1 +
\frac{1}{2}(\tilde{q}x)^2)\frac{1}{x^6}
\left(\int_a e^{-x^2/2a} \right)^3 \nonumber
\end{eqnarray}
which gives (up to the factor $-\tilde{g}_{4,l}^2(N+2)/18$), the last term
in (\ref{gamma2_on_gene}). Using $1/x^6 = 1/2\int_0^{\infty} dt t^2
e^{-tx^2}$, one can compute the integral over $x$ in
$\chi^{(2)}(\tilde{q})$:
\begin{eqnarray}
&&\chi^{(2)}(\tilde{q}) = \frac{2}{\tilde{q}^2} \int_{a,b,c}
\int_0^{\infty} dt \frac{t^2}{(t+\alpha_3)^2}
(e^{-\frac{\tilde{q}^2}{4(t+\alpha_3)}} - 1 \nonumber \\
&& + \frac{\tilde{q}^2}{4(t+\alpha_3)}) \nonumber
\end{eqnarray}
with $\alpha_3 = \frac{1}{2a} + \frac{1}{2b} + \frac{1}{2c}$   from
which we easily obtain the
asymptotic behavior
\begin{eqnarray}
&&\chi^{(2)}(\tilde{q}) \sim \tilde{q}^2 \int_{a,b,c} \frac{1}{48
\alpha_3} \quad \tilde{q} \ll 1  \nonumber \\
&&\chi^{(2)}(\tilde{q}) \sim \ln{\tilde{q}} \quad \tilde{q} \gg 1 \nonumber
\end{eqnarray}
as annouced in the text (\ref{fonc_gamma2}). This yields
a universal result for the $\eta$ exponent. In addition
$\chi^{(2)}(\tilde{q})$ gives the scaling function of the
two point correlator $\chi^{(2)}(\tilde{q}) = 2 \tilde{q}^2 Q(\tilde{q}^2)$
where $Q(y)$ was computed in \cite{aharony_fisher} in the particular
case of a IR ''massive'' cutoff function of the form (\ref{Pauli}).
Although our expression is more general we have checked through series
expansion that it coincides with the expression given in \cite{aharony_fisher}
for that choice of the cutoff.

Performing two integrations by part one can rewrite:
\begin{eqnarray}
&&\chi^{(2)}(\tilde{q}) =
\int_{a,b,c>0} \frac{\hat C(a) \hat C(b)}{a^2 b^2} \hat c(c)
\frac{4}{q^2} (\frac{4}{q^2} (1 - e^{- q^2/(4 \alpha_3)} ) \nonumber \\
&& - \frac{1}{\alpha_3}
+ \frac{q^2}{8 \alpha_3^2} )
\end{eqnarray}
with $\hat C(a)=\int_a^\infty d a' \hat c(a')$.

\subsection{Quartic contribution to $\Gamma_l(\phi)$}

The quartic term in (\ref{bilocal_On}) is obtained by inserting
(\ref{polyexp}) in (\ref{resultbiloc2}) and expanding the exponential
in (\ref{resultbiloc2}) to order zero. One gets, using
(\ref{double_somme})
\begin{eqnarray}
&&F_l(\phi_1,\phi_2,x) = - \int_0^l dl' \partial G^x_{l'} G^x_{l'}
(\sum_{i} \partial^1_i \partial^2_i)^2 U_l(\phi_1)
U_l(\phi_2)  \nonumber \\
&&= - ( \frac{N+4}{(3!)^2}\phi_1^2 \phi_2^2 +
\frac{4}{(3!)^2} (\phi_1 \cdot \phi_2)^2) \int_0^l dl' \partial_{l'}
G^x_{l'}  G^x_{l'}{{\tilde{g}^2_{4l'}}} \Lambda_{l'}^{2 \epsilon}
\nonumber \\
&& \label{bilocal_On_quart}
\end{eqnarray}
which is the last line in (\ref{bilocal_On}) (here again we have
dropped terms of the form $f(\phi_i,x)$ coming from
(\ref{double_somme})).
The integral over $l'$ in (\ref{bilocal_On_quart}) is then
treated as previously (\ref{IPP}). Then, when computing the Fourier
transform, one obtains (\ref{gamma4_on}), with, using (\ref{propag_x})
\begin{eqnarray}
&&\chi^{(4)}_l(q) = \int_x (e^{iqx} -1)(G^x_l)^2 \nonumber \\
&& =  \frac{1}{16 \pi^4}\int_x (e^{iqx} -1) \frac{1}{x^4}\left(\int_a
e^{-x^2 \Lambda_l^2/2a} -e^{-x^2 \Lambda_0^2/2a}   \right)^2 \nonumber
\end{eqnarray}
For any $\Lambda_l, \Lambda_0$ finite, this function is well defined,
and we see that the limit $\Lambda_0 \to \infty$ is also well defined,
thus
\begin{eqnarray}
&&\chi^{(4)}_l(q) = \chi^{(4)}(q/\Lambda_l) \nonumber \\
&&\chi^{(4)}(\tilde{q}) = \frac{1}{16 \pi^4}\int_x (e^{i\tilde{q}x} -1)
\frac{1}{x^4}(\int_a e^{-x^2/2a})^2 +
O\left(\frac{\Lambda_l^2}{\Lambda_0^2}\right) \nonumber
\end{eqnarray}
the integral over $x$ can be computed using $1/x^4 = \int_0^{\infty}
dt t e^{-tx^2}$, one obtains
\begin{eqnarray}
\chi^{(4)}(\tilde{q}) = \frac{1}{16 \pi^2} \int_{ab}
\int_{0}^{\infty} dt \frac{t}{(t + \alpha_2)^2}(
e^{-\frac{\tilde{q}^2}{4(t+ \alpha_2)}} - 1) \nonumber
\end{eqnarray}
with $\alpha_2 = 1/2a + 1/2b$, from which we extract the following
asymptotic behaviors
\begin{eqnarray}
&&\chi^{(4)}(\tilde{q}) \sim -\tilde{q}^2 \int_{a,b,c} \frac{1}{128
\alpha_2} \quad \tilde{q} \ll 1 \nonumber \\
&&\chi^{(4)}(\tilde{q}) \sim -\frac{1}{16\pi^2}\ln{\tilde{q}^2}
\quad \tilde{q} \gg 1
\nonumber
\end{eqnarray}
as announced in the text (\ref{gamma4_petit},\ref{gamma4_grand}).

\section{detailed calculations for the CO Model - statics.} \label{app:CO}

\subsection{$\beta_{g_l}$-function}

The $\beta$-function for the coupling constant $g^K_l$ is obtained by
inserting (\ref{COint}) in (\ref{RGorder2_Fourier}). This gives straightforwardly
(\ref{beta_co_gene}) using $\partial_l G_l^{x=0} = - \frac{T}{2 \pi} \int_0^\infty  du c'(u) =
\frac{T}{2 \pi}$. One has also $G^{x=0}_{l' l} = \frac{T}{2 \pi} (l'-l)$.
Considering specifically $g^{1,-1}_l = g_l$,
 we first consider the
possible fusion rules such that $P+Q = K_{-1,1}$~:
\begin{eqnarray}
&&\quad P \quad \hspace{0.1cm} + \quad \hspace{0.1cm} Q \quad = \quad
K_{1,-1} \\
&&\begin{pmatrix}
. \\
. \\
1 \\
. \\
. \\
. \\
-1 \\
. \\
\end{pmatrix}
+
\begin{pmatrix}
. \\
. \\
-1 \\
.\\
1 \\
. \\
. \\
. \\
\end{pmatrix}
=
\begin{pmatrix}
. \\
. \\
. \\
.\\
1 \\
. \\
-1 \\
. \\
\end{pmatrix}
\end{eqnarray}
where $. \equiv 0$,
and there are $2(n-2)$ different ways to choose $P,Q$ like that,
notice $P.Q = -1$.
 Other possible
fusions rules involve charges of higher modulus, for instance we could
consider

\begin{eqnarray}
&&\quad P \quad \hspace{0.1cm} + \quad \hspace{0.1cm} Q \quad = \quad
K_{1,-1} \\
&&\begin{pmatrix}
. \\
. \\
1 \\
. \\
-2 \\
. \\
1 \\
. \\
\end{pmatrix}
+
\begin{pmatrix}
-1 \\
. \\
-1 \\
.\\
2 \\
. \\
. \\
. \\
\end{pmatrix}
=
\begin{pmatrix}
-1 \\
. \\
. \\
.\\
. \\
. \\
1 \\
. \\
\end{pmatrix}
\\ \nonumber
\end{eqnarray}
with $P^2 = Q^2 = 6$.

It is then useful to write the integrals $\tilde{J}^{(1)}_l$ and
$\tilde{J}^{(2)}_{l,l'}$
in (\ref{beta_co_gene})
in terms of the variables $\tilde{x}=\Lambda_l x $ and $\mu = l-l'$.
Using (\ref{rescal_propagCO}), and specifying to $g_l$ one has~:
\begin{equation}
\frac{\tilde{J}^{(1)}_l}{2 T^2} {\sum_{P,Q}} ' g^P_l g^Q_l (P.Q)^2 =
(n-2)g_l^2 \int_{\tilde{x}} \partial \gamma_0(\tilde{x})
\gamma_l(\tilde{x}) \label{J1_co}
\end{equation}
and
\begin{widetext}
\begin{eqnarray}
\frac{-1}{2 T^2} {\sum_{P,Q}}' (P.Q)^3 \int_0^l dl'
\tilde{J}^{2}_{l,l'} g^P_{l'} g^Q_{l'} =
(n-2) T \int_{\tilde{x}} (\partial
\gamma_0(\tilde{x}) - \partial \gamma_0(0))\int_0^l d\mu \partial
\gamma_{\mu}(\tilde{x}) (\gamma_l(\tilde{x}) - \gamma_{\mu}(\tilde{x}))
e^{(4 - \frac{T}{\pi})\mu} e^{T \gamma_{\mu}(\tilde{x})}g_{l-\mu}^2 \label{J2_co}
\end{eqnarray}
\end{widetext}
with ${\sum_{P,Q}}' \equiv \sum_{P,Q,P+Q=K}$. We study the flow near $T_c = 4 \pi$, and
as (\ref{J2_co}) is already of order $g_{l-\mu}^2$, we can evaluate
the integral over $\mu$ exactly at $T_c$ : in particular $e^{(4
-T/\pi)\mu} =1 + O(\tau)$.
Moreover, as the integral is convergent, it is dominated by the vicinity of
the fixed point $\mu = 0$. We can then substitute in (\ref{J2_co})
$g_{l - \mu}$ by $g_l$.
The remaining integral over $\mu$ is then straightforwardly computed
by
integration by parts. (\ref{J1_co}) together with (\ref{J2_co}),
integrated over
$\mu$ and using (\ref{magic_rel}), then lead in the limit $n \to 0$ to (\ref{RGlocco}):
\begin{eqnarray}
&&\partial_l g_l = (2 - \frac{T}{2 \pi})g_l -
2g_l^2 \partial \gamma_0(0)\int_{\tilde{x}} \gamma_l(\tilde{x}) \nonumber \\
&&- \frac{2g_l^2}{T_c}
\int_{\tilde{x}} (\partial \gamma_0(\tilde{x}) - \partial \gamma_0(0)) (e^{T_c \gamma_l(\tilde{x})} -1)
\label{RGloccoapp}
\end{eqnarray}
To compute the integrals over $\tilde{x}$ in (\ref{RGloccoapp})
in the limit $l \to \infty$ at $T = T_c$
we first quote some useful relations.
Using the decomposition of the cut-off function (\ref{decomp_cutoff}), we have
\begin{eqnarray}
&& \partial \gamma_\mu(x) =
\frac{1}{2 \pi} \int_a e^{- x^2 e^{2 \mu}/(2 a)} \nonumber \\
&& \gamma_\mu(x) =
\frac{1}{4\pi} \int_a \int_{x^2/2a}^{x^2 e^{2 \mu}/(2 a)}
\frac{dy}{y} e^{- y} \label{propag_app}
\end{eqnarray}
and the following identities:
\begin{eqnarray}
&& \partial_{\mu} \gamma_\mu(x) = \partial \gamma_\mu(x)
\nonumber \\
&& 2 x^2 \partial_{x^2}  \gamma_\mu(x)
= \partial \gamma_\mu(x)  - \partial \gamma_{\mu=0}(x) \label{magic_rel}
\end{eqnarray}
We first compute these integrals in the semi-bounded domain $|x| >
\epsilon$ and then take the limit $\epsilon \rightarrow 0$, in order
to avoid problems of convergence (the limit $l \rightarrow \infty$ does not
introduce any problem). Let us decompose the integrals over $\tilde{x}$
in the following way, writing $\frac{{\cal B}_\infty}{2}$ as:
\begin{eqnarray}
&& \partial \gamma_0(0) \int_{x}' \gamma_l(x)
+ \frac{1}{T_c} \int_x' (\partial \gamma_0(x) - \partial \gamma_0(0)
(e^{T_{c} \gamma_{\infty}(x)} -1)
 \nonumber \\
&& = \frac{1}{T_c} \int_x' \partial \gamma_{0}(x) e^{T_c \gamma_{\infty}(x)}
- \frac{1}{T_c} \int_x' \partial \gamma_0(0)(e^{T_c \gamma_{\infty}(x)} - 1) \nonumber \\
&&+\int_x' \partial \gamma_0(0) \gamma_{\infty}(x)   - \frac{\partial \gamma_0 (x)}{T_c} \nonumber \\
\end{eqnarray}
with $\int_x' \equiv \int_{|x| > \epsilon}$.
Using the previous formula (\ref{magic_rel}) for $l \to \infty$,
$2x^2 \partial_{x^2}\gamma_{\infty}(x) = - \partial \gamma_{\mu = 0}(x) , x>0$
since $\partial \gamma_{\infty}(x) = 0, x>0$, together with $\partial
\gamma_0(0) = 2/T_c$
we are left with (performing the change of variable $u = x^2$), and denoting
$\gamma_{\infty}(x)=\tilde{\gamma}_{\infty}(x^2)$
\begin{eqnarray}
&& \partial \gamma_0(0) \int_{x}' \gamma_l(x)
+ \frac{1}{T_c} \int_x' (\partial \gamma_0(x) - \partial \gamma_0(0) (e^{T_{c} \gamma_{\infty}(x)} -1)
 \nonumber \\
&& = \frac{-2\pi}{T_c^2} \int_{\epsilon}^{\infty} du (u T_c \partial_u
\tilde{\gamma}_{\infty}(u)
e^{T_c \tilde{\gamma}_{\infty}(u)} + (e^{T_c
\tilde{\gamma}_{\infty}(u)} - 1)) \nonumber \\
&& + \frac{2\pi}{T_c} \int_{\epsilon}^{\infty} du (\tilde{\gamma}_{\infty}(u) +
u \partial_u \tilde{\gamma}_{\infty}(u))
\nonumber \\
&&= \frac{-2 \pi}{T_c^2} [u(e^{T_c \tilde{\gamma}_{\infty}(u)}-1 - T_c
\tilde{\gamma}_{\infty}(u))]_{\epsilon}^{\infty} \label{beta_integr}
\end{eqnarray}
as one recognizes total derivatives in the integrands.
Using explicitly (\ref{propag_app}):
\begin{eqnarray}
\tilde{\gamma}_{\infty}(u) = \frac{1}{4 \pi} \int_a E_1(u/(2 a))
\end{eqnarray}
where $E_1(z)=-Ei(-z)=\int_z^{+\infty} e^{-z}/z$, with $Ei(x)$ the exponential
integral function, behaves asymptotically as
\begin{eqnarray}
&&E_1(z) \sim -\gamma_E - \ln{z} + O(z) \quad z \ll 1 \label{E1_small_arg} \\
&&E_1(z) \sim \frac{e^{-z}}{z} (1 + O(1/z)) \quad z \gg 1 \label{E1_large_arg}
\end{eqnarray}
where $\gamma_E$ is the Euler constant,
the limit $\epsilon \to 0$ in (\ref{beta_integr}) can be taken safely
to obtain
\begin{eqnarray}
&& {\cal B}_\infty = \frac{4 \pi}{T_c^2} e^{-(\gamma_E - \int_a \ln{2a} )}
\end{eqnarray}
which leads, together with (\ref{RGlocco}) to the fixed point value $g^*$
(\ref{co_fp}).

\subsection{Bi-local term for CO model}\label{bilocalco}

We compute in this section the bilocal part in the effective action
given by (\ref{Fl_co_gen}). Performing in (\ref{Fl_co_gen}) the change
of variable $l'
\to \mu = l-l'$ and using the notations (\ref{rescal_propagCO}) and
$G_{l'}^x = - T (\gamma_\mu(\tilde x) - \gamma_l(\tilde x))$,  one gets
\begin{eqnarray}
&&\hat V^{KPq}_{l} = \frac{1}{2}\int_x (e^{iqx} -1)\hat F_{l}^{KPx}
\label{Vl_app}\\
&&\hat F_{l}^{KPx} = \frac{\Lambda_l^4}{T^2} (K.P)^2 \int_0^l d\mu \partial
\gamma_{\mu}(\Lambda_l x) (\gamma_{\mu}(\Lambda_l x) -
\gamma_{l}(\Lambda_l x)  ) \nonumber \\
&&\times e^{- TK.P\gamma_{\mu}(\Lambda_l x)}
e^{(4-\frac{T}{\pi})\mu} g_{l-\mu}^2 \nonumber
\end{eqnarray}
As previously, this integral is already of order $g_{l-\mu}^2$, so it
can be evaluated at $T_c$, in particular $e^{(4-\frac{T}{\pi})\mu} = 1
+ O(\epsilon)$. Besides the integral over $\mu$ is convergent and
dominated by $\mu = 0$, so that we substitute $g_{l-\mu}^2$ by
$g_{l}^2$. The remaining integral over $\mu$ is then straightforwardly
computed to obtain
\begin{equation}
\hat F_l^{KPx} = - \frac{\Lambda_l^4}{T^2} g_l^2
\left(\frac{1}{T^2} (e^{-T_c
K.P \gamma_l(\Lambda_l x)} -1) + K.P \frac{\gamma_l(\Lambda_l x)}{T}
\right)
\end{equation}
where (\ref{propag_app}) can be written as
\begin{equation}
\gamma_l(\Lambda_l x) = \frac{1}{4 \pi} \int_a E_1(x^2\Lambda_l^2/2a)
- E_1(x^2\Lambda_0^2/2a)
\end{equation}
whith the asymptotic behaviors of $E_1(z)$ given in
(\ref{E1_small_arg}, \ref{E1_large_arg}). For any $\Lambda_l$,
$\Lambda_0$ the integral over $x$ in (\ref{Vl_app}) is well defined,
but we see that in the
limit $\Lambda_0 \to \infty$ (i.e. $\Lambda_l, q \ll \Lambda_0$),
there is a logarithmic divergence (for
small $x$) and only for charges such that $K.P = -2$. Indeed, at small
$x$, using (\ref{E1_small_arg}),
$-T_c K.P \gamma_l(\Lambda_l x) \sim K.P (\gamma_E + \ln{(\Lambda_l^2
x^2)})$, leading to
$e^{-T_c K.P \gamma_l(\Lambda_l x)} \sim x^{2 K.P}$. This implies that
the limit
$\Lambda_0 \to \infty$ only diverges for $K.P = -2$ (there is no
problem with the large $x$ behavior as the integrand decay exponentially for
any couple of charges we consider here).

\subsubsection{The case of charges $K.P = 1$ or $2$.}

For these charges, the limit $\Lambda_0 \to \infty$ can be taken
directly on (\ref{Vl_app}). This leads to, performing the change of
variable $x \to \Lambda_l x$ and the integral over the angular
variable on (\ref{Vl_app})
\begin{eqnarray}
&&\hat V^{KPq}_l = - q^2 g_l^2 \frac{\pi}{T_c^4} \int_0^{\infty} dr
\frac{r}{\tilde{q}^2}
(J_0(|\tilde{q}| r) - 1) \\
&&(e^{-K.P \int_a E_1(r^2/2a)} - 1 + K.P \int_a E_1(r^2/2a)) \nonumber
\end{eqnarray}
where $\tilde{q} = q/\Lambda_l$
and $J_0(z)$ is a Bessel function of the first kind. This defines
the function $\chi^{K.P}(k)$ in that case
\begin{eqnarray}\label{chi_1}
&&\chi^{K.P}(k) = 4 e^{2(\gamma_E - \int_a \ln{2a})} \int_0^{\infty} dr
\frac{r}{k^2}
(J_0(|k| r) - 1) \nonumber \\
&&(e^{-K.P \int_a E_1(r^2/2a)} - 1 + K.P \int_a E_1(r^2/2a))
\end{eqnarray}
The small $k$ behavior (the first line of (\ref{gamma_co_small}))
 is straighforwardly obtained as
\begin{eqnarray}\label{chi_1_small}
&&\chi^{K.P}(k) \sim a_{K,P} + O(k^2) \\
&&a_{K,P} = - \frac{e^{2(\gamma_E - \int_a \ln{2a})}}{2}
\int_{u>0} u (e^{-K.P \int_a E_1(u/2a)} - 1 \nonumber \\
&&+ K.P \int_a E_1(u/2a)) \nonumber
\end{eqnarray}
where we perfomed the change of variable $u=r^2$.

For $K.P = 1$ or $2$, $r e^{-K.P \int_a E_1(r^2/2a)}
\sim r^{2 K.P + 1}$ when $r \ll 1$ is analytic in $0$ and using
$J_0(k) \sim k^{-1/2} \cos(k-\frac{\pi}{4})$ one finds for $k \gg 1$
\begin{equation}
\int_0^{\infty} dr
\frac{r}{k^2}J_0(|k| r)(e^{-K.P \int_a E_1(r^2/2a)} - 1) \sim O(\frac{1}{k^{5/2}})
 \label{int_reg_1}
\end{equation}
We have moreover
\begin{eqnarray}
&&\int_0^{\infty} dr
\frac{r}{k^2}
(J_0(|k| r) - 1)\int_a E_1(r^2/2a) \nonumber \\
&&= \int_a
\frac{2 - 2e^{-ak^2/2} - a k^2}{k^4} \label{BesselEi}
\end{eqnarray}
Using (\ref{BesselEi}) together with (\ref{int_reg_1})
one obtains the leading behavior of $\chi^{K,P}(k)$
(\ref{chi_1}) in the large $k$ limit, i.e. the first line of
(\ref{gamma_co_large})
\begin{eqnarray}
&&\chi^{K.P}(k) \sim b_{K.P} \frac{1}{k^2} \\
&&b_{K.P} =  - 2 e^{2(\gamma_E - \int_a \ln{2a})}(\int_{u>0}
(e^{-K.P \int_a E_1(\frac{u}{2a})} - 1) \nonumber \\
&&- 2 K.P c'(0)  ) \nonumber
\end{eqnarray}
where we made the change of variable $u=r^2$ and used $c'(0) = - \int_a a$.

\subsubsection{The case of charges $K.P=-1$}

In that case, $\chi^{K.P}(k)$ is formally obtained as previously
(\ref{chi_1}),
the small $k$ behavior being still given by
(\ref{chi_1_small}). However the large $k$
behavior is dominated by the small $r$ region and as noticed previously
for $r \ll 1$, $r (e^{-K.P \int_a E_1(r^2/2a)} - 1)
\sim r^{2 K.P + 1} = r^{-1}$, which leads to a logarithmic divergence in the
large $k$ limit. It can be obtained by computing
\begin{eqnarray} \label{int_div_-1}
&&\int_0^{\infty} dr r (J_0(k r) - 1) (e^{\int_a E_1(r^2/2a)} -1 ) \nonumber \\
&&\sim e^{-\gamma_E + \int_a \ln{2a}}
\int_0^{\infty} \frac{dr}{r} (J_0(k r) - 1) \nonumber \\
&&\sim - e^{-\gamma_E + \int_a \ln{2a}}\ln{k} \quad k \gg 1
\end{eqnarray}
The last term in (\ref{chi_1})
has the same behavior (\ref{BesselEi}) independently of
$K.P$ and (\ref{int_div_-1}) together with (\ref{chi_1}) for $K.P =-1$
lead to the second line of (\ref{gamma_co_large})
\begin{eqnarray}
&&\chi^{K,P}(k) \sim b_{-1} \frac{\ln k}{k^2} \nonumber \\
&&b_{-1} = - 4 e^{\gamma_E - \int_a \ln{2a}}
\end{eqnarray}

\subsubsection{The case of charges $K.P = -2$}

As pointed out previously, there is in that case a logarithmic divergence
when $\Lambda_0 \gg 1$.
We isolate this divergence by
writing
\begin{eqnarray}
&&\hat V_l^{KPq} =  \frac{-1}{4} \int_x (qx)^2 \hat F_l^{KPx} \nonumber \\
&&+ \frac{1}{2} \int_x (e^{iqx} - 1 + \frac{1}{2}(qx)^2)\hat F_l^{KPx}
\label{Vl_isol}
\end{eqnarray}
the second term being well defined in the limit $\Lambda_0 \to
\infty$. We focus now on the first part, using the explicit expression of
$\gamma_l(\Lambda_l x)$ (\ref{propag_app})
\begin{eqnarray}
&&- \frac{1}{4} \int_x (qx)^2 \hat F_l^{K-Kx} = \\
&& \frac{1}{8 T_c^4} g_l^2
q^2 \int_x x^2 \Big \{e^{2 \int_a E_1(\frac{x^2}{2a}) -
E_1(\frac{x^2 \Lambda_0^2}{2a \Lambda_l^2})} -1     \nonumber \\
&&- 2 \left(\int_a E_1\left(\frac{x^2}{2a}\right) - E_1\left(\frac{x^2
\Lambda_0^2}{2a
\Lambda_l^2}\right) \right)\Big \} = {\cal
H}\left(\frac{\Lambda_0^2}{\Lambda_l^2}\right)
\nonumber
\end{eqnarray}
where we made the change of variable $x \to \Lambda_l x$.
To anlyse the large argument behavior of ${\cal H}(\lambda)$, we take
the derivative w.r.t $\lambda$
\begin{eqnarray}
&&{\cal H}'(\lambda) = \frac{\pi}{4 T_c^4} g_l^2 q^2 \int_0^{\infty}
dx x^3  \\
&&\int_a \frac{2}{\lambda} e^{- \frac{\lambda x^2}{2a}}
(e^{2 (\int_a E_1(\frac{x^2}{2a}) -
E_1(\frac{\lambda x^2}{2a}))} -1) \nonumber
\end{eqnarray}
where we have used $E_1'(z) = - e^{-z}/z$.
Making the change of variable $u = \lambda x^2$ in ${\cal
H}'(\lambda)$ one obtains:
\begin{eqnarray}
&&{\cal H}'(\lambda) =  \frac{\pi}{8 T_c^4} g_l^2 q^2 \int_{0}^{\infty}
 \frac{du}{\lambda^3}u \nonumber \\
&&\int_a 2 e^{-\frac{u}{2a}} (e^{2(\int_a E_1(\frac{u}{2a \lambda})
 - E_1(\frac{u}{2a}))} - 1) \nonumber
\end{eqnarray}
using the large $\lambda$ behavior $E_1(u/(2a\lambda)) \sim -\gamma_E +
\int_a \ln{(2a)}
- \ln{(u/\lambda)} + O(1/\lambda)$ one gets
\begin{eqnarray}
&&{\cal H}'(\lambda) = \frac{\pi}{8 T_c^4} g_l^2 q^2 e^{2\int_a
(-\gamma_E + \ln{2a})}
\frac{1}{\lambda} \nonumber \\
&&\int_0^{\infty} du \int_a \frac{2}{u}e^{-\frac{u}{2a}}e^{-2 \int_a
E_1(\frac{u}{2a})}
+ O(\lambda^{-2}) \nonumber \\
&& =\frac{\pi}{8 T_c^4} g_l^2 q^2 e^{2\int_a (-\gamma_E +
\ln{2a})}\frac{1}{\lambda}
 [e^{-2\int_a E_1(\frac{u}{2a})}]_0^{\infty} + O(\lambda^{-2}) \nonumber \\
&&= \frac{\pi}{8 T_c^4} g_l^2 q^2 e^{2\int_a (-\gamma_E +
\ln{2a})}\frac{1}{\lambda}
+ O(\lambda^{-2})
\end{eqnarray}
where we have used the asymptotic behaviors
(\ref{E1_small_arg},\ref{E1_large_arg}).
This leads finally to
\begin{eqnarray}
&&- \frac{1}{4} \int_x (qx)^2 \hat F_l^{KPx} = - \delta_{K,-P}
A_l q^2 \ln{(\frac{\Lambda_l}{\Lambda_0})}
+ O(\Lambda_l^2/\Lambda_0^2) \nonumber \\
&& A_l = \frac{\pi}{4 T_c^4} g_l^2 e^{2\int_a (-\gamma_E + \ln{2a})}\
\end{eqnarray}
which is the first term in (\ref{Vl_CO}) with the amplitude $A_l$ given in
(\ref{amplitude_CO}).

In the second line of (\ref{Vl_isol}), we perform the change of
variable $x \to \Lambda_l x$ and the integral over
the angular variable to get
\begin{eqnarray}
&&\frac{1}{2} \int_x (e^{iqx} - 1 + \frac{1}{2}(qx)^2)\hat F_l^{KPx}
\nonumber \\
&&= - q^2 g_l^2 \frac{\pi}{T_c^4} \int_0^{\infty} dr \frac{r}{\tilde{q}^2}
(J_0(|\tilde{q}| r) - 1 + \frac{1}{4}\tilde{q}^2 r^2) \nonumber \\
&&(e^{2 \int_a E_1(r^2/2a)} - 1 -2 \int_a E_1(r^2/2a))
\end{eqnarray}
where $J_0(z)$ is a Bessel function of the first kind, from which we
get the function $\chi^{K.P}(k)$ defined in the text for $K.P = -2$
\begin{eqnarray}\label{chi_-2}
&&\chi^{K.P}(k) = 4 e^{2(\gamma_E - \int_a \ln{2a})} \int_0^{\infty} dr
\frac{r}{k^2}
(J_0(|k| r) - 1 + \frac{1}{4} k^2 r^2) \nonumber \\
&&(e^{2 \int_a E_1(r^2/2a)} - 1 - 2 \int_a E_1(r^2/2a))
\end{eqnarray}
the small $k$ behavior (i.e. the second line of (\ref{gamma_co_small})
in the text) is easily obtained
\begin{eqnarray}
&&\chi^{K.P}(k) \sim a_{-2} k^2 \quad k \ll 1 \\
&&a_{-2} = \frac{e^{2(\gamma_E - \int_a \ln{2a})}}{32}
\int_{0}^{\infty} du u^2 \nonumber \\
&&(e^{2 \int_a E_1(u/2a)} - 1 - 2 \int_a E_1(u/2a)) \nonumber
\end{eqnarray}
where we made the change of variable $u = r^2$. The large $k$
behavior is governed by the small $r$ region in the integral
(\ref{chi_-2}), where $r(e^{2 \int_a E_1(r^2/2a)} - 1 - 2 \int_a
E_1(r^2/2a))  \sim e^{- 2 \gamma_E + \int_a \ln{2a}} r^{-3}$, which implies for $k \gg 1$
\begin{eqnarray}
&&\chi^{K.P}(k) \sim \frac{4}{k^2} \int_0^{\infty}
\frac{dr}{r^3}(J_0(|k| r) - 1 + \frac{1}{4} k^2 r^2) + O(1) \nonumber \\
&&\sim \ln{k} + O(1)
\end{eqnarray}
which is the last line of (\ref{gamma_co_large}) in the text.

\section{Detailed calculations for the CO model : Equilibrium
dynamics.}\label{app_dyncoeq}

\subsection{Derivation of the RG flow.}

We restrict our analysis to order one $O(U_l)$, and at this order the RG flow
reads (\ref{RGorder2})
\begin{eqnarray} \label{app_rgdyn}
\partial_l U_l(u,i\hat u) =  \frac{1}{2} \partial G^{x=0}_{l,ij} \partial_i
\partial_j U_l(u,i\hat u)
\end{eqnarray}
where $U_l(u,i\hat u)$ is given by (\ref{local_dyn})
and the indices $i,j$ formally refer to the components of the vector $\phi$
(\ref{vector_dyn}) and the time dependence, i.e.
$\partial_i \equiv \frac{\delta}{\delta u_t},
\frac{\delta}{\delta i \hat u_t}$. From (\ref{propag_dyn}), the matrix
$G^q_{l,tt'}$ has the following expression
\begin{eqnarray}
&&G_l^q =
\left(
\begin{array}{c c}
C_{l}^{q} & R_l^{q} \\
R_l^{q\dagger} & 0
\end{array}
\right) \nonumber \\
\end{eqnarray}
With these notations, we have
\begin{eqnarray}
&& \frac{1}{2} \partial G^{x=0}_{l,ij} \partial_i
\partial_j = \frac{1}{2} \frac{\delta}{\delta u} \cdot \partial C_l^{x=0} \cdot
\frac{\delta}{\delta u} + \frac{\delta}{\delta u} \cdot \partial R_l^{x=0} \cdot
\frac{\delta}{\delta i\hat u} \nonumber \\
&&
\end{eqnarray}
where we will often use the matrix notation for time, i.e
$u \cdot v = \int_t u_t v_t$.
Acting with this operator on $U_l(u,i \hat u)$, one gets
\begin{eqnarray}
&&\frac{1}{2} \partial G^{x=0}_{l,ij} \partial_i
\partial_j (\int_t i\hat u_{xt} F_{lt}(u) - \frac{1}{2} \int_{tt'} i\hat
u_{xt} i\hat u_{xt'} \Delta_{ltt'}(u) ) \nonumber \\
&& = - \frac{1}{4} \int_{tt'}i\hat u_{t} i\hat u_{t'}   \int_{t_1
t'_1}\frac{\delta}{\delta u_{t_1}} \partial C_{lt_1 t'_1}^{x=0}
\frac{\delta}{\delta u_{t'_1}}
\Delta_{ltt'}(u) \nonumber \\
&&  + \frac{1}{2} \int_{t}i\hat u_{t} \int_{t_1
t'_1} \frac{\delta}{\delta u_{t_1}}\partial C_{lt_1t'_1}^{x=0}
\frac{\delta}{\delta u_{t'_1}} F_{lt}(u) \nonumber \\
&& - \int_t i\hat u_{t} \int_{t_1>t'_1} \partial R_{l t_1
t'_1}^{x=0} \frac{\delta}{\delta u_{t_1}} \Delta_{l t t'_1}
\nonumber \\
&&+ \int_{t_1 > t} \partial R_{l t_1
t}^{x=0} \frac{\delta}{\delta u_{t_1}} F_t(u)
\end{eqnarray}
The last term vanishes by causality since
$F_t(u)$ depends on $u_{t_1}$ with $t_1<t$ only.
Identifying in (\ref{app_rgdyn})
the coefficient of the powers in the field $i\hat u$ one gets
(\ref{ERGE_dyn_order_one}) in the text.

The first equation of (\ref{ERGE_dyn_order_one}) is easily solved, and
it gives
\begin{eqnarray}\label{solrg_delta}
&&\Delta_{ltt'}(u) = e^{\frac{1}{2}\int_{t_1t'_1} \frac{\delta}{\delta u_{t_1}}
C^{x=0}_{lt_1t'_1} \frac{\delta}{\delta u_{t'_1}} }
\Delta_{l=0tt'}(u) \\
&& = 2 e^{- C^{x=0}_{l0} + C^{x=0}_{lt-t'}} \Lambda_0^2 g_0
\cos(u_{xt} - u_{xt'}) \nonumber
\end{eqnarray}
where we have used $C_{l=00}^{x=0} =0$.
From the previous study of the statics, one has that
\begin{eqnarray}
e^{- C^{x=0}_{l0}} \Lambda_0^2 g_0 = \Lambda_l^2 g_l + O(g_l^2)
\end{eqnarray}
which leads together with (\ref{solrg_delta}) to the first line in
(\ref{loc_co_dyn_eq}).
By taking the functional derivative w.r.t. $u_{xt'}$ in the
second line of (\ref{ERGE_dyn_order_one}) and using
the same manipulations one gets the second line in (\ref{loc_co_dyn_eq}).

\subsection{Computation of the dynamical exponent $z$.}

Here we compute the self energy $\Sigma_{l\omega}$ (\ref{def_sigma})
given by
\begin{eqnarray} \label{app_expr_sigma}
&&\Sigma_{l\omega} = \int_0^{\infty} dt e^{i \omega t} \Sigma_{lt} \\
&&\Sigma_{lt} = - 2 \Lambda_l^2 g_l\left(   R^{x=0}_{lt} e^{C_{lt}^{x=0}} -
\delta(t) \int_{0}^{\infty}dt' R^{x=0}_{lt'}
e^{C_{lt'}^{x=0}}\right)   \nonumber
\end{eqnarray}
Notice the terms proportional to $\delta(t)$ in $\Sigma_{lt}$ (not
given in the text (\ref{loc_co_dyn_eq}) for clarity)
which guarantees that $\Sigma_{l\omega = 0} = 0$, and
with the explicit expressions for the bare correlation and response
functions (\ref{barefunc}) computed with the cut-off decompostion
(\ref{decomp_cutoff})
\begin{eqnarray} \label{expressions}
&&C_{lt}^{x=0} = \frac{T}{4\pi} \int_a \ln{\left(\frac{|t| +
\frac{a}{2\Lambda_l^2}}{|t| + \frac{a}{2\Lambda_0^2}}\right)} \\
&&R_{lt}^{x=0} = \frac{\theta(t)}{4 \pi} \int_a \frac{1}{t +
\frac{a}{2\Lambda_0^2}} - \frac{1}{t +
\frac{a}{2\Lambda_l^2}}
\end{eqnarray}
After an integration by part in (\ref{app_expr_sigma}) and using those
explicit expressions one gets to
order $\tau$
\begin{equation}
\Sigma_{l\omega} = -\frac{2 \Lambda_l^2 g_l}{T_c} i \omega
\int_0^{\infty} dt e^{i\omega t}
(e^{\int_a \ln{\left(\frac{t +
\frac{a}{2\Lambda_l^2}}{t + \frac{a}{2\Lambda_0^2}}\right)}  } -1)
\end{equation}
This expression is logarithmically divergent for $\Lambda_0 \to
\infty$ (the integrand behaves as $1/t$ at small $t$ in this limit),
and a way to isolate this divergence is to decompose this
integral in the following way (and performing the change of variable
$t \to t/\Lambda_l^2$)
\begin{eqnarray}\label{app_decomp_sigma}
&&\Sigma_{l\omega} =  -\frac{2 g_l}{T_c} i \omega
\int_1^{\infty} dt e^{i\tilde{\omega} t}
(e^{\int_a \ln{\left(\frac{t +
\frac{a}{2}}{t + \frac{\lambda a}{2}}\right)}  } -1)
\nonumber \\
&& - \frac{2 g_l}{T_c} i \omega \int_0^{1} dt (e^{i \tilde{\omega}
t} - 1)
(e^{\int_a \ln{\left(\frac{t +
\frac{a}{2}}{t + \frac{\lambda a}{2}}\right)}  }
-1)\nonumber \\
&& - \frac{2 g_l}{T_c} i \omega \int_0^{1} dt
(e^{\int_a \ln{\left(\frac{t +
\frac{a}{2}}{t + \frac{\lambda a}{2}}\right)}  }
-1)\nonumber \\
\end{eqnarray}
where $\tilde{\omega} = \omega/\Lambda_l^2$ and $\lambda =
\Lambda_l^2/\Lambda_0^2$.
In the first two lines we can take safely the limit $\lambda \to
0$ and we focus now on the divergent part of the last term
\begin{eqnarray}
\mathbb{H}(\lambda) = - \frac{2 g_l}{T_c} i\omega
\int_0^{1} dt
e^{\int_a \ln{\left(\frac{t +
\frac{a}{2}}{t + \frac{\lambda a}{2}}\right)}  } \nonumber \\
\end{eqnarray}
Taking the derivative w.r.t. $\lambda$, one has
\begin{eqnarray} \label{deriv_h}
&&\mathbb{H}'(\lambda) = - \frac{ 2g_l}{T_c}i\omega \int_0^{1} dt
\int_a \frac{-a}{2} \frac{1}{t + \frac{\lambda a}{2}}e^{\int_a
\ln{\left(\frac{t +
\frac{a}{2}}{t + \frac{\lambda a}{2}}\right)}  } \nonumber \\
&& =
-\frac{\mathbb{H}(\lambda)}{\lambda} -  \frac{2
g_l}{T_c}\frac{i\omega}{\lambda} \int_0^1 dt t
\int_a \frac{1}{t + \frac{a}{2}} e^{\int_a \ln{\left(\frac{t +
\frac{a}{2}}{t + \frac{\lambda a}{2}}\right)}  } \nonumber \\
&& - \frac{2 g_l}{T_c} \frac{i\omega}{\lambda} \int_0^1 dt t
\int_a (\frac{1}{t + \frac{\lambda a}{2}}
-\frac{1}{t + \frac{a}{2}})e^{\int_a \ln{\left(\frac{t +
\frac{a}{2}}{t + \frac{\lambda a}{2}}\right)}  }  \nonumber \\
\end{eqnarray}
In the integral of the second line, we can take the limit $\lambda
\to 0$, it gives
\begin{eqnarray}
&&-\frac{2 g_l}{T_c}\frac{i\omega}{\lambda}\int_0^1 dt t
\int_a \frac{1}{t + \frac{a}{2}} e^{\int_a \ln{\left(\frac{t +
\frac{a}{2}}{t + \frac{\lambda a}{2}}\right)}  }) \nonumber \\
&&\sim -\frac{2 g_l}{T_c}\frac{i\omega}{\lambda}\int_0^1 \int_a
\frac{1}{t + \frac{a}{2}} e^{\int_a \ln{t +
\frac{a}{2}}  } + O(1) \nonumber \\
&&\sim  -\frac{2 g_l}{T_c} \frac{i\omega}{\lambda} (e^{\int_a
\ln{(1+\frac{a}{2})}} -e^{\int_a
\ln{(\frac{a}{2})}} ) + O(1) \nonumber
\end{eqnarray}
The last term in (\ref{deriv_h}) can be integrated by parts, to get
\begin{eqnarray}
&& -\frac{2 g_l}{T_c} \frac{i\omega}{\lambda} \int_0^1 dt t
\int_a \frac{1}{t + \frac{\lambda a}{2}}
-\frac{1}{t + \frac{a}{2}}e^{\int_a \ln{\left(\frac{t +
\frac{a}{2}}{t + \frac{\lambda a}{2}}\right)}  }  \nonumber \\
&& = \frac{\mathbb{H}(\lambda)}{\lambda} + \frac{2 g_l}{T_c}
\frac{i\omega}{\lambda}e^{\int_a
\ln{(1+\frac{a}{2})}}  + O(1)
\end{eqnarray}
Finally $\mathbb{H}'(\lambda)$ in (\ref{deriv_h}) can be written as
\begin{eqnarray}
&&\mathbb{H}'(\lambda) \sim i\omega (\frac{g_l}{2 T_c} e^{\int_a \ln{2a}}
\frac{1}{\lambda} + O(1)) \nonumber \\
&&\mathbb{H}(\lambda) \sim i\omega(\frac{g_l}{2 T_c} e^{\int_a \ln{2a}}
\ln{\lambda} + O(1))
\end{eqnarray}
which gives together with the last line of (\ref{app_decomp_sigma})
the first term in (\ref{expr_sigma}) with the amplitude $B_l
=\frac{g_l}{2 T_c} e^{\int_a \ln{2a}} $. The first two lines of
(\ref{app_decomp_sigma}) where we take the limit $\lambda \to 0$
define the function $\chi^{\text{(dyn)}}(\nu)$ of
(\ref{expr_sigma}):
\begin{eqnarray}\label{app_def_chi_dyn}
&&\chi^{\text{(dyn)}}(\nu) = -4 e^{-\int_a \ln{2a}} \Big\{
\int_1^{\infty} dt e^{i\nu t}
(\frac{1}{t} e^{\int_a \ln{(t + \frac{a}{2})}} -1)
\nonumber \\
&&+ \int_0^{1} dt (e^{i \nu t} - 1)
(\frac{1}{t}e^{\int_a \ln{(t +
\frac{a}{2}})}-1) \Big \}
\end{eqnarray}
The small argument behavior of $\chi^{\text{(dyn)}}(\nu)$ is
dominated by the large $t$ region of the integrand (i.e. the first
line of (\ref{app_def_chi_dyn})). Using that $(\frac{1}{t} e^{\int_a
\ln{(t + \frac{a}{2})}} -1) \sim \int_a \frac{a}{2t}$ for $t \gg 1$,
one gets
\begin{eqnarray}
&&\chi^{\text{(dyn)}}(\nu) \sim -4 e^{-\int_a \ln{2a}} \int_a
\frac{a}{2} \int_1^{\infty} e^{i\nu t}\frac{1}{t} \quad \nu \ll 1
\nonumber \\
&&\sim 4 e^{-\int_a \ln{2a}} \int_a
\frac{a}{2} \ln{\nu} \quad \nu \ll 1
\end{eqnarray}
which is the asymptotic behavior announced in the text
(\ref{chi_dyn_small}) with the non universal amplitude
$a_{\text{dyn}} =  4 e^{-\int_a \ln{2a}} \int_a
\frac{a}{2}$.
The large $\nu$ behavior of $\chi^{\text{(dyn)}}(\nu)$ is governed by
the small $t$ region of the integrand, i.e. the second line of
(\ref{app_def_chi_dyn}):
\begin{eqnarray}
&&\chi^{\text{(dyn)}}(\nu) \sim -4 e^{-\int_a \ln{2a}} e^{\int_a
\ln{\frac{a}{2}}} \int_0^1 dt
(e^{i\nu t} - 1)\frac{1}{t} \nonumber \\
&& \sim \ln{\nu} \quad \nu \gg 1
\end{eqnarray}
which is the asymptotic behavior annouced in the text (\ref{chi_dyn_large}).

We show here how to take directly, in a cruder way,
 the limit $l \to \infty$ in
$\Sigma_{lt}$ (\ref{app_expr_sigma}).
Indeed, using the explicit expression of $C_{lt}^{x=0}$ and
$R_{lt}^{x=0}$ (\ref{expressions}), one has
\begin{eqnarray}
&&
C_{lt}^{x=0} \sim - \frac{T}{4 \pi} \int_a (\ln{(4\Lambda_0^2 t + 2a)}
- \ln{(2a)} - 2l) + O(e^{-2l}) \nonumber \\
&& R_{lt}^{x=0} \sim \frac{1}{4\pi} \int_a \frac{1}{t +
\frac{a}{2\Lambda_0^2}} +  O(e^{-2l})
\end{eqnarray}
This allows to take the large $l$ limit in $\Sigma_{lt}$ at $T_c$ (as
it is already of order $\tau$)
\begin{equation} \label{sigma_t_fp}
\lim_{l\to \infty} \Sigma_{lt} = - \frac{\Lambda_0^2}{2 \pi} g^*
e^{\int_a \ln{(2a)}}\int_a \frac{1}{t+\frac{a}{2\Lambda_0^2}}
e^{-\int_a \ln{(4\Lambda_0^2t + 2a)}}
\end{equation}
for $t>0$.
We then obtain directly $\Sigma_{l\omega}$ in the limit $l \to \infty$
as
\begin{eqnarray}
&&\lim_{l \to \infty} \Sigma_{l\omega} = -\frac{g^* \Lambda_0^2 e^{\int_a \ln{(2a)}} }{2 \pi}
i\omega \int_0^{\infty} dt
e^{i\omega t} e^{- \int_a \ln{4\Lambda_0^2 t + 2a}} \nonumber \\
&& = -\frac{g^*e^{\int_a \ln{(2a)}}}{2 \pi} i\omega \int_0^{\infty} dt
e^{i\frac{\omega}{\Lambda_0^2} t} e^{- \int_a \ln{4t + 2a}}
\end{eqnarray}
The small $\omega/\Lambda_0^2$ behavior is governed by the large $t$
region of the integrand, which gives
\begin{eqnarray}
&&\lim_{l \to \infty} \Sigma_{l\omega} \sim -\frac{g^*e^{\int_a
\ln{(2a)}}}{2 \pi} i\omega \int_1^{\infty}
e^{i\frac{\omega}{\Lambda_0^2} t}\frac{1}{4t} \nonumber \\
&& \sim  B^* i\omega
\ln{\left(\frac{\omega}{\Lambda_0^2}\right)} +
O\left(\frac{i\omega}{\Lambda_0^2}\right) \quad \frac{i\omega}{\Lambda_0^2} \ll 1
\end{eqnarray}
which gives the same result obtained by the previous analysis
(\ref{reponse_fp}).

\subsection{Scaling function at equilibrium}

In this section, we show how to solve the equation for the response
function (\ref{eqResp}). First, it is natural to search for a solution under
the form ${\cal R}^q_{t} = e^{-q^2 t} {\cal G}_{t}^q$. Then,
performing the change of variable $u = t-t_1$ and using the explicit
expression (\ref{sigma_t_fp}), one gets the following equation for
${\cal G}_{t}^q$:
\begin{eqnarray}\label{Resp_eq_app}
&&\partial_{\tilde t} {\cal G}_{\tilde t}^{\tilde q} = 4 B^* \int_0^{\tilde t}
du \int_a \frac{1}{u+\frac{a}{2}}
e^{-\int_a \ln{(4u +2a)}}e^{{\tilde q}^2 u} \nonumber \\
&& - 4B^* \int_0^{\infty} du \int_a \frac{1}{u+\frac{a}{2}}
e^{-\int_a \ln{(4u +2a)}}
\end{eqnarray}
where $\tilde q = q/\Lambda_0$ and $\tilde t = \Lambda_0^2 t$,
with the intial conditions:
\begin{eqnarray}
&&{\cal G}^{\tilde{q}}_{0^+} = 1 \\
&&{\cal G}^{\tilde{q}}_{0} = 0
\end{eqnarray}
The second term in the l.h.s. is a total derivative and can be
integrated. Performing an integration by part on the first term one
gets
\begin{eqnarray}
&&{\cal G}_{\tilde t}^{\tilde q} = 1 +  4 B^* ( {\tilde
q}^2 \int_0^{\tilde t}dv \int_0^{v} du e^{-\int_a \ln{(4u
+2a)}}e^{q^2u} \nonumber \\
&&-  \int_0^{\tilde t}dv e^{q^2 {v}}
e^{-\int_a \ln{(4{v} +2a)}}     )
\end{eqnarray}
Performing an integration by part in the integral over $v$ on the
first integral and peforming the change of variable $u' = q^2 u$ in
the remaining integrals one gets
\begin{eqnarray}
&&{\cal G}_{\tilde t}^{\tilde q} = 1 + 4B^* ( ( {\tilde q}^2 \tilde{t}
- 1) \frac{1}{{\tilde q}^2} \int_0^{\tilde{q}^2 \tilde{t}} du
e^{-\int_a \ln{(\frac{4 u}{\tilde{q}^2} + 2a)}} (e^u-1)  \nonumber \\
&& - \frac{1}{{\tilde q}^2} \int_0^{\tilde{q}^2 \tilde{t}} du u e^u
e^{-\int_a \ln{(\frac{4 u}{\tilde{q}^2} + 2a)}}   \\
&& + ( {\tilde q}^2 \tilde{t}
- 1) \frac{1}{{\tilde q}^2} \int_0^{\tilde{q}^2 \tilde{t}} du
e^{-\int_a \ln{(\frac{4 u}{\tilde{q}^2} + 2a)}}    )
\end{eqnarray}

We now want to find the scaling function, i.e. the asymptotic behavior when
$\tilde{q} \to 0$ ($\Lambda_0 \to \infty$), keeping $\tilde{q}^2
\tilde{t} = y$ fixed.
In the two first lines of the above expression, the limit $\tilde q
\to 0$ can be
taken safely, although the last term is divergent in this limit. Thus
one has
\begin{eqnarray}
&&{\cal G}_{\tilde t}^{\tilde q} = 1 + 4B^* (( {\tilde q}^2 \tilde{t}
- 1)  \int_0^{\tilde{q}^2 \tilde{t}} du
\frac{e^u -1}{4 u} - \frac{1}{4} \int_0^{\tilde{q}^2 \tilde{t}} du e^u
\nonumber \\
&& + ( {\tilde q}^2 \tilde{t}
- 1) \frac{1}{{\tilde q}^2} \int_0^{\tilde{q}^2 \tilde{t}} du
e^{-\int_a \ln{(\frac{4 u}{\tilde{q}^2} + 2a)}}    ) + O(\tilde{q}^2)
\end{eqnarray}
To find the asymptotic behavior of the last term we
write
\begin{eqnarray}
&&\frac{1}{{\tilde q}^2} \int_0^{y} du
e^{-\int_a \ln{(\frac{4 u}{\tilde{q}^2} + 2a)}} = \\
&&\frac{1}{\tilde{q}^2} \int_0^{y} du
(e^{-\int_a \ln{(\frac{4 u}{\tilde{q}^2} + 2a)}} - \int_a
\frac{1}{\frac{4 u}{\tilde{q}^2} + 2a} ) \\
&& + \frac{1}{\tilde{q}^2} \int_0^{y} du \int_a
\frac{1}{\frac{4 u}{\tilde{q}^2} + 2a}
\end{eqnarray}
In the integral on the second line, we can take the limit $\tilde{q}
\to 0$ by making the change of variable $\lambda = u/\tilde{q}^2$, and
the second can be done exactly. We thus have
\begin{eqnarray}
&&\frac{1}{{\tilde q}^2} \int_0^{\tilde{q}^2 \tilde{t}} du
e^{-\int_a \ln{(\frac{4 u}{\tilde{q}^2} + 2a)}} =
\frac{1}{4}\ln{\frac{y}{\tilde{q}^2}} -  \int_a \ln{\frac{a}{2}} \nonumber \\
&& + \int_0^{\infty} d\lambda e^{-\int_a \ln{(4\lambda + 2a)}} -
\int_a \frac{1}{4 \lambda + 2a} + O(\tilde{q}^2) \nonumber
\end{eqnarray}
Finally, using
\begin{eqnarray}
\int_0^y du \frac{e^u - 1}{4u} =\frac{1}{4}( - \gamma_E + Ei(y) - \ln{y})
\end{eqnarray}
one has, up to terms of order $\tilde{q}^2$
\begin{eqnarray}
&&{\cal G}_{\tilde t}^{\tilde q} = 1 + B^*( (y-1) Ei(y) + 1 - e^y +
(1-y) (\ln{\tilde{q}^2} + \rho)) \nonumber \\
&& \rho = \gamma_E + \int_a \ln{\frac{a}{2}} - 4 \int_0^{\infty}
d\lambda e^{-\int_a \ln{(4\lambda + 2a)}} + 4
\int_a \frac{1}{4 \lambda + 2a} \nonumber \\
\label{rho2}
\end{eqnarray}
which yields the scaling function given in the text.

\section{Nonequilibrium dynamics of the CO model.}\label{app_dyncononeq}

\subsection{Some useful expressions}

To begin with, we give the explicit expression of $\Delta_{ltt'}$ and
$\Sigma_{ltt'}$ and their limiting expression when $l \to \infty$ in
the case of nonequilibrium dynamics. The general expression of
$\Delta_{ltt'}(u)$, i.e. the first line of
(\ref{solrg_delta}) is still valid for non equilibrium dynamics. To
evaluate it, we only need the expression of $C_{ltt'}^{x=0}$
that we compute from (\ref{Dirichlet}) using the same cutoff function
$c(z)$ as previously (\ref{decomp_cutoff}):
\begin{eqnarray}\label{autocorrel_noneq}
&&C_{ltt'}^{x=0} = \frac{T}{4\pi}\int_a\Big(
\ln{(t+t'+\frac{a}{2\Lambda_0^2})} -
\ln{(|t-t'|+\frac{a}{2\Lambda_0^2})} \nonumber \\
&&-  \frac{T}{4\pi}
\ln{(t+t'+\frac{a}{2\Lambda_l^2})} +
\ln{(|t-t'|+\frac{a}{2\Lambda_l^2})}\Big)
\end{eqnarray}
Notice that the response function $R^{x=0}_{tt'}$ has its equilibrium
expression. From (\ref{solrg_delta}) and (\ref{autocorrel_noneq}) one
obtains
\begin{eqnarray}
\Delta_{ltt'}(u) =
e^{-\frac{1}{2}C_{ltt}^{x=0}-\frac{1}{2}C_{lt't'}^{x=0} +
C_{ltt'}^{x=0}}\Delta_{l=0 tt'}(u)
\end{eqnarray}
Using (\ref{autocorrel_noneq}) one has, using 
$T = T_c = 4 \pi$ to this order:
\begin{eqnarray}
&&\lim_{l \to \infty} -\frac{1}{2}C_{ltt}^{x=0}-\frac{1}{2}C_{lt't'}^{x=0} +
C_{ltt'}^{x=0} \nonumber \\
&& = \int_a -\ln{(\Lambda_0^2 |t-t'| + \frac{a}{2})} +
\ln{(\Lambda_0^2 (t+t') +
\frac{a}{2})} \nonumber \\
&&- \frac{1}{2} \ln{(\Lambda_0^2 t + \frac{a}{4})} -\frac{1}{2}
\ln{(\Lambda_0^2 t' + \frac{a}{4})} + \ln{\frac{a}{4}}
\end{eqnarray}
and using the definition (\ref{def_sigma_d}), one obtains finally
\begin{eqnarray}\label{d_noneq}
&&D_{tt'} = \lim_{l \to \infty} D_{ltt'} \\
&& = \frac{\Lambda_0^2 T_c B^*}{2} e^{\int_a
-\ln{(\Lambda_0^2 |t-t'| + \frac{a}{2})} + \ln{(\Lambda_0^2 (t+t')
+\frac{a}{2} )}}
\nonumber \\
&& \times e^{\int_a -
\frac{1}{2} \ln{(\Lambda_0^2 t + \frac{a}{4})} -\frac{1}{2}
\ln{(\Lambda_0^2 t' +
\frac{a}{4}})} \nonumber
\end{eqnarray}
where we have used the expression of $B^*$ given in (\ref{reponse_fp}).
The expression for $\Sigma_{ltt'}$
can be obtained
in a very similar way~:
\begin{eqnarray}\label{sigma_noneq}
&&\Sigma_{tt'} = \lim_{l \to \infty} \Sigma_{ltt'}  \\
&& = \frac{-\Lambda_0^4 B^*}{2} \int_a \Big \{ \frac{\theta(t-t')}{\Lambda_0^2
(t-t') + \frac{a}{2}}  e^{\int_a
-\ln{(\Lambda_0^2 |t-t'| +\frac{a}{2} )}}
\nonumber \\
&& \times e^{\int_a \ln{(\Lambda_0^2 (t+t') + \frac{a}{2})} -
\frac{1}{2} \ln{(\Lambda_0^2 t + \frac{a}{4})} -\frac{1}{2}
\ln{(\Lambda_0^2 t' +
\frac{a}{4}})} \nonumber \\
&& -\delta(t-t') \int_0^t dt_1 \frac{1}{\Lambda_0^2
(t-t_1) + \frac{a}{2}}  e^{\int_a
-\ln{(\Lambda_0^2 |t-t_1| + \frac{a}{2})}}
\nonumber \\
&& \times e^{\int_a \ln{(\Lambda_0^2 (t+t_1) + a/2)} -
\frac{1}{2} \ln{(\Lambda_0^2 t + \frac{a}{4})} -\frac{1}{2}
\ln{(\Lambda_0^2 t_1 +
\frac{a}{4}})} \Big \} \nonumber
\end{eqnarray}
These expressions (\ref{d_noneq},\ref{sigma_noneq}) will be very
useful to determine explicit expressions for ${\cal R}^{q}_{tt'} =
\lim_{l\to\infty}{\cal R}^{q}_{ltt'}$ and  ${\cal C}^{q}_{tt'} =
\lim_{l\to\infty}{\cal C}^{q}_{ltt'}$ by solving perturbatively
(\ref{eqR},\ref{eqC}).

\subsection{Nonequilibrium response function : detailed calculations.}

The starting point of our analysis is the equation (\ref{eqR}) that we
solve perturbatively by replacing, in the rhs of this equation, ${\cal
R}^{q}_{ltt'}$ by its bare value. One obtains, in the limit $l \to
\infty$, using (\ref{sigma_noneq}), and in terms of the rescaled
variables $\tilde{t} = \Lambda_0^2 t$, $\tilde{q} = q/\Lambda_0$:
\begin{eqnarray}\label{eqR_noneq_app}
&&\partial_{\tilde{t}} {\cal R}^{\tilde{q}}_{\tilde{t}\tilde{t'}} +
\tilde{q}^2 {\cal R}^{\tilde{q}}_{\tilde{t}\tilde{t'}} = \\
&&\frac{B^*}{2} \int_0^{\tilde{t}} dt_1  \frac{1}{
(\tilde{t}-t_1) + \frac{a}{2}}  e^{\int_a
-\ln{(\tilde{t}-t_1 + \frac{a}{2})}}
\nonumber \\
&& \times e^{\int_a \ln{((\tilde{t}+t_1) +\frac{a}{2} )} -
\frac{1}{2} \ln{(\tilde{t} + \frac{a}{4})} -\frac{1}{2} \ln{(t_1 +
\frac{a}{4}})} \nonumber \\
&& \times \left(\theta(t_1-\tilde{t'})e^{-\tilde{q}^2(t_1 - \tilde{t'})} -
e^{-\tilde{q}^2(\tilde{t} - \tilde{t'})}\right) \nonumber
\end{eqnarray}
Let us first focus on the last term in the rhs of
(\ref{eqR_noneq_app}) where we make the change of variable $t_1 = u
\tilde{t}$ and analysing the limit $\tilde{t} \gg 1$:
\begin{eqnarray}
&&-\frac{B^*}{2}e^{-\tilde{q}^2(\tilde{t}-\tilde{t'})} \int_0^1 du \int_a
 \frac{1}{1-u+ \frac{a}{2\tilde{t}} } e^{\int_a
 -\ln{(1-u+ \frac{a}{2\tilde{t}} )}}
 \nonumber \\
&&\times e^{\int_a
 \ln{(1+u+ \frac{a}{2\tilde{t}} )} - \frac{1}{2}\ln{(u+
 \frac{a}{4\tilde{t}} )} -
 \frac{1}{2}\ln{(\tilde{t}+ \frac{a}{4})} - \frac{1}{2}\ln{\tilde{(t)}}  }
 \nonumber \\
&&\sim -\frac{B^*}{2}e^{-\tilde{q}^2(\tilde{t}-\tilde{t'})} \int_0^1
 \frac{du}{\tilde{t}} \int_a
 \frac{1}{1-u+ \frac{a}{2\tilde{t}} } e^{\int_a -\ln{(1-u+
 \frac{a}{2\tilde{t}})}}
 \nonumber \\
&&\times e^{\int_a
 \ln{(1+u+ \frac{a}{2\tilde{t}} )} - \frac{1}{2}\ln{(u+
 \frac{a}{4\tilde{t}} )}} = Q
\end{eqnarray}
In the integrand one can not directly take the limit
$\tilde{t}$~$\to$~$\infty$ because it generates a divergence of the
integral when $u \to
1$. Therefore we substract the divergent term in the following way
\begin{eqnarray}
&& Q = -\frac{B^*}{2}e^{-\tilde{q}^2(\tilde{t}-\tilde{t'})} \int_0^1
 \frac{du}{\tilde{t}} \int_a
 \frac{2}{1-u+ \frac{a}{2\tilde{t}}} e^{\int_a
 -\ln{(1-u+\frac{a}{2\tilde{t}}) }}
 \nonumber \\
&& -\frac{B^*}{2}e^{-\tilde{q}^2(\tilde{t}-\tilde{t'})} \int_0^1
 \frac{du}{\tilde{t}} \int_a
 \frac{1}{1-u+ \frac{a}{2\tilde{t}}} e^{\int_a -\ln{(1-u+
 \frac{a}{2\tilde{t}} )}}
 \nonumber \\
&&\times \left( e^{\int_a
 \ln{(1+u+ \frac{a}{2\tilde{t}})} - \frac{1}{2}\ln{(u+
 \frac{a}{4\tilde{t}} )}} -
 2\right)
\end{eqnarray}
Interestingly except in the first line, one can take directly the
limit $\tilde{t} \to \infty$ in the integrand of the two last lines
using
\begin{eqnarray}\label{trick}
&&\lim_{\tilde{t} \to \infty} \frac{1}{1-u+ \frac{a}{2\tilde{t}}} e^{\int_a
-\ln{(1-u+ \frac{a}{2\tilde{t}} )}} \nonumber \\
&&\times \left( e^{\int_a
 \ln{(1+u+ \frac{a}{2\tilde{t}} )} - \frac{1}{2}\ln{(u+
\frac{a}{4\tilde{t}} )}} -
 2\right) \nonumber \\
&& = \frac{1}{(1-u)^2} (\frac{1+u}{\sqrt{u}} -2) =
\frac{1}{\sqrt{u}(1+\sqrt{u})^2}
\end{eqnarray}
and the divergence for $u \to 1$ is cured. Then all the remaining
integrals can be performed exactly, giving finally
\begin{equation} \label{firstterm}
Q = B^* e^{-\tilde{q}^2(\tilde{t} -\tilde{t'})}\left(- e^{- \int_a \ln{\frac{a}{2}}} +
 \frac{1}{2\tilde{t}} + O(\tilde{t}^{-2}) \right) 
\end{equation}
We now perform exactly the same manipulations on the first term in the
rhs of
(\ref{eqR_noneq_app}). Performing the change of variable $t_1 = u
\tilde{t}$ and considering the limit $\tilde{t} \gg 1$ (keeping
$\tilde{q}^2 \tilde{t}$,$\tilde{q}^2 \tilde{t'}$ and
$\tilde{t}/\tilde{t'}$ fixed) one obtains
\begin{eqnarray}\label{secondterm}
&&\frac{B^*}{2} \int_{\tilde{t'}/\tilde{t}}^1 \frac{du}{\tilde{t}} \int_a
\frac{e^{-\tilde{q}^2(u\tilde{t} -\tilde{t'})}}{1-u+ \frac{a}{2\tilde{t}}}
\nonumber \\
&&\times e^{\int_a -\ln{(1-u+ \frac{a}{2\tilde{t}})}+\ln{(1+u+
\frac{a}{2\tilde{t}} )}
- \frac{1}{2}\ln{(u+\frac{a}{4\tilde{t}} )}} \nonumber \\
&& = {B^*} \int_{\tilde{t'}/\tilde{t}}^1 \frac{du}{\tilde{t}} \int_a
\frac{e^{-\tilde{q}^2(u\tilde{t} -\tilde{t'})}}{1-u+ \frac{a}{2\tilde{t}}}
e^{\int_a -\ln{(1-u+ \frac{a}{2\tilde{t}} )}} \nonumber \\
&&+\frac{B^*}{2}\int_{\tilde{t'}/\tilde{t}}^1 \frac{du}{\tilde{t}}
\frac{e^{-\tilde{q}^2(u\tilde{t}-\tilde{t'})}}{\sqrt{u}(1+\sqrt{u})^2}
+ O(\tilde{t}^{-2})
\end{eqnarray}
where we have used the same trick (\ref{trick}) as previously. Using
(\ref{secondterm}) together with (\ref{firstterm}) on can write
(\ref{eqR_noneq_app}) in a rather simple way
\begin{eqnarray}
&&\partial_{\tilde{t}} {\cal R}^{\tilde{q}}_{\tilde{t}\tilde{t'}} +
\tilde{q}^2 {\cal R}^{\tilde{q}}_{\tilde{t}\tilde{t'}} = \nonumber \\
&&4B^* \int_{\tilde{t'}}^{\tilde{t}} dt_1 \int_a \frac{1}{t-t_1+ \frac{a}{2}}
e^{-\int_a
\ln{(4(\tilde{t}-t_1)+2a)}} e^{-\tilde{q}^2(t_1-\tilde{t'})}
\nonumber \\
&&-  B^*e^{-\tilde{q}^2(\tilde{t} -\tilde{t'})} e^{- \int_a \ln{(\frac{a}{2})}}
\nonumber \\
&& + \frac{B^*}{2} \left(\int_{\tilde{t'}}^{\tilde{t}}
\frac{dt_1}{\sqrt{\tilde{t}t_1}}
\frac{e^{-\tilde{q}^2(t_1-\tilde{t'})}}{(\sqrt{\tilde{t}} +
\sqrt{t_1})^2} +
\frac{e^{-\tilde{q}^2(\tilde{t}
-\tilde{t'})}}{\tilde{t}} \right)
\end{eqnarray}
The two first lines correspond to equilibrium fluctuations
(\ref{Resp_eq_app}) and their
contribution to the response function has already been computed
(\ref{perturbscaleq}). The
last term does not depend any more on the cutoff function and
characterizes the contributions coming from nonequilibrium
fluctuations. The linearity of this equation suggests then to look for
a solution under the form ${\cal R}^{\tilde{q}}_{\tilde{t}\tilde{t'}}
={\cal R}^{\tilde{q} \text{eq}}_{\tilde{t}\tilde{t'}} + {\cal
R}^{\tilde{q}\text{noneq}}_{\tilde{t}\tilde{t'}} $, where ${\cal
R}^{\tilde{q} \text{eq}}_{\tilde{t}\tilde{t'}} = \tilde{q}^{z-2} {
F_R^{\text{eq}}}(\tilde q^{z} (\tilde t - \tilde t'))$ (\ref{scalingf}) and  ${\cal
R}^{\tilde{q}\text{noneq}}_{\tilde{t}\tilde{t'}} = e^{-\tilde{q^2}(\tilde{t}
- \tilde{t'})} H^q_{\tilde{t}\tilde{t'}}$ with $
H^{\tilde{q}}_{\tilde{t}\tilde{t'}}$
determined by
(\ref{eqR_noneq_app})
\begin{eqnarray}\label{eqHapp}
&&\partial_{\tilde{t}} H^{\tilde{q}}_{\tilde{t}\tilde{t'}} =
 \frac{B^*}{2} \left(\int_{\tilde{t'}}^{\tilde{t}}
\frac{dt_1}{\sqrt{\tilde{t}t_1}}
\frac{e^{-\tilde{q}^2(t_1-\tilde{t})}}{(\sqrt{\tilde{t}} +
\sqrt{t_1})^2} +
\frac{1}{\tilde{t}} \right) \nonumber \\
&&H^{\tilde{q}}_{\tilde{t}\tilde{t}} =
 H^{\tilde{q}}_{\tilde{t}\tilde{t-}} = 0
\end{eqnarray}
This allows to write a close expression for the perturbative
expansion of
${\cal R}^{\tilde{q} \text{noneq}}_{\tilde{t}\tilde{t'}}$ in terms of the
scaling variables
$v' = \tilde{q^2}(\tilde{t}-\tilde{t'}) $,
$u=\frac{\tilde{t}}{\tilde{t'}}$
\begin{eqnarray}\label{startR}
&&{\cal R}^{\tilde{q} \text{noneq}}_{\tilde{t}\tilde{t'}} = \frac{B^*
e^{-v'}}{2}\Big(
\int_{\frac{v'}{u-1}}^{\frac{u v'}{u-1}} \frac{dt_2}{\sqrt{t_2}}
\int_{\frac{v'}{u-1}}^{t_2}  \frac{dt_1}{\sqrt{t_1}}
\frac{e^{t_2-t_1}}{(\sqrt{t_1} +
\sqrt{t_2})^2} \nonumber \\
&& + \ln u \Big)
\end{eqnarray}
Unfortunately it is quite difficult to extract directly the asymptotic
behaviors from this double integral. However one can perform
straightforward (although tedious) manipulations to
obtain a quasi-explicit expression for
${\cal R}^{\tilde{q} \text{noneq}}_{\tilde{t}\tilde{t'}}$. Performing the
natural change of
variables $\alpha = \sqrt{t_2} - \sqrt{t_1}$, $\beta = \sqrt{t_2} +
\sqrt{t_1}$ one is left with integrals over one variable
\begin{eqnarray}
&&\frac{B^*}{2} e^{-v}
\int_{\frac{v}{u-1}}^{\frac{uv}{u-1}} \frac{dt_2}{\sqrt{t_2}}
\int_{\frac{v}{u-1}}^{t_2}  \frac{dt_1}{\sqrt{t_1}}
\frac{e^{t_2-t_1}}{(\sqrt{t_1} +
\sqrt{t_2})^2} \nonumber \\
&& = B^* e^{-v} \left(\frac{u-1}{8vu} - \frac{u-1}{8v} + {\cal
Q}(\frac{v}{u-1},u) + {\cal
Q}(\frac{-vu}{u-1},\frac{1}{u}) \right) \nonumber \\
&& {\cal Q}(x,y) = \frac{1}{x}\int_1^{\sqrt{y}} d\beta
\frac{e^{x(\beta^2-1)}}{(\beta+1)^3}
\end{eqnarray}
Peforming further manipulations we find that one can write:
\begin{widetext}
\begin{eqnarray}\label{exprF1R}
&&{\cal R}^{\tilde{q} \text{noneq}}_{\tilde{t}\tilde{t'}} = \theta \ln {u} e^{-v} +
\tau F_R^{1 \text{noneq}}(v',u) + O(\tau^2) \\
&& \theta = B^*  + O(\tau^2)
\end{eqnarray}
where the logarithmic behavior determining $\theta$ has been extracted such
that the function $F_R^{1 \text{noneq}}(v,u)$ has a good limit for
$u \to \infty$, as will be shown below. A useful expression
for this function is found as:
\begin{eqnarray}\label{exprF1RR}
&&{F}_R^{1 \text{noneq}}(v,u) = e^{\gamma_E} \Big\{ 1-e^{-v} -
\sqrt{\pi}\sqrt{\frac{vu}{u-1}}e^{\frac{v}{u-1}}
\left(\Erf{\sqrt{\frac{vu}{u-1}}}
-
\Erf{\sqrt{\frac{v}{u-1}}}     \right) \nonumber \\
&&-\sqrt{\pi}\sqrt{\frac{v}{u-1}}e^{-\frac{vu}{u-1}}
\left(\Erfi{\sqrt{\frac{vu}{u-1}}} -
\Erfi{\sqrt{\frac{v}{u-1}}}\right) + e^{-v}(1-v)(Ei(v) - \ln{v} - \gamma_E)
\nonumber \\
&&+ 2
e^{-v}\frac{vu}{u-1}\left( \frac{1}{u}\left(\frac{v}{u-1}+\frac{1}{2}\right)
 \quad_2F_2(\{1,1\},\{\frac{3}{2},2\},
-\frac{v}{u-1}) - (-\frac{vu}{u-1} +
\frac{1}{2}) _2F_2(\{1,1\},\{\frac{3}{2},2\},\frac{vu}{u-1}) \right)
\nonumber \\
&&+e^{-v}\pi\left(\frac{1}{2}-\frac{vu}{u-1}\right) \Erf\sqrt{\frac{v}{u-1}}
\Erfi\sqrt{\frac{vu}{u-1}}  - 2\sqrt{\pi} (1 - v)\frac{e^{-v}}{\sqrt{u}}
\int_0^{\sqrt{\frac{vu}{u-1}}} ds e^{-\frac{s^2}{u}} \Erfi{(s)}
\nonumber \\
&& + 2(1-v)e^{-v}\ln{\frac{1+\frac{1}{\sqrt{u}}}{2}} +
\frac{ve^{-v}}{u-1}\ln{u}
  \Big \}
\end{eqnarray}
\end{widetext}
where $\Erf{z}$ is the error function, $\Erfi{z}$ is
the imaginary error function:
\begin{eqnarray}
&& \Erf{z} = \frac{2}{\sqrt{\pi}} \int_0^z ds e^{- s^2} \\
&& \Erfi{z} = \frac{2}{\sqrt{\pi}} \int_0^z ds e^{ s^2}
\end{eqnarray}
with $\Erfi{z}=- i \Erf{i z}$. One has the
following asymptotic behaviors
\begin{eqnarray}
&&\Erf{z} \sim  2z/\sqrt{\pi} \quad z \ll 1 \label{erf_asymp_small} \\
&&\Erf{z} \sim 1 - e^{-z^2}/(\sqrt{\pi} z) \quad z \gg 1 \label{erf_asymp_large}
\end{eqnarray}
and
\begin{eqnarray}
&&\Erfi{z} \sim  2z/\sqrt{\pi} \quad z \ll 1 \label{erfi_asymp_small} \\
&&\Erfi{z} \sim e^{z^2}/(\sqrt{\pi} z) \quad z \gg 1 \label{erfi_asymp_large}
\end{eqnarray}
and $_2F_2(\{1,1\},\{\frac{3}{2},2\},z)$ is a generalized
hypergeometric series which has the following asymptotic behaviors
\begin{eqnarray}
&&_2F_2(\{1,1\},\{\frac{3}{2},2\},z) \sim 1 + O(z)
\label{hypergeo_asymp_small}\\
&&_2F_2(\{1,1\},\{\frac{3}{2},2\},z) \sim_{z \to + \infty} \frac{\sqrt{\pi}}{2}
\frac{e^z}{z^{3/2}} (1+O(z^{-1})) \nonumber \\
&&_2F_2(\{1,1\},\{\frac{3}{2},2\},z) \sim_{z \to - \infty} - \frac{\ln(-z)}{2 z}  \nonumber \\
&&  \label{hypergeo_asymp_large}
\end{eqnarray}
Under this form, asymptotic behaviors are more easily obtained.
Note that we have also performed numerical checks that
(\ref{exprF1RR}) and the starting integral (\ref{startR})
do indeed coincide.

Note some simple formulae for the same point response:
\begin{eqnarray}
&& {\cal R}^{x=0}_{\tilde{t}\tilde{t'}}
= \frac{1}{2 \pi z (t-t')} h(t/t') \\
&& h(u) = u^\theta \int_0^\infty dv F_R(v,u) 
\end{eqnarray}

\subsubsection{Expansion at large $u$, $v$ fixed.}

The asymptotic behavior of $F^1_R(v,u)$ is easily obtained in this
limit. From (\ref{expr_FR_gen}), one has $\lim_{u\to\infty} F^1_R(v,u) =
F^{1\text{eq}}_R(v) + \lim_{u\to\infty} F^{1 \text{noneq}}_R(v,u)$, where
$F^{1\text{eq}}_R(v)$ is given in (\ref{perturbscaleqexpl}). On the expressions
(\ref{exprF1R}, \ref{exprF1RR}) together with the asymptotic behaviors
(\ref{erf_asymp_small}, \ref{erfi_asymp_small}, \ref{hypergeo_asymp_small})
we see  that all terms vanish in this limit except the following ones
\begin{eqnarray}
&&\lim_{u\to \infty, v \text{fixed}} F^{1\text{noneq}}_R(v,u) =  -
F^{1 \text{eq}}_R(v) \nonumber \\
&& + e^{\gamma_E}\Big \{ -
\sqrt{\pi v}\Erf{\sqrt{v}}
- e^{-v}\Big((1-v)\ln{(4ve^{\gamma_E})} \nonumber \\
&& - 2 v(v-\frac{1}{2})_2F_2(\{1,1\},\{\frac{3}{2},2\},v)\Big)\Big \} \nonumber
\end{eqnarray}
which leads to (\ref{FRlarge_u}) in the text.

\subsubsection{Expansion at large $v$, $u$ fixed.}

Although one can extract more rigorously
the large $v$ behavior at $u$ fixed from the complete expression
(\ref{exprF1R}), it is easier to compute it from the starting integral
in (\ref{startR}). Indeed, in the large $v$ limit, the integral will
be dominated by the region $t_2 - t_1 \sim v$, i.e. one can replace in
the integrand (except of course in the term $e^{t_2-t_1}$ ) $t_2$ by
$vu/(u-1)$ and $t_1$ by $v/(u-1)$:
\begin{eqnarray}
&&\frac{B^*
e^{-v}}{2}\Big(
\int_{\frac{v}{u-1}}^{\frac{uv}{u-1}} \frac{dt_2}{\sqrt{t_2}}
\int_{\frac{v}{u-1}}^{t_2}  \frac{dt_1}{\sqrt{t_1}}
\frac{e^{t_2-t_1}}{(\sqrt{t_1} +
\sqrt{t_2})^2} \\
&& \sim \frac{B^*}{2} \frac{1}{v^2} \frac{(\sqrt{u} -1)^2}{\sqrt{u}}
e^{-v} \int_{\frac{v}{u-1}}^{\frac{vu}{u-1}} dt_2 e^{t_2}
\int_{\frac{v}{u-1}}^{\frac{vu}{u-1}} dt_1 e^{-t_1} \nonumber
\end{eqnarray}
which leads finally to
\begin{eqnarray}\label{FRnoneq_large_v}
F_R^{1\text{noneq}}(v,u) \sim \frac{B^*}{2 \tau} \frac{1}{v^2} \frac{(\sqrt{u}
-1)^2}{\sqrt{u}} + O(v^{-3})
\end{eqnarray}
We have checked that we obtain the same result by performing this
expansion on (\ref{exprF1R}).
Finally, using the large $v$ behavior of $F_R^{1\text{eq}}$
(\ref{FReq_asymp_large}) and the value of
$B^*$ (\ref{reponse_fp}), one obtains
\begin{eqnarray}
F_R^1(v,u) \sim e^{\gamma_E} \frac{1}{2 v^2} \frac{u+1}{\sqrt{u}} + O(v^{-3})
\end{eqnarray}
which gives (\ref{FRlarge_v}) in the text.

\subsubsection{The limit of vanishing momentum.}

The limit $\tilde{q} \to 0$ is easily obtained by looking for the
leading term in $F_R(v,u)$ when $v = \tilde{q}^2(\tilde{t}-\tilde{t'})
\to 0$. Using (\ref{exprF1RR}) together with (\ref{expr_FR_gen}) and
(\ref{perturbscaleqexpl}) one has
\begin{eqnarray}
F_R^1(v) \sim  - e^{\gamma_E} (\ln{v} +
\gamma_E -2 \ln{\frac{1+\frac{1}{\sqrt{u}}}{2}})
\end{eqnarray}
This logarithmic behavior together with (\ref{expz}) cancels the
$\log{\tilde{q}}$ divergence in (\ref{pertresp}) and
allows to take the limit of vanishing
momentum.
We also give here the expression of $F_R^{1{\text{noneq}}}(0,u)$, obtained
from (\ref{exprF1R})
\begin{eqnarray}\label{FR1noneq_q=0}
F_R^{1{\text{noneq}}}(0,u) = 2 e^{\gamma_E}
\ln{\frac{1+\frac{1}{\sqrt{u}}}{2}}
\end{eqnarray}
this will be useful for further applications.

\subsection{Nonequilibrium correlation function : detailed calculations.}

The starting point of our analysis is the following expression given
in the text (\ref{start_correl}), for $\tilde{t} > \tilde{t'}$:
\begin{eqnarray}
&&{\cal C}^{\tilde{q}}_{\tilde{t}\tilde{t'}} = \lim_{l \to \infty}
{\cal C}^{\tilde{q}}_{l\tilde{t}\tilde{t'}} \\
&& = 2T\int_0^{\tilde{t'}} dt_1 {{\cal R}}^{\tilde{q}}_{\tilde{t}t_1}{{\cal
R}}^{\tilde{q}}_{\tilde{t'}t_1}
+ \int_0^{\tilde{t}} dt_1 \int_0^{\tilde{t'}} dt_2{{\cal
R}}^{\tilde{q}}_{\tilde{t}t_1} D_{t_1t_2}
{{\cal R}}^{\tilde{q}}_{\tilde{t'} t_2} \nonumber \\
\end{eqnarray}
where $D_{t_1t_2} = \lim_{l \to \infty} D_{lt_1t_2}$ is given in
(\ref{d_noneq}),
that we expand perturbatively using the expression we obtained for
${\cal R}^q_{tt'}$ (\ref{pertresp}). As we did previously for the
response function we could keep the complete cut-off dependence in
(\ref{d_noneq}). However given the complexity of these manipulations
and the experience we acquired before, we know that the only cutoff
dependence is contained in an overall nonuniversal scale $\tilde{q}
\to \lambda \tilde{q}$. For these reasons we will perform the
computation using a simplified cutoff $\hat c(a) = \delta(a-a_0)$ and we
will choose $a_0 = 2$ for simplicity. $D_{t_1t_2}$ can then be written
as (\ref{d_noneq})
\begin{eqnarray}
D_{t_1 t_2} = \frac{1}{2} e^{\gamma_E} T_c \tau \frac{t_1 +
t_2}{(|t_1-t_2|+1)\sqrt{t_1 t_2}} +
O(\tau^2)
\end{eqnarray}
where we have dropped the $a_0$ dependence where it turns
out to be unimportant. 

Performing the integrals that do not involve $F_R^1(v,u)$ one has
\begin{eqnarray}\label{exprgenC}
&&{\cal C}^q_{tt'} = \frac{T}{q^2} F_C^0(v,u) + \frac{T}{q^2}
\theta \ln{u}  F_C^0(v,u)  \\
&&+ \frac{T}{q^2} (z-2)
\ln{q} \left(v
\frac{\partial F_C^0(v,u)}{\partial v} + F_C^0(v,u)\right)  \nonumber \\
&&+ \frac{2 T \theta}{q^2} e^{-v \frac{1+u}{u-1}}
\left(Ei(\frac{2v}{u-1}) - \ln{\left(\frac{2v}{u-1}\right)} - \gamma_E
\right)
\nonumber \\
&& + \frac{2 T\tau}{q^2} \frac{v}{u-1} \int_0^1 ds
F_R^0\left(\frac{u-s}{u-1}v\right)
F^1_R\left(\frac{1-s}{u-1}v,\frac{1}{s}\right)  \nonumber \\
&& + F^0_R\left(\frac{1-s}{u-1}v\right)
F^1_R\left(\frac{u-s}{u-1}v,\frac{u}{s}\right) \nonumber \\
&&+\frac{e^{\gamma_E} T_c \tau}{q^2} e^{-v\frac{1+u}{u-1}}
\int_0^{\frac{uv}{u-1}} dt_1 \int_0^{\frac{v}{u-1}} dt_2
e^{(t_1+t_2)} \nonumber \\
&&\times \left( \frac{1}{|t_1-t_2|+q^2} +
\frac{1}{2}\frac {\sqrt{t_1}-\sqrt{t_2}}{\sqrt{t_1 t_2} (\sqrt{t_1} +
\sqrt{t_2})}   \right) \nonumber
\end{eqnarray}
where we have used the trick (\ref{trick}), and dropped the prime
in $v'=\tilde q^2 (\tilde t - \tilde t')$ for simplicity. 
A natural way to perform this computation is to use for $F^1_R(v,u)$
the decomposition in an equilibrium and a nonequilibrium contributions
(\ref{expr_FR_gen}). Parts of (\ref{exprgenC}) can then be computed
analytically:
\begin{eqnarray}\label{correl_easy}
&&\frac{2 T\tau}{q^2} \frac{v}{u-1} \int_0^1 ds
F_R^0\left(\frac{u-s}{u-1}v\right)
{F}_R^{1\text{eq}}\left(\frac{1-s}{u-1}v \right) \nonumber  \\
&&+ F^0_R\left(\frac{1-s}{u-1}v\right)
{F}_R^{1\text{eq}}\left(\frac{u-s}{u-1}v\right) \nonumber \\
&&+ \frac{ e^{\gamma_E} T_c\tau}{q^2} e^{-v\frac{1+u}{u-1}}
\int_0^{\frac{uv}{u-1}} dt_1
\int_0^{\frac{v}{u-1}} dt_2
\frac{e^{(t_1+t_2)}}{|t_1-t_2|+q^2} \nonumber \\
&& = - \frac{ e^{\gamma_E} \tau}{q^2} \ln{(q^2 e^{\gamma_E})} T F_C^0(v,u) \nonumber \\
&& + \frac{1}{2} e^{\gamma_E}\tau T \Big \{ \frac{e^{-\frac{v(u+1)}{u-1}}}{q^2} \Big(-4
- 2\frac{uv}{u-1}Ei\left(\frac{uv}{u-1}\right) \nonumber \\
&&-
2\frac{v}{u-1}Ei\left(\frac{v}{u-1}\right) \Big) +
\frac{2}{q^2}(e^{-\frac{vu}{u-1}} + e^{-\frac{v}{u-1}}) \nonumber \\
&& + \frac{2}{q^2} \left(e^{-v} -1 + ve^{-v}Ei(v)   \right)
\Big \}
\end{eqnarray}
The expressions (\ref{exprgenC}) together with (\ref{correl_easy})
allows to identify the following perturbative scaling behavior
(\ref{janssenscalingcorr})
\begin{eqnarray}\label{exprF1C}
&&{\cal C}^q_{tt'} = \frac{T}{q^2} (F_C^0(v,u) + (z-2) v \ln{q}
\frac{\partial F_C^0(v,u)}{\partial v}  \nonumber  \\
&&+ \theta \ln{u}F_C^0(v,u) + \tau F_C^1(v,u)) + O(\tau^2) \\
&&F_C^1(v,u) = 2 e^{\gamma_E} e^{-v \frac{1+u}{u-1}}
\left(Ei(\frac{2v}{u-1}) - \ln{\left(\frac{2v}{u-1}\right)}
-\gamma_E \right) \nonumber \\
&& + \frac{1}{2} e^{\gamma_E}  \Big \{ e^{-\frac{v(u+1)}{u-1}} \Bigg ( -4
- 2\frac{uv}{u-1}Ei\left(\frac{uv}{u-1}\right) \nonumber \\
&&-2\frac{v}{u-1}Ei\left(\frac{v}{u-1}\right) \Bigg)
+2(e^{-\frac{vu}{u-1}} + e^{-\frac{v}{u-1}}) \nonumber \\
&& + 2 \left(e^{-v} -1 + ve^{-v}Ei(v)   \right) \Big \} \nonumber \\
&& + 2 \frac{v}{u-1} \int_0^1 ds \Big \{
F_R^0\left(\frac{u-s}{u-1}v\right)
{F}_R^{1\text{noneq}}\left(\frac{1-s}{u-1}v,\frac{1}{s} \right)  \nonumber \\
&& + F^0_R\left(\frac{1-s}{u-1}v\right)
{F}_R^{1\text{noneq}}\left(\frac{u-s}{u-1}v,\frac{u}{s}\right) \Big \}
\nonumber \\
&& + \frac{1}{2} e^{\gamma_E} e^{-v\frac{1+u}{u-1}}\int_0^{\frac{vu}{u-1}}
dt_1\int_0^{\frac{v}{u-1}}dt_2 \frac
{(\sqrt{t_1}-\sqrt{t_2})e^{t_1+t_2}}{\sqrt{t_1 t_2} (\sqrt{t_1} +
\sqrt{t_2})} \nonumber
\end{eqnarray}
with the exponents $z$ and $\theta$ given in (\ref{expz}) and
(\ref{expr_FR_gen}). The scaling functions are universal up to
a cutoff dependent additive constant. It was explicitly computed for
the equilibrium response 
in (\ref{rho2}). Here, we do not determine it and thus we
can drop some multiplicative factors of momentum in the $\log \tilde{q}$ term.

\subsubsection{Expansion at large $u$, $v$ fixed.}

First, one has
\begin{eqnarray}\label{Fc0_large_u}
F_C^0(v,u) = e^{-v} - e^{-v\frac{u+1}{u-1}} \sim \frac{2e^{-v}v}{u} +
O(u^{-2})
\end{eqnarray}

We now focus on the asymptotic behavior of $F^1_C(v,u)$
for large $u$, keeping $v$ fixed. Using the small argument behavior of
$Ei(z) \sim \ln{z} + \gamma_E + O(z)$ one has for the first line of
(\ref{exprF1C}) in this limit
\begin{eqnarray}
&& 2 e^{\gamma_E} e^{-v \frac{1+u}{u-1}}
\left(Ei(\frac{2v}{u-1}) - \ln{\left(\frac{2v}{u-1} \right)}
-\gamma_E \right) \nonumber \\
&& \sim O(u^{-1})
\end{eqnarray}
Again using the small argument behavior of $Ei(z)$, one has
\begin{eqnarray}\label{log_u_term}
&& + \frac{1}{2} e^{\gamma_E}  \Big \{ e^{-\frac{v(u+1)}{u-1}} \Bigg ( -4
- 2\frac{uv}{u-1}Ei\left(\frac{uv}{u-1}\right) \nonumber \\
&&-2\frac{v}{u-1}Ei\left(\frac{v}{u-1}\right) \Bigg)
+2(e^{-\frac{vu}{u-1}} + e^{-\frac{v}{u-1}}) \nonumber \\
&& + 2 \left(e^{-v} -1 + ve^{-v}Ei(v)   \right) \Big \} \nonumber \\
&&\sim e^{\gamma_E} e^{-v} v \frac{\ln{u}}{u} + O(u^{-1})
\end{eqnarray}

One then analyses the integrals involving $F_R^{1 \text{noneq}}(v,u)$~:
\begin{eqnarray}
&&\frac{2 T\tau}{q^2} \frac{v}{u-1} \int_0^1 ds
\Big \{ F_R^0\left(\frac{u-s}{u-1}v\right)
F^{1\text{noneq}}_R\left(\frac{1-s}{u-1}v,\frac{1}{s}\right)  \nonumber \\
&&+ F^0_R\left(\frac{1-s}{u-1}v\right)
F^{1\text{noneq}}_R\left(\frac{u-s}{u-1}v,\frac{u}{s}\right) \Big \} \\
&&\sim \frac{2 T\tau}{q^2} \frac{1}{u} \Big \{ vF_R^0(v)\int_0^1 ds
{F}_R^{1\text{noneq}}(0,\frac{1}{s}) \nonumber \\
&&+ vF_R^0(0) {F}_R^{1\text{noneq}}(v,\infty) +
O(u^{-1})\Big \}
\end{eqnarray}
and the remaining integral in (\ref{exprF1C}) where we perform the
natural change of variable $\alpha = \sqrt{t_1}$, $\beta = \sqrt{t_2}$
\begin{eqnarray}
&&\frac{e^{\gamma_E} T_c
\tau}{2 q^2}e^{-v\frac{1+u}{u-1}}\int_0^{\frac{vu}{u-1}}
dt_1\int_0^{\frac{v}{u-1}}dt_2 \frac
{(\sqrt{t_1}-\sqrt{t_2})e^{t_1+t_2}}{\sqrt{t_1 t_2} (\sqrt{t_1} +
\sqrt{t_2})} \nonumber \\
&& = \frac{2 e^{\gamma_E}T_c
\tau}{q^2}e^{-v\frac{1+u}{u-1}}\int_{\sqrt{\frac{v}{u-1}}}^{
\sqrt{\frac{vu}{u-1}}} d\alpha
\int_0^{\sqrt{\frac{v}{u-1}}} d\beta e^{\alpha^2 + \beta^2} \nonumber \\
&& - \frac{4 e^{\gamma_E} T_c
\tau}{q^2}e^{-v\frac{1+u}{1-u}}\int_{\sqrt{\frac{v}{u-1}}}^{
\sqrt{\frac{vu}{u-1}}} d\alpha
\int_0^{\sqrt{\frac{v}{u-1}}} d\beta e^{\alpha^2 + \beta^2}
\frac{\beta}{\alpha+\beta} \nonumber
\end{eqnarray}
The first double integral can be performed exactly
\begin{eqnarray}\label{D_large_u_1}
&&\frac{2 e^{\gamma_E}T_c
\tau}{q^2}e^{-v\frac{1+u}{1-u}}\int_{\sqrt{\frac{v}{u-1}}}^{
\sqrt{\frac{vu}{u-1}}} d\alpha
\int_0^{\sqrt{\frac{v}{u-1}}} d\beta e^{\alpha^2 + \beta^2} \nonumber \\
&&= \frac{\pi e^{\gamma_E}T_c
\tau}{2 q^2}e^{-v\frac{1+u}{u-1}}
\left(\Erfi{\sqrt{\frac{uv}{u-1}}}-\Erfi{\sqrt{\frac{v}{u-1}}} \right)
\nonumber \\
&& \times \Erfi{\sqrt{\frac{v}{u-1}}} \nonumber \\
&& \sim \frac{\sqrt{\pi}  e^{\gamma_E} T_c \tau}{q^2}e^{-v} \sqrt{\frac{v}{u}}
\Erfi{\sqrt{v}} + O(u^{-1})
\end{eqnarray}
And we expand the
second one in the following way
\begin{eqnarray}\label{D_large_u_2}
&& - \frac{4 e^{\gamma_E} T_c
\tau}{q^2}e^{-v\frac{1+u}{u-1}}\int_{\sqrt{\frac{v}{u-1}}}^{
\sqrt{\frac{vu}{u-1}}} d\alpha
\int_0^{\sqrt{\frac{v}{u-1}}} d\beta e^{\alpha^2 + \beta^2}
\frac{\beta}{\alpha+\beta} \nonumber \\
&&= - 2\frac{2 e^{\gamma_E} T_c
\tau}{q^2}e^{-v\frac{1+u}{u-1}}\int_{\sqrt{\frac{v}{u-1}}}^{
\sqrt{\frac{vu}{u-1}}} d\alpha
e^{\alpha^2} \Big(\frac{1}{2 \alpha} \frac{v}{u} \nonumber \\
&&+ \sum_{n > 2}
\left(\frac{v}{u}\right)^{n/2}\frac{a_n}{\alpha^{n-1}}(1+O(\alpha))
\Big)  \nonumber \\
&&\sim -\frac{2 e^{\gamma_E} T_c \tau}{q^2}e^{-v}\frac{v}{u}
\int_{\sqrt{\frac{v}{u}}}^{\sqrt{v}} \frac{d\alpha}{\alpha}e^{\alpha^2} +
O(u^{-1}) \nonumber \\
&&\sim -\frac{ e^{\gamma_E T_c} \tau}{q^2}e^{-v}\frac{v}{u} \ln{u} +
O(u^{-1})
\end{eqnarray}

Finally,
(\ref{Fc0_large_u},\ref{log_u_term},\ref{D_large_u_1},\ref{D_large_u_2})
lead to the asymptotic following form for
$F_C(v,u)$  in the limit $u
\to \infty$, $v$ fixed,
\begin{eqnarray}\label{corr_vfi_uinf}
&&\lim_{v\to \infty} F_C(v,u) = \frac{2ve^{-v}}{u} +
 \tau \frac{F^1_{C,\infty}(v)}{\sqrt{u}} +
 O(u^{-2},\tau u^{-1},\tau^2) \nonumber \\
&&F^1_{C,\infty}(v) = e^{\gamma_E} e^{-v} \sqrt{\pi v} \Erfi{\sqrt{v}}
\nonumber \\
\end{eqnarray}
notice that the subdominant terms in $\ln{u}/u$ cancel between
(\ref{log_u_term}) and (\ref{D_large_u_2}) so that the leading
corrections are of order $u^{-1}$.  (\ref{corr_vfi_uinf}) gives the
asymptotic behavior given in the text (\ref{FC_large_u}).

\subsubsection{Expansion at large $v$, $u$ fixed.}

In this limit, the terms in the four first lines of (\ref{exprF1C})
decay exponentially in this limit. The fifth line however (which
corresponds to the equilibrium contribution) decays like a power
law. Indeed, using the large $v$ behavior of $Ei(v) \sim e^{v}(1/v +
1/v^2 + 0(v^{-3}))$ one has
\begin{eqnarray}\label{FC_large_v_eq}
e^{\gamma_E} (e^{-v} -1 + ve^{-v}Ei(v)) \sim \frac {e^{\gamma_E}}{v}
+ O (v^{-2})
\end{eqnarray}
We now analyse the behavior of the terms involving $F^{1\text{noneq}}_R$
in (\ref{exprF1C}). Using the large $v$ behavior of
$F^{1\text{noneq}}_R(v,u)$ (\ref{FRnoneq_large_v}), one has
\begin{eqnarray}
&&2 \frac{v}{u-1} \int_0^1 ds \Big \{
F_R^0\left(\frac{u-s}{u-1}v\right)
{F}_R^{1\text{noneq}}\left(\frac{1-s}{u-1}v,\frac{1}{s} \right)  \nonumber \\
&& + F^0_R\left(\frac{1-s}{u-1}v\right)
{F}_R^{1\text{noneq}}\left(\frac{u-s}{u-1}v,\frac{u}{s}\right) \Big \}
\\
&&\sim \frac{v}{u-1} \int_0^1 ds \Big \{
e^{-\left(\frac{u-s}{u-1}v\right)} \frac{(u-1)^2}{v^2 (1-s)^2}
\frac{(\sqrt{1/s}-1)^2}{\sqrt{1/s}} \nonumber \\
&&+
e^{-\left(\frac{1-s}{u-1}v\right)}\frac{(u-1)^2}{v^2 (u-s)^2}
\frac{(\sqrt{u/s}-1)^2}{\sqrt{u/s}}
\Big \} \nonumber
\end{eqnarray}
Notice first on this expression that we are left with convergent
integrals over $s$. Moreover, in the large $v$ limit, due to the
exponential prefactors the first term decays also exponentially (for
$u >1$), and the second one is dominated by $s=1$, which leads to a
power law decay
\begin{eqnarray}
&&\sim \frac{1}{v(u-1)} \frac{(\sqrt{u}-1)^2}{\sqrt{u}} \int_0^1 ds
e^{-\left(\frac{1-s}{u-1}v\right)} \sim O(v^{-2}) \nonumber
\end{eqnarray}
We are now left with the double integral in (\ref{exprF1C}), which is
dominated - also due to the exponential prefactor - by $t_1 \sim
vu/(u-1)$ and $t_2 \sim v/(u-1)$. Therefore to get the leading
behavior, we substitute $t_1$ and $t_2$ by these values
in the integrand (except of course in the exponential
$e^{t_1+t_2}$). This yields
\begin{eqnarray}
&&  \frac{1}{2} e^{\gamma_E} e^{-v\frac{1+u}{u-1}}\int_0^{\frac{vu}{u-1}}
dt_1\int_0^{\frac{v}{u-1}}dt_2 \frac
{\sqrt{t_1}-\sqrt{t_2}e^{t_1+t_2}}{\sqrt{t_1 t_2} (\sqrt{t_1} +
\sqrt{t_2})} \nonumber \\
&& = \frac{1}{2} e^{\gamma_E} \frac{(\sqrt{u}-1)^2}{\sqrt{u}} \frac{1}{v}
e^{-v\frac{u+1}{u-1}} \int_0^{\frac{vu}{u-1}} dt_1 e^{t_1}
\int_0^{\frac{v}{u-1}} dt_2 e^{t_2} \nonumber \\
&&+ O(\frac{1}{v^2}) \nonumber \\
&& =  \frac{1}{2}  e^{\gamma_E} \frac{(\sqrt{u}-1)^2}{\sqrt{u}} \frac{1}{v} + O(\frac{1}{v^2})
\end{eqnarray}
which together with the other term in $v^{-1}$ (\ref{FC_large_v_eq})
yields (\ref{FC_large_v}) in the text.

\subsubsection{The limit of vanishing momentum.}

To obtain the limit of vanishing momentum $\tilde{q} \to 0$ of the
correlation function, we look at the behavior of $F_C(v,u)$ when $v
\to 0$, up to order $O(v)$ terms (due to the $q^{-2}$ prefactor in
(\ref{exprF1C})). This is done in the following way
\begin{eqnarray}\label{Cq=0part1}
&&2 e^{\gamma_E} e^{-v \frac{1+u}{u-1}}
\left(Ei(\frac{2v}{u-1}) - \ln{\left(\frac{2v}{u-1} \right)}
-\gamma_E \right)  \\
&& + \frac{1}{2}  e^{\gamma_E}  \Big \{ e^{-\frac{v(u+1)}{u-1}} \Bigg ( -4
- 2\frac{uv}{u-1}Ei\left(\frac{uv}{u-1}\right) \nonumber \\
&&-2\frac{v}{u-1}Ei\left(\frac{v}{u-1}\right) \Bigg)
+2(e^{-\frac{vu}{u-1}} + e^{-\frac{v}{u-1}}) \nonumber \\
&& + 2 \left(e^{-v} -1 + ve^{-v}Ei(v)   \right) \Big \} \nonumber \\
&&= \frac{e^{\gamma_E} v}{u-1}(6 - 2 \ln{ve^{\gamma_E}} -
u\ln{u} + (u+1)\ln{(u-1)} \nonumber \\
&&+ O(v^2) \nonumber
\end{eqnarray}

Then using the expression of $F_R^{1\text{noneq}}(0,u)$
(\ref{FR1noneq_q=0}), one has
\begin{eqnarray}\label{Cq=0part2}
&& 2 \frac{v}{u-1} \int_0^1 ds \Big \{
F_R^0\left(\frac{u-s}{u-1}v\right)
{F}_R^{1\text{noneq}}\left(\frac{1-s}{u-1}v,\frac{1}{s} \right)  \nonumber \\
&& + F^0_R\left(\frac{1-s}{u-1}v\right)
{F}_R^{1\text{noneq}}\left(\frac{u-s}{u-1}v,\frac{u}{s}\right) \Big \}
 \\
&&= \frac{2v}{u-1} \int_0^1 ds \Big \{
{F}_R^{1\text{noneq}}\left(0,\frac{1}{s} \right)   +
{F}_R^{1\text{noneq}}\left(0,\frac{u}{s}\right) \Big \} \nonumber \\
&&+ O(v^2) \nonumber \\
&&= 4 e^{\gamma_E}\frac{v}{u-1} \int_0^1
\ln{\frac{1+\sqrt{s}}{2}} +  \ln{\frac{1+\sqrt{\frac{s}{u}}}{2}} +
O(v^2) \nonumber \\
&& = \frac{4 e^{\gamma_E} v}{u-1} (\sqrt{u} -
(u-1)\ln{(1+\frac{1}{\sqrt{u}})} - 2 \ln{4} ) + O(v^2) \nonumber
\end{eqnarray}
To treat the double integral in the last line of (\ref{exprF1C}) we
come back to the variables $\tilde{t},\tilde{t'},\tilde{q}$
\begin{eqnarray}
&&\frac{1}{2}  \frac{T}{{\tilde{q}}^2}\tau e^{\gamma_E}
e^{-v\frac{1+u}{u-1}}\int_0^{\frac{vu}{u-1}}
dt_1\int_0^{\frac{v}{u-1}}dt_2 \frac
{(\sqrt{t_1}-\sqrt{t_2})e^{t_1+t_2}}{\sqrt{t_1 t_2} (\sqrt{t_1} +
\sqrt{t_2})} \nonumber \\
&& = \frac{1}{2}  T \tau e^{\gamma_E}
e^{-\tilde{q}^2(\tilde{t}+\tilde{t'})}\int_0^{\tilde{t}}
dt_1\int_0^{\tilde{t'}}dt_2 \frac
{(\sqrt{t_1}-\sqrt{t_2})e^{q^2(t_1+t_2)}}{\sqrt{t_1 t_2} (\sqrt{t_1} +
\sqrt{t_2})} \nonumber \\
\end{eqnarray}
Under this form, the limit $\tilde{q} \to$ is very simply obtained:
\begin{eqnarray}\label{Cq=0part3}
&&\lim_{\tilde{q} \to 0} \frac{1}{2}  \frac{T}{{\tilde{q}}^2}\tau e^{\gamma_E}
e^{-v\frac{1+u}{u-1}}\int_0^{\frac{vu}{u-1}}
dt_1\int_0^{\frac{v}{u-1}}dt_2 \frac
{(\sqrt{t_1}-\sqrt{t_2})e^{t_1+t_2}}{\sqrt{t_1 t_2} (\sqrt{t_1} +
\sqrt{t_2})} \nonumber \\
&& =\frac{1}{2}  T \tau e^{\gamma_E}\int_0^{\tilde{t}}
dt_1\int_0^{\tilde{t'}}dt_2 \frac
{\sqrt{t_1}-\sqrt{t_2}}{\sqrt{t_1 t_2} (\sqrt{t_1} +
\sqrt{t_2})} \nonumber \\
&& = 2 T \tau e^{\gamma_E} \tilde{t'}\left( (u-1)\ln{(1+\sqrt{u})}
-\frac{u}{2}\ln{u} \right)
\end{eqnarray}
Finally (\ref{Cq=0part1}, \ref{Cq=0part2}, \ref{Cq=0part3}), together
with (\ref{exprF1C}) and the complete expression of the correlation
function ${\cal C}^{\tilde{q}}_{\tilde{t}\tilde{t'}}$
(\ref{pert_corr}) lead to
\begin{eqnarray}
&&{\cal C}^{\tilde{q}=0}_{\tilde{t}\tilde{t'}} = 2T_c\tilde{t'} \Big( 1+\tau
 - \frac{z-2}{2} (\ln{(\tilde{t} - \tilde{t'})} + \gamma_E) + \theta
\ln{\frac{\tilde{t}}{\tilde{t'}}} \nonumber \\
&&+ \tau F_C^{\text{diff}1}\left(\frac{\tilde{t}}{\tilde{t'}}\right)
 \Big) \nonumber \\
&& F_C^{\text{diff}1}(u) = \frac{1}{2}  e^{\gamma_E} \Big( 4 \sqrt{u} + (u+1)
\ln{(u-1)} \nonumber \\
&&- 2(u-1)\ln{(1+\sqrt{u})} - 2 \ln{u} + 6 - 8\ln{4} \Big)
\end{eqnarray}
where we have used $F_C^0(v,u) = v/(u-1) + O(v^2)$, $\partial_v
F_C^0(v,u) = 1/(u-1) + O(v)$ and $v/(u-1) = \tilde{q}^2 \tilde{t'}$ :
this gives the scaling form (\ref{Cq=0}) given in the text.

\begin{acknowledgments}
We thank Pascal Chauve for useful discussions on the ERG method at the
earliest stage 
of this work. We thank Leticia Cugliandolo, Daniel Dominguez, Antoine
Georges and 
Alejandro Kolton for discussions and pointing out references. 
\end{acknowledgments}

\end{document}